\documentclass[journal,draftcls,onecolumn,12pt,romanappendices]{IEEEtran-v17}
\pdfoutput=1

\usepackage{xspace}
\usepackage{url}
\usepackage[cmex10]{amsmath}
\usepackage{bbm}
\usepackage{graphicx}

\usepackage{fancyref} 
\usepackage{amsmath}
\usepackage{amssymb}
\usepackage{mathtools}
\usepackage{dsfont}
\usepackage{footnote}
\usepackage[nosort]{cite}
\usepackage[british]{babel}

\usepackage{color}
\definecolor{links}{rgb}{0.7,0,0}   
\definecolor{urls}{rgb}{0,0,0.8}    
\definecolor{cites}{rgb}{0,0,0.8}   

\usepackage[colorlinks,hyperindex,linkcolor=links,citecolor=cites,urlcolor=urls]{hyperref}

\usepackage{bm}
\usepackage{upgreek}
\usepackage{amssymb}
\usepackage{amsfonts}
\usepackage{mathrsfs}
\usepackage{xspace}

\makeatletter
\def\amsbb{\use@mathgroup \M@U \symAMSb}
\makeatother

\newcommand{\lefto}{\mathopen{}\left}
\newcommand{\safemath}[2]{\newcommand{#1}{\ensuremath{#2}\xspace}}
\safemath{\opE}{\amsbb{E}}
\newcommand{\Ex}[2]{\ensuremath{\amsbb{E}_{#1}\mathopen{}\left[#2\right]}} 	
\safemath{\prob}{\amsbb{P}}
\safemath{\bigO}{\mathcal{O}}
\safemath{\littleo}{\mathit{o}}

\safemath{\extendreal}{\overline{\realset}}

\newcommand{\tp}[1]{\ensuremath{#1^{\mathrm{T}}}} 		
\newcommand{\herm}[1]{\ensuremath{#1^{\mathrm{H}}}} 	
\newcommand{\tr}{\mathrm{tr}}

\newcommand{\fnorm}[1]{\ensuremath{\left\|#1\right\|_{\mathsf{F}}}}

\newtheorem{thm}{Theorem}
\newtheorem{lemma}[thm]{Lemma}
\newtheorem{dfn}{Definition}
\newtheorem{rem}{Remark}

\newtheorem{cor}[thm]{Corollary}

\safemath{\matA}{\mathsf{A}}
\safemath{\matB}{\mathsf{B}}
\safemath{\matC}{\mathsf{C}}
\safemath{\matD}{\mathsf{D}}
\safemath{\matE}{\mathsf{E}}
\safemath{\matF}{\mathsf{F}}
\safemath{\matG}{\mathsf{G}}
\safemath{\matH}{\mathsf{H}}
\safemath{\matI}{\mathsf{I}}
\safemath{\matJ}{\mathsf{J}}
\safemath{\matK}{\mathsf{K}}
\safemath{\matL}{\mathsf{L}}
\safemath{\matM}{\mathsf{M}}
\safemath{\matN}{\mathsf{N}}
\safemath{\matO}{\mathsf{O}}
\safemath{\matP}{\mathsf{P}}
\safemath{\matQ}{\mathsf{Q}}
\safemath{\matR}{\mathsf{R}}
\safemath{\matS}{\mathsf{S}}
\safemath{\matT}{\mathsf{T}}
\safemath{\matU}{\mathsf{U}}
\safemath{\matV}{\mathsf{V}}
\safemath{\matW}{\mathsf{W}}
\safemath{\matX}{\mathsf{X}}
\safemath{\matY}{\mathsf{Y}}
\safemath{\matZ}{\mathsf{Z}}
\safemath{\matSigma}{\mathsf{\Sigma}}
\safemath{\matPhi}{\mathsf{\Phi}}
\safemath{\matLambda}{\mathsf{\Lambda}}
\safemath{\matDelta}{\mathsf{\Delta}}

\safemath{\randveca}{\bm{A}}
\safemath{\randvecb}{\bm{B}}
\safemath{\randvecc}{\bm{C}}
\safemath{\randvecd}{\bm{D}}
\safemath{\randvece}{\bm{E}}
\safemath{\randvecf}{\bm{F}}
\safemath{\randvecg}{\bm{G}}
\safemath{\randvech}{\bm{H}}
\safemath{\randveci}{\bm{I}}
\safemath{\randvecj}{\bm{J}}
\safemath{\randveck}{\bm{K}}
\safemath{\randvecl}{\bm{L}}
\safemath{\randvecm}{\bm{M}}
\safemath{\randvecn}{\bm{N}}
\safemath{\randveco}{\bm{O}}
\safemath{\randvecp}{\bm{P}}
\safemath{\randvecq}{\bm{Q}}
\safemath{\randvecr}{\bm{R}}
\safemath{\randvecs}{\bm{S}}
\safemath{\randvect}{\bm{T}}
\safemath{\randvecu}{\bm{U}}
\safemath{\randvecv}{\bm{V}}
\safemath{\randvecw}{\bm{W}}
\safemath{\randvecx}{\bm{X}}
\safemath{\randvecy}{\bm{Y}}
\safemath{\randvecz}{\bm{Z}}

\safemath{\randvecLambda}{\bm{\Lambda}}

\safemath{\randmatA}{\amsbb{A}}
\safemath{\randmatB}{\amsbb{B}}
\safemath{\randmatC}{\amsbb{C}}
\safemath{\randmatD}{\amsbb{D}}
\safemath{\randmatE}{\amsbb{E}}
\safemath{\randmatF}{\amsbb{F}}
\safemath{\randmatG}{\amsbb{G}}
\safemath{\randmatH}{\amsbb{H}}
\safemath{\randmatI}{\amsbb{I}}
\safemath{\randmatJ}{\amsbb{J}}
\safemath{\randmatK}{\amsbb{K}}
\safemath{\randmatL}{\amsbb{L}}
\safemath{\randmatM}{\amsbb{M}}
\safemath{\randmatN}{\amsbb{N}}
\safemath{\randmatO}{\amsbb{O}}
\safemath{\randmatP}{\amsbb{P}}
\safemath{\randmatQ}{\amsbb{Q}}
\safemath{\randmatR}{\amsbb{R}}
\safemath{\randmatS}{\amsbb{S}}
\safemath{\randmatT}{\amsbb{T}}
\safemath{\randmatU}{\amsbb{U}}
\safemath{\randmatV}{\amsbb{V}}
\safemath{\randmatW}{\amsbb{W}}
\safemath{\randmatX}{\amsbb{X}}
\safemath{\randmatY}{\amsbb{Y}}
\safemath{\randmatZ}{\amsbb{Z}}
\safemath{\randmatSigma}{\mathbb{\Sigma}}
\safemath{\randmatPhi}{\mathbb{\Phi}}

\safemath{\pdff}{f}
\safemath{\pdfp}{p}
\safemath{\pdfq}{q}

\safemath{\cdfF}{F}
\safemath{\cdfP}{P}
\safemath{\cdfQ}{Q}

\safemath{\veca}{\bm{a}}
\safemath{\vecb}{\bm{b}}
\safemath{\vecc}{\bm{c}}
\safemath{\vecd}{\bm{d}}
\safemath{\vece}{\bm{e}}
\safemath{\vecf}{\bm{f}}
\safemath{\vecg}{\bm{g}}
\safemath{\vech}{\bm{h}}
\safemath{\veci}{\bm{i}}
\safemath{\vecj}{\bm{j}}
\safemath{\veck}{\bm{k}}
\safemath{\vecl}{\bm{l}}
\safemath{\vecm}{\bm{m}}
\safemath{\vecn}{\bm{n}}
\safemath{\veco}{\bm{o}}
\safemath{\vecp}{\bm{p}}
\safemath{\vecq}{\bm{q}}
\safemath{\vecr}{\bm{r}}
\safemath{\vecs}{\bm{s}}
\safemath{\vect}{\bm{t}}
\safemath{\vecu}{\bm{u}}
\safemath{\vecv}{\bm{v}}
\safemath{\vecw}{\bm{w}}
\safemath{\vecx}{\bm{x}}
\safemath{\vecy}{\bm{y}}
\safemath{\vecz}{\bm{z}}

\safemath{\veclambda}{\bm{\lambda}}
\safemath{\vecpi}{\bm{\pi}}
\safemath{\vecsigma}{\bm\sigma}              			

\safemath{\setA}{\mathcal{A}}
\safemath{\setB}{\mathcal{B}}
\safemath{\setC}{\mathcal{C}}
\safemath{\setD}{\mathcal{D}}
\safemath{\setE}{\mathcal{E}}
\safemath{\setF}{\mathcal{F}}
\safemath{\setG}{\mathcal{G}}
\safemath{\setH}{\mathcal{H}}
\safemath{\setI}{\mathcal{I}}
\safemath{\setJ}{\mathcal{J}}
\safemath{\setK}{\mathcal{K}}
\safemath{\setL}{\mathcal{L}}
\safemath{\setM}{\mathcal{M}}
\safemath{\setN}{\mathcal{N}}
\safemath{\setO}{\mathcal{O}}
\safemath{\setP}{\mathcal{P}}
\safemath{\setQ}{\mathcal{Q}}
\safemath{\setR}{\mathcal{R}}
\safemath{\setS}{\mathcal{S}}
\safemath{\setT}{\mathcal{T}}
\safemath{\setU}{\mathcal{U}}
\safemath{\setV}{\mathcal{V}}
\safemath{\setW}{\mathcal{W}}
\safemath{\setX}{\mathcal{X}}
\safemath{\setY}{\mathcal{Y}}
\safemath{\setZ}{\mathcal{Z}}
\safemath{\emptySet}{\varnothing}

\safemath{\veczero}{\mathbf{0}} 

\safemath{\diag}{\mathrm{diag}}

\safemath{\jpg}{\mathcal{CN}}			
\safemath{\complexset}{\amsbb{C}}
\safemath{\realset}{\amsbb{R}}
\safemath{\natunum}{\mathbb{N}}
\safemath{\posrealset}{\realset_{+}}
\safemath{\integerset}{\amsbb{N}}

\newcommand{\given}{\,\vert\,}				
\safemath{\define}{\triangleq}			

\usepackage{bm}
\usepackage{upgreek}
\usepackage{amssymb}
\usepackage{amsfonts}
\usepackage{mathrsfs}
\usepackage{xspace}

\safemath{\mi}{I}
\safemath{\difent}{h}

\safemath{\constrm}{\mathrm{const}}

\safemath{\expdist}{\mathrm{Exp}}

\safemath{\NonnegReal}{\mathbb{R}^{+}}
\safemath{\re}{\mathrm{re}}
\safemath{\Real}{\mathrm{Re}} 
\safemath{\gradient}{\nabla}
\safemath{\genericpdf}{f}

\safemath{\surform}{\mathrm{d}S}

\safemath{\nocsi}{\mathrm{no}}
\safemath{\csir}{\mathrm{CSIR}}

\safemath{\bl}{n} 

\safemath{\error}{\epsilon} 

\safemath{\cohtime}{n_\mathrm{c}}
\safemath{\rxant}{m_\mathrm{r}}
\safemath{\txant}{m_\mathrm{t}}

\safemath{\snr}{\rho}
\safemath{\const}{k}

\safemath{\spanm}{\mathrm{span}}

\def\sumlog{l_a}
\def\weightsum{L}
\def\pdfa{f}

\def\infden{\imath}

\safemath{\eb}{E_{\mathrm{b}}}
\safemath{\ebr}{\eb^{\mathrm{r}}}

\begin{document}
\IEEEoverridecommandlockouts

\title{Minimum Energy to Send $k$ Bits Over Multiple-Antenna Fading Channels}

\author{Wei Yang,~\IEEEmembership{Member,~IEEE}, Giuseppe Durisi,~\IEEEmembership{Senior Member,~IEEE},\\ and Yury Polyanskiy,~\IEEEmembership{Senior Member,~IEEE}
\thanks{This work was supported by the Swedish Research Council (VR) under grant no. 2012-4571, and by the National Science Foundation CAREER award under grant agreement CCF-12-53205. The material of this paper was presented in part at the IEEE Information Theory Workshop (ITW), Jerusalem, Israel, April 2015 and at the IEEE International Symposium on Information Theory (ISIT), Hong Kong, China, June 2015.}
\thanks{W. Yang and G. Durisi are with the Department of Signals and Systems, Chalmers University of Technology, 41296,  Gothenburg, Sweden (e-mail: \{ywei, durisi\}@chalmers.se).}
\thanks{Y. Polyanskiy is with the Department of Electrical Engineering and Computer Science, MIT, Cambridge, MA, 02139 USA (e-mail: yp@mit.edu).}
}

\maketitle
\begin{abstract}
%

This paper investigates the minimum energy required to transmit $k$ information bits  with a given reliability over a multiple-antenna Rayleigh block-fading channel, with and without channel state information (CSI) at the receiver. No feedback is assumed.
It is well known that the ratio between the minimum energy per bit and the noise level converges to $-1.59$ dB as $k$ goes to infinity, regardless of whether CSI is available at the receiver or not.
This paper shows that lack of CSI at the receiver causes a slowdown in the speed of convergence to $-1.59$ dB as $k\to\infty$ compared to the case of perfect receiver CSI.
Specifically, we show that, in the no-CSI case, the gap to $-1.59$ dB is proportional to $((\log k) /k)^{1/3}$, whereas when perfect CSI is available at the receiver, this gap is proportional to  $1/\sqrt{k}$.
In both cases, the gap to $-1.59$ dB is independent of the number of transmit antennas and of the channel's coherence time.
Numerically, we observe that, when the receiver is equipped with a single antenna, to achieve an energy per bit of $ - 1.5$ dB in the no-CSI case, one needs to transmit at least $7\times 10^7$ information bits, whereas $6\times 10^4$ bits suffice for the case of perfect CSI at the receiver.

\end{abstract}

\section{Introduction}
A classic result in information theory is that, for a wide range of channels including AWGN channels and fading channels, the minimum  energy per  bit $\eb$ required for reliable communication satisfies~\cite{shannon49,verdu02-06}
\begin{IEEEeqnarray}{rCl}
\frac{\eb}{N_0}_{\min} = \log_e 2 =- 1.59\, \mathrm{dB}.
\label{eq:intro-1.6db}
\end{IEEEeqnarray}
Here, $N_0$ is the noise power per complex degree of freedom.
%
%
%
%
For fading channels, \eqref{eq:intro-1.6db} holds regardless of whether the instantaneous fading realizations are known to the receiver or not~\cite[Th.~1]{verdu02-06},\cite{kennedy69}.\footnote{Knowledge of the fading realizations at the transmitter may improve~\eqref{eq:intro-1.6db}, because it enables the transmitter to signal on the channel maximum-eigenvalue eigenspace~\cite{verdu02-06}.}

The expression in~\eqref{eq:intro-1.6db} is asymptotic in several aspects:
\begin{itemize}
\item the blocklength $n$ of each codeword is infinite;
\item the number of information bits $k$, or equivalently, the number of messages $M =2^k$  is infinite;
\item the error probability $\error$ vanishes;
\item the total energy $E$ is infinite;
\item $E/n$ vanishes.
\end{itemize}
For many channels, the limit in~\eqref{eq:intro-1.6db} does not change if we allow the error probability to be positive.
However, keeping any of the other parameters fixed results in a backoff from~\eqref{eq:intro-1.6db}~\cite{verdu02-06,gallager88,verdu90-09,zheng07-03,polyanskiy10-05,polyanskiy11-08b}.
%

%
In this paper, we study the maximum number of information bits~$k$ that can be transmitted with a finite energy~$E$ and a fixed error probability $\error > 0$ over a multiple-input multiple-output (MIMO) Rayleigh block-fading channel, when there is no constraint on the  blocklength~$n$.
Equivalently, we determine the minimum energy~$E$ required to transmit~$k$ information bits with error probability~$\error$.
We consider two scenarios:
\begin{enumerate}
\item neither the transmitter nor the receiver have \emph{a priori} channel state information (CSI);
\item perfect CSI is available at the receiver (CSIR) and no CSI is available at the transmitter.
\end{enumerate}
Throughout the paper, we shall refer to these two scenarios as no-CSI case and perfect-CSIR case, respectively.

\paragraph*{Related work}
%
%
For nonfading AWGN channels with unlimited blocklength, Polyanskiy, Poor, and Verd\'{u}~\cite{polyanskiy11-08b} showed that the maximum number of codewords $M^*(E,\error) $ that can be transmitted with energy $E$ and error probability $\error$ satisfies\footnote{Unless otherwise indicated, the $\log$ and the $\exp$ functions are taken with respect to an arbitrary fixed base.}
\begin{IEEEeqnarray}{rCl}
\log M^*(E,\error) &=& \frac{E}{N_0} \log e - \sqrt{\frac{2E}{N_0}}Q^{-1}(\error)\log e
+ \frac{1}{2} \log \frac{E}{N_0} + \bigO(1),\quad E\to\infty.
\label{eq:intro-second-order-awgn}\IEEEeqnarraynumspace
\end{IEEEeqnarray}
Here, $Q^{-1}(\cdot)$ denotes the inverse of the Gaussian $Q$-function.
The first term on the right-hand side (RHS) of~\eqref{eq:intro-second-order-awgn} gives the $-1.59$ dB limit.
The second term captures the penalty due to the stochastic variations of the channel.
This term plays the same role as the \emph{channel dispersion} in finite-blocklength analyses~\cite{polyanskiy10-05,tan14}.
In terms of the minimum energy per bit $\eb^*(k,\error)$ necessary to transmit $k$ bits with error probability $\error$,~\eqref{eq:intro-second-order-awgn} implies that, for large~$E$,
\begin{IEEEeqnarray}{rCl}
\frac{\eb^*(k,\error)}{N_0} \approx \log_e 2 + \sqrt{\frac{2 \log_e 2}{k}}Q^{-1}(\error) 
\label{eq:minimum-energy-per-bit-awgn}
\end{IEEEeqnarray}
i.e., that the gap to $-1.59$ dB is proportional to $1/\sqrt{k}$.
The asymptotic expansion~\eqref{eq:intro-second-order-awgn} is established in~\cite{polyanskiy11-08b} by showing that in the limit $E\to\infty$ a nonasymptotic achievability bound and a nonasymptotic converse bound match up to third order.
The achievability bound is obtained by computing the error probability under maximum-likelihood decoding of a codebook consisting of $M$ orthogonal codewords (e.g., uncoded $M$-ary pulse-position modulation (PPM)). The converse bound follows from the meta-converse theorem~\cite[Th.~27]{polyanskiy10-05} with auxiliary distribution chosen equal to the noise distribution.
Kostina, Polyanskiy, and Verd\'{u}~\cite{kostina15-01} generalized~\eqref{eq:intro-second-order-awgn} to the setting of joint source and channel coding, and characterized the minimum energy required to reproduce~$k$ source samples with a given fidelity after transmission over an AWGN channel.

Moving to fading channels,  for the case of no CSI, \emph{flash signalling}~\cite[Def.~2]{verdu02-06} (i.e., peaky signals)  must be used to reach the $-1.59$ dB limit~\cite{verdu02-06}.
In the presence of a peak-power constraint,~\eqref{eq:intro-1.6db} can not be achieved~\cite{telatar00-04,medard02-04,sethuraman05-09,durisi10-01}.
Verd\'{u}~\cite{verdu02-06} studied the rate of convergence of the minimum energy per bit to $-1.59$~dB as the spectral efficiency vanishes.
He showed that, differently from the perfect-CSIR case, in the no-CSI case the $-1.59$ dB limit is approached with \emph{zero wideband slope}. Namely, the slope of the spectral-efficiency versus energy-per-bit function at $-1.59$ dB is zero.
 This implies that operating close to the $-1.59$ dB limit is very expensive in terms of bandwidth in the no-CSI case.
For the scenario of finite blocklength $n$, fixed energy budget~$E$, and fixed probability of error~$\error$, bounds and approximations on the maximum channel coding rate over fading channels (under various CSI assumptions) are reported in~\cite{polyanskiy11-08a,yang12-09,hoydis15-12,yang14-07a,collins14-07,yang15-09,durisi16-02}.
%
%

%
%
%

%

\paragraph*{Contributions} Focusing on the regime of unlimited blocklength, but finite energy~$E$, and finite error probability~$\error$, we provide upper and lower bounds on the maximum number of codewords $M^*(E,\error)$ that can be transmitted over an $\txant \times \rxant$ MIMO Rayleigh block-fading channel with channel's coherence interval of $\cohtime$ symbols.
For the no-CSI case, we show that for every $\error\in(0,1/2)$
\begin{IEEEeqnarray}{rCl}
\log M^*(E,\error) &=& \frac{\rxant E}{N_0} \log e - V_0\cdot \left(\frac{\rxant E }{N_0}Q^{-1}(\error)\right)^{2/3} \left(\log\frac{\rxant E}{N_0}\right)^{1/3} + \bigO\lefto(\frac{E^{2/3} \log\log E}{(\log E)^{2/3}}\right),\IEEEeqnarraynumspace \notag\\
&&\hfill \quad  E\to\infty \quad\IEEEeqnarraynumspace
\label{eq:logM-E23-intro}
\end{IEEEeqnarray}
where
\begin{equation}
V_0 = \bigg(12^{-1/3} + \Big(\frac{2}{3}\Big)^{1/3}\bigg)\left(\log e\right)^{2/3}.
\label{eq:def-V0-main-thm}
\end{equation}
Note that the asymptotic expansion~\eqref{eq:logM-E23-intro} does not depend on the number of transmit antennas $\txant$ and the channel's coherence interval $\cohtime$.
The fact that the first term does not depend on $\cohtime$ and $\txant$ follows directly from~\cite[Eq.~(52)]{verdu02-06} by noting that an $\txant\times\rxant$ block-fading MIMO channel with coherence interval $\cohtime$ is equivalent to an $\txant\cohtime \times \rxant\cohtime$ memoryless MIMO fading channel with block-diagonal channel matrix~\cite[p.~1339]{verdu02-06}.
Our result~\eqref{eq:logM-E23-intro} shows that the same holds for the second term in the expansion of $\log M^*(E,\error)$ for $E\to\infty$.
In terms of minimum (received) energy per bit $E_{\mathrm{b}}^*(k,\error)$,~\eqref{eq:logM-E23-intro} implies that, for large $E$,\footnote{By considering the received energy per bit instead of the transmit energy per bit, we account for the array gain resulting from the use of multiple receive antennas.}
\begin{IEEEeqnarray}{rCl}
\frac{E_{\mathrm{b}}^*(k,\error)}{N_0} \approx \log_e 2+ V_0 \cdot \left(\frac{\log_e k}{ k}\right)^{1/3}\left(Q^{-1}(\error)\right)^{2/3}(\log_e 2)^{4/3}
\label{eq:min-energy-per-bit-fading}
\end{IEEEeqnarray}
i.e., the gap to~$-1.59$ dB is proportional to~$((\log_e k) / k)^{1/3}$.

We establish~\eqref{eq:logM-E23-intro} by analyzing in the limit $E\to\infty$ an achievability bound and a converse bound. The achievability bound follows from a nonasymptotic extension of Verd\'{u}'s capacity-per-unit-cost achievability scheme~\cite[pp.~1023--1024]{verdu90-09}. This scheme relies on a codebook consisting of the concatenation of uncoded PPM and a repetition code, and on a decoder that performs binary hypothesis testing. The converse bound relies on the meta-converse theorem~\cite[Th.~31]{polyanskiy10-05} with auxiliary distribution chosen as in the AWGN case. The resulting bound involves an optimization over the infinite-dimensional space of input codewords (recall that in our setup there is no constraint on the blocklength $n$).
By exploiting the Gaussianity of the fading process, we show that this infinite-dimensional optimization problem can be reduced to a three-dimensional one.
The tools needed to establish this result are the ones developed by Abbe, Huang, and Telatar~\cite{abbe13-05} to prove Telatar's minimum outage probability conjecture for multiple-input single-output (MISO) Rayleigh-fading channels. Indeed, both problems involve the optimization of quantiles of a weighted convolution of exponential distributions.

The asymptotic analysis of  achievability and converse bounds
reveals the following tension: on the one hand, one would like to make the codewords peaky to overcome lack of channel knowledge; on the other hand, one would like to spread the energy of the codewords uniformly over multiple coherence intervals to mitigate the stochastic variations in the received signal energy due to the fading.

For the case of perfect CSIR, we prove that for every $\error \in (0,1/2)$
\begin{IEEEeqnarray}{rCl}
\log M^*(E,\error) &=& \frac{\rxant E}{N_0} \log e - \sqrt{\frac{2\rxant E}{N_0}}Q^{-1}(\error)\log e
+ \frac{1}{2} \log \frac{\rxant E}{N_0} + \bigO\mathopen{}\big(\sqrt{\log E}\big), \, \,\,E\to\infty.\IEEEeqnarraynumspace
\label{eq:intro-second-order-csir}
\end{IEEEeqnarray}
Note that the asymptotic expansion~\eqref{eq:intro-second-order-csir} is also independent of the number of transmit antennas $\txant$ and the channel's coherence interval $\cohtime$.
Furthermore, apart from an energy normalization resulting from the array gain, this asymptotic expansion coincides with the one given in~\eqref{eq:intro-second-order-awgn} for the AWGN case up to a $\bigO\mathopen{}\big(\sqrt{\log E}\big)$ term. In terms of minimum (received) energy per bit,~\eqref{eq:intro-second-order-csir} implies that~\eqref{eq:minimum-energy-per-bit-awgn} holds also for the perfect-CSIR case.

To establish~\eqref{eq:intro-second-order-csir}, we show that every code for the AWGN channel can be transformed into a code for the MIMO block-memoryless Rayleigh-fading channel having the same probability of error. This is achieved by concatenating the AWGN code with a rate $1/N$ repetition code, by performing maximum ratio combining at the receiver, and then by letting $N\to\infty$. We obtain a converse bound that matches the achievability bound up to third order as $E\to\infty$ by using again the meta-converse theorem and then by optimizing over all input codewords. The asymptotic analysis of the converse bound reveals that spreading the energy of the codewords uniformly across many coherence intervals is necessary to mitigate the stochastic variations in the energy of the received signal due to fading.

In both the no-CSI and the perfect-CSIR case, the asymptotic analysis of the achievability bound is based on a standard application of   Berry-Esseen central-limit theorem (see, e.g.,~\cite[Ch.~XVI.5]{feller70b}). 
The asymptotic analysis of the converse part in both cases is not as straightforward.  
The main difficulty  is that, unlike for discrete memoryless channels and AWGN channels,  we can not directly invoke the central-limit theorem to evaluate the information density, because the central-limit theorem may not hold if the energy of a codeword is concentrated on few of its symbols. 
To solve this problem, we develop new tools that rely explicitly on  the Gaussianity of the fading process.
Specifically, for the no-CSI case, we exploit the log-concavity of the information density to lower-bound its cumulative distribution function (cdf). The resulting bound allows us to eliminate the codewords for which the central-limit theorem does not apply. 
For the perfect-CSIR case, we show that the distribution of the information density is unimodal and \emph{right-skewed} (i.e., its mean is greater than its mode). Using this result, we then prove that to optimize the cdf of the information density,  it is necessary to reduce its ``skewness'', thereby showing that the optimized information density  must converge as $E\to\infty$ to a (non-skewed) Gaussian distribution.

By comparing~\eqref{eq:intro-second-order-csir} with~\eqref{eq:logM-E23-intro}, we see that, although the minimum (received) energy per bit approaches~\eqref{eq:intro-1.6db} as~$k$ increases regardless of whether CSIR is available or not, the convergence is slower for the no-CSI case.
For the case $\rxant =1$,  our nonasymptotic bounds reveal that to achieve an energy per bit of $-1.5$ dB, one needs to transmit at least $7\times 10^7$ information bits in the no-CSI case, whereas $6\times 10^4$ bits suffice in the perfect-CSIR case.
Furthermore, the bounds also reveal that it takes~$2$~dB more of energy to transmit $1000$ information bits in the no-CSI case compared to the perfect-CSIR case.
As a possible application, our results may be relevant for the design of wireless sensor networks, where energy constraints are often more stringent than bandwidth constraints, and where data packets are usually short.

\paragraph*{Notation}
Upper case letters such as $X$ denote scalar random variables and their realizations are written in lower case, e.g., $x$.
We use boldface upper case letters to denote random vectors, e.g., $\randvecx$, and boldface lower case letters for their realizations, e.g., $\vecx$.
Upper case letters of two special fonts are used to denote deterministic matrices (e.g.,~$\matY$) and random matrices (e.g.,~$\randmatY$).
The symbol $\natunum$ denotes the set of natural numbers, and $\posrealset$ denotes the set of nonnegative real numbers.
The superscripts $\tp{}$ and $\herm{}$ stand for transposition and Hermitian transposition, respectively, and $\bar{\cdot}$ stands for the complex conjugate.
We use $\tr(\matA)$ and $\det(\matA)$ to denote the trace and determinant of the matrix $\matA$, respectively, and use $\fnorm{A}\define \sqrt{\tr(\matA\herm{\matA})}$ to designate the Frobenius norm of~$\matA$.
For an infinite-dimensional complex vector $\vecx\in\complexset^\infty$, we use $\|\vecx\|_p$ to denote the $\ell_p$-norm of $\vecx$, i.e., $\|\vecx\|_p \define \big(\sum\nolimits_{i=1}^\infty |x_i|^p\big)^{1/p}$. The $\ell_\infty$-norm of $\vecx$ is defined as $\|\vecx\|_{\infty} \define \sup\limits_{i}|x_i|$.
We use $\vece_j $ to denote the infinite dimensional vector that has $1$  in the $j$th entry and $0$ elsewhere, and  use $\matI_{a}$ to denote the identity matrix of size $a\times a $.
The distribution of a circularly symmetric Gaussian random vector with covariance matrix~$\matA$ is denoted by $\jpg(\mathbf{0}, \matA)$.
We use $\expdist(\mu)$ to denote the exponential distribution with mean $\mu$, and use $\mathrm{Gamma}(a,b)$ to denote the Gamma distribution with shape parameter~$a$ and scale parameter~$b$~\cite[Ch.~17]{johnson95-1}.
For two functions~$f$ and $g$, we use $f\star g$ to denote the convolution of $f$ and $g$.
Furthermore, the
notation~$f(x) = \bigO(g(x))$, $x\to \infty$, means that
$\lim \sup_{x\to\infty}\bigl|f(x)/g(x)\bigr|<\infty$, and
$f(x) = \littleo(g(x))$, $x\to \infty$, means that $\lim_{x\to\infty}\bigl|f(x)/g(x)\bigr|=0$.
For two measures $\mu$ and $\nu$, we write $\mu \ll \nu$ if $\mu$ is absolutely continuous~\cite[p.~88]{folland99} with respect to $\nu$.
Finally, $|\cdot|^{+} \define \max\{0,\cdot \}$.

Next, we introduce two definitions related to the performance of optimal hypothesis testing.
Given two probability distributions $P$ and $Q$ on a common measurable space $\setW$, we define a randomized test between $P$ and $Q$ as a random transformation $P_{Z\given W}: \setW\to\{0,1\}$ where $0$ indicates that the test chooses $Q$.
We shall need the following performance metric for the test between~$P$ and~$Q$:
\begin{IEEEeqnarray}{rCl}
\label{eq:def-beta}
\beta_\alpha(P,Q) \define \!\!  \!\! \min\limits_{P_{Z|W}: \int\! P_{Z\!\given\! W} (1\!\given \!w) P(dw)\geq \alpha}\! \int \!\!P_{Z\!\given\! W}(1\!\given \! w)  Q(d w)\,\,\,\,\quad
\end{IEEEeqnarray}
where the minimum is over all probability distributions $P_{Z\given W}$
satisfying
\begin{IEEEeqnarray}{rCl}
 \int P_{Z\given W} (1\given w) P(dw)\geq \alpha.
\end{IEEEeqnarray}
The minimum in~\eqref{eq:def-beta} is guaranteed to be achieved by the Neyman-Pearson lemma~\cite{neyman33a}.
For an arbitrary set $\setF$, we define the following performance metric for the composite hypothesis testing between $Q_{Y}$ and the collection $\{P_{Y|X=x}\}_{x\in \setF}$:
\begin{IEEEeqnarray}{rCl}
\label{qe:def-kappa}
\kappa_{\tau}(\setF, Q_{Y}) \define \inf \int P_{Z|Y}(1|y)Q_{Y}(dy).
\end{IEEEeqnarray}
Here, the infimum is over all conditional probability distributions $P_{Z|Y}: \setW \to \{0,1\}$ satisfying
\begin{IEEEeqnarray}{rCl}
\inf_{x\in\setF}\int P_{Z|Y}(1|y) P_{Y|X=x}(dy) \geq \tau.
\end{IEEEeqnarray}

\section{Problem Formulation}
\label{sec:setup}
\subsection{Channel Model and Codes}
We consider a MIMO Rayleigh block-fading channel with $\txant$ transmit antennas and $\rxant$ receive antennas that stays constant over a block of $\cohtime$ channel uses (coherence interval) and changes independently from block to block.
The channel input-output relation within the $i$th coherence interval is given by
\begin{IEEEeqnarray}{rCl}
\randmatV_{i} =  \matU_i \randmatH_i + \randmatZ_i.
\label{eq:MIMO-channel-io}
\end{IEEEeqnarray}
Here, $\matU_i \in \complexset^{\cohtime \times \txant}$ and $\randmatV_i\in  \complexset^{\cohtime \times \rxant} $ are the transmitted and received signals, respectively, expressed in matrix form;
$\randmatH_i \in\complexset^{\txant\times\rxant}$ is the channel matrix, which is assumed to have i.i.d. $\jpg(0,1)$ entries;
$\randmatZ_{i} \in \complexset^{\cohtime \times \rxant}$ is the additive noise matrix, also with i.i.d. $\jpg(0,N_0)$ entries.
We assume that $\{\randmatH_{i}\}$ and $\{\randmatZ_{i}\}$ are mutually independent, and take on independent realizations over successive coherence intervals (block-memoryless assumption).
%
In the remainder of the paper, we shall set $N_0 = 1$, for notational convenience.

We are interested in the scenario where the  blocklength is unlimited, and we aim at characterizing the minimum energy required to transmit $k$ information bits over the channel~\eqref{eq:MIMO-channel-io} with a given reliability.
We shall use $\matU^\infty$ and $\randmatV^\infty$ to denote the infinite sequences $\{\matU_i\}$ and $\{\randmatV_i\}$, respectively.
At times, we shall interpret $\matU^\infty$ as the infinite-dimensional matrix obtained by stacking the matrices $\{\matU_i\}$, $i\in\natunum$, on top of each other.
In this case, the matrix $\matU^\infty$ has $\txant$ columns and infinitely many rows, and its $t$th column vector represents the signal sent from the $t$th transmit antenna.
The energy of the input matrix $\matU^\infty$ is measured as follows
\begin{equation}
\fnorm{\matU^\infty}^2 = \sum\limits_{i=1}^\infty \fnorm{\matU_i}^2.
\end{equation}
Furthermore, we denote the set of all input matrices $\matU^\infty$ by $\setA$ and the set of all output matrices $\matV^\infty$ by $\setB$. Finally, we let $\setH$ be the set of channel matrices $\matH^\infty$.

Next, we define channel codes for the channel~\eqref{eq:MIMO-channel-io} for both the no-CSI and the perfect-CSIR case.

\begin{dfn}
\label{eq:def-code-inf-bw-siso}
An $(E,M,\error)$-code for the channel~\eqref{eq:MIMO-channel-io} for the no-CSI case consists of a set of codewords $\{\matC_1,\ldots,\matC_M\} \in \setA^M$ satisfying the energy constraint
\begin{equation}
\fnorm{\matC_j}^2 \leq E,\quad j\in\{1,\ldots,M\}
\label{eq:power-constraint}
\end{equation}
and a decoder $g: \setB \to \{1,\ldots,M\}$ satisfying the maximum error probability constraint
\begin{IEEEeqnarray}{rCl}
\max\limits_{j\in\{1,\ldots,M\} }\prob[ g(\randmatV^\infty) \neq j \given \randmatU^\infty= \matC_j] \leq \error.
\label{eq:max-error-prob-eee}
\end{IEEEeqnarray}
Here, $\randmatV^\infty$ is the output induced by the codeword $\randmatU^\infty= \matC_j$ according to~\eqref{eq:MIMO-channel-io}.
The maximum number of messages that can be transmitted with energy $E$ and maximum error probability~$\error$ is
\begin{IEEEeqnarray}{rCl}
M^*(E,\error) \define \max\mathopen{}\big\{M: \exists\, (E,M,\error)\text{-code}\big\}.
\label{eq:max-message-def}
\end{IEEEeqnarray}
Similarly, the minimum energy per bit is defined as
\begin{IEEEeqnarray}{rCl}
\eb^*(k,\error) \define \frac{1}{k} \inf\lefto\{E: \, \exists \,(E,2^k,\error)\text{-code}\right\}.
\label{eq:mini-energy-bit-def-rr}%
\end{IEEEeqnarray}%
\end{dfn}%

\begin{dfn}
\label{eq:def-code-inf-bw-siso}
An $(E,M,\error)$-code for the channel~\eqref{eq:MIMO-channel-io} for the perfect-CSIR case consists of a set of codewords $\{\matC_1,\ldots,\matC_M\} \in \setA^M$ satisfying the energy constraint~\eqref{eq:power-constraint},
and a decoder $g: \setB\times\setH \to \{1,\ldots,M\}$ satisfying the maximum error probability constraint
\begin{IEEEeqnarray}{rCl}
\max\limits_{j\in\{1,\ldots,M\} }\prob[ g(\randmatV^\infty,\randmatH^\infty) \neq j \given \randmatU^\infty= \matC_j] \leq \error.
\end{IEEEeqnarray}
The maximum number of messages that can be transmitted with energy $E$ and maximum error probability~$\error$ for the perfect-CSIR case is defined as in~\eqref{eq:max-message-def}.
\end{dfn}

As we shall show in the next section, one can derive tight bounds on $M^*(E,\error)$ (for both the no-CSI and the perfect-CSIR case) by focusing exclusively on the memoryless single-input multiple-output (SIMO) scenario $\cohtime =\txant =1$.
Therefore, we shall next develop a specific notation to address this setup.
In the SIMO case, the input-output relation reduces to
\begin{equation}
V_{r,i} = H_{r,i} u_i + Z_{r,i},\,\, r\in\{1,\ldots,\rxant\}, \,\, i\in\natunum \, .
\label{eq:SIMO-channel-io}
\end{equation}
Here, $V_{r,i}\in\complexset$ denotes the received symbol at the $r$th receive antenna on the $i$th channel use, and $H_{r,i}$ and $Z_{r,i}$ denote the fading coefficient and the additive noise, respectively.
We shall set $\vecu \define [u_1,u_2,\ldots]$ and $\randvecv_r \define [V_{r,1},V_{r,2},\ldots]$.

\subsection{An Equivalent Channel Model for the no-CSI case}

Focusing on the no-CSI case, we define next a channel model that is equivalent to~\eqref{eq:SIMO-channel-io}.
Observe that, given $\randvecu = \vecu$, the output vectors $\randvecv_1,\ldots,\randvecv_{\rxant}$ are i.i.d.  Gaussian, i.e.,
\begin{equation}
P_{\randvecv_r\given \randvecu =\vecu} = \prod\limits_{i=1}^{\infty} \jpg(0,(1+|u_i|^2)),\quad r \in\{1,\ldots,\rxant\}.
\label{eq:cond-dist-simo-nocsi}
\end{equation}
Since the $\{\randvecv_r\}$ depend on the input symbols $\{u_i\}$ only through their squared magnitude $\{|u_i|^2\}$, we can reduce without loss of generality the input space to $\posrealset^{\infty}$.
We also note that, given $\randvecu = \vecu$, the joint conditional probability distribution of the random variables $\{V_{r,i}\}$  in~\eqref{eq:SIMO-channel-io}  does not change if we multiply   $\{V_{r,i}\}$ with arbitrary deterministic phases.
This means that the $\{|V_{r,i}|^2\}$ are a sufficient statistics for the detection of $\vecu$ from $\{\randvecv_{r}\}$.
Letting $x_i \define |u_i|^2$ and $Y_{r,i} \define |V_{r,i}|^2$, $r\in\{1,\ldots,\rxant\}$, $i\in\natunum$, we obtain the following  input-output relation, which is equivalent to~\eqref{eq:SIMO-channel-io}:
\begin{IEEEeqnarray}{rCl}
Y_{r,i} = (1+ x_i)S_{r,i}, \quad r\in\{1,\ldots,\rxant\},\quad i\in\natunum.
\label{eq:channel-io}
\end{IEEEeqnarray}
Here, the input $x_i$ and the output~$ Y_{r,i}$ are nonnegative real numbers, and $\{S_{r,i}\}$ are i.i.d. $\expdist(1)$-distributed. %
We shall denote the input of the channel~\eqref{eq:channel-io} by $\vecx \define [x_1,x_2,\ldots]\in\posrealset^{\infty}$ and denote the output by the matrix $\randmatY$, whose entry on the $r$th row and the $i$th column is  $Y_{r,i}$.
Since $\|\vecx\|_1 = \|\vecu\|_2^2$ and since $\|\vecx\|_{\infty} = \|\vecu\|_{\infty}^2$, we shall measure the energy and the peakiness of an input codeword $\vecx$ for the channel~\eqref{eq:channel-io} by its $\ell_1$-norm $\|\vecx\|_1$, and by its $\ell_\infty$-norm $\|\vecx\|_{\infty}$, respectively.

%
%

\section{Minimum Energy Per Bit}
\label{sec:minimum-energy-per-bit}
We shall now characterize $M^*(E,\error)$ for both the no-CSI and the perfect-CSIR case.
The organization of this section is as follows.
In Section~\ref{sec:general-ach-bound}, we first present nonasymptotic achievability and converse bounds on $M^*(E,\error)$ for general channels subject to a cost constraint.
In Section~\ref{sec:nonasy-bounds}, we then particularize these bounds to the channel~\eqref{eq:MIMO-channel-io} for the no-CSI case.
Both the converse and achievability bounds in Section~\ref{sec:nonasy-bounds} are derived by reducing the MIMO channel~\eqref{eq:MIMO-channel-io} to the SIMO channel~\eqref{eq:channel-io}.
We then show in Section~\ref{sec:asy-analy-energy-bit} that these bounds match asymptotically as $E\to\infty$ up to second order, thus establishing~\eqref{eq:logM-E23-intro}.
In Section~\ref{sec:perfect-csir}, we derive bounds on $M^*(n,\error)$ for the perfect-CSIR case and prove the asymptotic expansion~\eqref{eq:intro-second-order-csir}.
Finally, the nonasymptotic bounds for both the no-CSI and the perfect-CSIR case are evaluated numerically in Section~\ref{sec:numerical-results}.

\subsection{General Nonasymptotic Bounds}
\label{sec:general-ach-bound}
We consider in this section general stationary memoryless channels $(\setX, P_{Y|X}, \setY)$ with input codewords subject to a cost constraint. As in~\cite{verdu90-09}, we use $b[x]$ to denote the cost of the symbol~$x$ in the input alphabet~$\setX$. We shall also assume that there exists a zero-cost symbol, which we label as ``$0$''.
With a slight abuse of notation, we use $E$ to denote the cost constraint imposed on a codeword.
An $(E,M,\error)$-code for this general channel consists of a set of  $M$ codewords $\vecc_j =[c_{j,1},c_{j,2},\ldots]$, $j=1,\ldots,M$ that satisfy the cost constraint
\begin{equation}
\sum\limits_{i=1}^{\infty} b[c_{j,i}] \leq E, \quad j=1,\ldots,M
\label{eq:cost-constraint}
\end{equation}
and has maximum error probability not exceeding~$\error$.
We next present two achievability bounds on $M^*(E,\error)$ that are finite-energy generalizations of Verd\'{u}'s lower bound~\cite[pp.~1023--1024]{verdu90-09} on the capacity per unit cost.\footnote{For stationary memoryless channels, the capacity per unit cost is given by $\lim\limits_{\error \to 0}\lim\limits_{E\to\infty} (\log M^*(E,\epsilon))/E$.}

\begin{thm}
\label{thm:kappa-beta-cpuc}
Consider a  stationary memoryless channel $(\setX, P_{Y|X}, \setY)$ that has a zero-cost input symbol.
For every $N \in \natunum$, every $0<\error<1$, and every input symbol~$x_0\in\setX$ satisfying $b[x_0]>0$, there exists an $(E,M,\error)$-code for which $E = b[x_0] N $ and
\begin{IEEEeqnarray}{rCl}
M -1 \geq \sup\limits_{0<\tau<\error} \frac{\tau }{\beta_{1-\error+\tau}(P^{\otimes N}_{Y|X=x_0} , P^{\otimes N}_{Y|X=0})}.
\label{eq:verdu-lb-general}
\end{IEEEeqnarray}
Here, $\beta_{(\cdot)}(\cdot,\cdot)$ is given in~\eqref{eq:def-beta}, and
\begin{IEEEeqnarray}{rCl}
P_{Y|X=x}^{\otimes N} \define \underbrace{P_{Y|X=x} \times \cdots \times P_{Y|X = x}}_{N \text{ times}}
\label{eq:define-p-otimes-yx}
\end{IEEEeqnarray}
for every $x\in\setX$.
\end{thm}

\begin{IEEEproof}
As in~\cite{verdu90-09}, we choose the codewords $\vecc_j \in \mathcal{X}^{\infty}$, $j=1,\ldots,M$, as follows:
\begin{IEEEeqnarray}{rCl}
\vecc_{j} \define [\underbrace{0,\ldots,0}_{(j-1)N},\underbrace{x_0,\ldots,x_0}_{N},0,\ldots].
\label{eq:def-codewords-verdu}
\end{IEEEeqnarray}
Fix an arbitrary $\tau \in (0,\error)$.
For a given received signal $\randvecy \in \mathcal{Y}^{\infty}$, the decoder runs $M$ parallel binary hypothesis tests~$Z_j$, $j=1,\ldots,M$, between $P_{\randvecy\given \randvecx = \mathbf{0}}$ and  $P_{\randvecy\given \randvecx =\vecc_j}$. Here, $Z_j =1$ indicates that the test selects $P_{\randvecy\given \randvecx =\vecc_j}$.
The tests $\{Z_j\}$, $j=1,\ldots,M$, are chosen to satisfy
\begin{IEEEeqnarray}{rCl}
\prob[ Z_j = 1 \given \randvecx = \vecc_j ] &\geq& 1 -\error + \tau \label{eq:def-test-verdu1}\\
\prob[Z_j=1 \given \randvecx =\mathbf{0}] &=& \beta_{1-\error + \tau}(P_{\randvecy\given \randvecx = \vecc_j}, P_{\randvecy\given \randvecx =\mathbf{0}}).\label{eq:def-test-verdu2}\IEEEeqnarraynumspace
\end{IEEEeqnarray}
The existence of tests that satisfy~\eqref{eq:def-test-verdu1} and~\eqref{eq:def-test-verdu2} is guaranteed by the  Neyman-Pearson lemma~\cite{neyman33a}.
The decoder outputs the index $m$ if $Z_m =1$ and $Z_j = 0$ for all $j\neq m$. It outputs $1$ if no such index can be found.
By construction, the maximum probability of error of the code just defined is upper-bounded~by
\begin{IEEEeqnarray}{rCl}
\error &\leq& \prob[ Z_1 = 0 \given \randvecx = \vecc_1 ] + (M-1) \prob[Z_1 = 1 \given \randvecx = \mathbf{0}] \label{eq:verdu-bound-1} \IEEEeqnarraynumspace\\
&\leq& \error-\tau + (M-1) \beta_{1-\error+\tau}(P_{\randvecy \given \randvecx = \vecc_1}, P_{\randvecy\given \randvecx = \mathbf{0}}).
\label{eq:verdu-bound}
\end{IEEEeqnarray}
Here,~\eqref{eq:verdu-bound-1} follows because for each test $Z_j$ ($j\neq 1$) satisfying~\eqref{eq:def-test-verdu1} and~\eqref{eq:def-test-verdu2},
\begin{IEEEeqnarray}{rCl}
\prob[Z_j =1|\randvecx = \vecc_1] &=& \prob[Z_j =1 | \randvecx = \mathbf{0}]\\
 &=& \prob[Z_1 = 1 | \randvecx = \mathbf{0}]
\end{IEEEeqnarray}
and~\eqref{eq:verdu-bound} follows by~\eqref{eq:def-test-verdu1} and~\eqref{eq:def-test-verdu2}.
From~\eqref{eq:verdu-bound}, we conclude~that
\begin{IEEEeqnarray}{rCl}
M-1 \geq \frac{\tau}{\beta_{1-\error+\tau}(P_{\randvecy \given \randvecx = \vecc_1}, P_{\randvecy\given \randvecx = \mathbf{0}}) }.
\label{eq:lb-M-verdu}
\end{IEEEeqnarray}
The proof is completed by noting that
\begin{IEEEeqnarray}{rCl}
\beta_{1-\error +\tau }(P_{\randvecy \given \randvecx = \vecc_1}, P_{\randvecy\given \randvecx = \mathbf{0}})  = \beta_{1-\error + \tau}(P^{\otimes N}_{Y|X=x_0} , P^{\otimes N}_{Y|X=0}) \IEEEeqnarraynumspace
\end{IEEEeqnarray}
and by maximizing the RHS of~\eqref{eq:lb-M-verdu} over $\tau\in( 0, \error)$.
\end{IEEEproof}

The proof of Theorem~\ref{thm:kappa-beta-cpuc} is based on the same binary hypothesis-testing decoder that is used in the proof of the $\kappa\beta$~bound~\cite[Th.~25]{polyanskiy10-05}.
In fact, if $P_{Y|X=x_0} \ll P_{Y|X=0}$, a slightly weakened version of~\eqref{eq:lb-M-verdu}, with $M-1$ replaced by $M$, follows directly from the $\kappa\beta$~bound~\cite[Th.~25]{polyanskiy10-05} upon setting $Q_{\randvecy} = P_{\randvecy \given \randvecx = \veczero}$ and choosing the set $\setF$ as
\begin{IEEEeqnarray}{rCl}
\setF &=& \Big\{\vecx \in \setX^{\infty}: \vecx = [\underbrace{0,\ldots,0}_{(j-1) N},\underbrace{x_0,\ldots, x_0}_{N}, 0,\ldots]
\text{ for some }j\in\natunum\Big\}. \IEEEeqnarraynumspace
\end{IEEEeqnarray}
Since $\beta_{1-\error + \tau}(P_{\randvecy\given\randvecx=\vecx} , Q_{\randvecy})$ takes the same value for all $\vecx\in\setF$, to establish this looser bound it is sufficient to show that (proof omitted)
\begin{equation}
\kappa_{\tau} (\setF ,Q_{\randvecy} ) = \tau
\label{eq:compute-tau}
\end{equation}
where $\kappa_{(\cdot)}(\cdot,\cdot)$ is given in~\eqref{qe:def-kappa}.
%

Using the same codebook as in Theorem~\ref{thm:kappa-beta-cpuc} together with a maximum likelihood decoder, we obtain a different achievability bound, which is stated in the following theorem.
\begin{thm}
\label{thm:RCU-cpuc}
Consider a  stationary memoryless channel $(\setX, P_{Y|X}, \setY)$ that has a zero-cost input symbol.
For every $N\in \natunum$, every $0<\error<1$, and every input symbol $x_0\in\setX$ satisfying $b[x_0]>0$, there exists an $(E,M,\error)$-code for which $E =b[x_0] N $ and
\begin{equation}
\error \leq  \Ex{}{ \min\mathopen{}\Big\{1, (M-1)\prob\lefto[  \imath_{N}(x_0; Y^N)  \leq \imath_{N}(x_0; \hat{Y}^N)\given Y^N\right]\! \Big\}}.
\label{eq:thm-rcu}
\end{equation}
Here, $P_{Y^N \hat{Y}^N}(y^N,\hat{y}^N) \define  P^{\otimes N}_{Y|X=x_0} ({y}^N) P^{\otimes N}_{Y|X=0}(\hat{y}^N)$ and
\begin{IEEEeqnarray}{rCl}
\imath_{N}(x; y^N) \define \log \frac{dP^{\otimes N}_{Y|X=x}}{dP^{\otimes N}_{Y|X=0}}(y^N)
\end{IEEEeqnarray}
with $P^{\otimes N}_{Y|X=x}$, $x\in\setX$, defined in~\eqref{eq:define-p-otimes-yx}.
\end{thm}
\begin{rem}
For AWGN channels with cost function $b[x]= x^2$, one can recover~\cite[Eq.~(15)]{polyanskiy11-08b} from~\eqref{eq:thm-rcu} by  setting $N=1$ and $x_0=\sqrt{E}$.
\end{rem}
\begin{IEEEproof}
We use the same codebook as in Theorem~\ref{thm:kappa-beta-cpuc}, together with a maximum likelihood decoder.
Let
\begin{IEEEeqnarray}{rCl}
\infden(\vecx,\vecy) &\define &\log\frac{dP_{\randvecy\given \randvecx =\vecx}}{dP_{\randvecy \given \randvecx = \mathbf{0}}} (\vecy).
\label{eq:def-info-density-infty}
\end{IEEEeqnarray}
Let $Y^N=[Y_1,\ldots,Y_N]$ denote the vector containing the first $N$ entries of $\randvecy$ and let $\hat{Y}^N\sim P_{Y|X=0}^{\otimes N}$ be independent of $\randvecy$.
The probability of error $\error$ is upper-bounded as follows:
\begin{IEEEeqnarray}{rCl}
\error &\leq & P_{\randvecy \given \randvecx = \vecc_1}\lefto[\, \bigcup\limits^{M}_{j= 2} \Big\{\infden(\vecc_1, \randvecy) \leq \infden(\vecc_j, \randvecy)\Big\}\right]\label{eq:ML-error-0}\\
& = & \Ex{}{\prob \lefto[\,\bigcup\limits_{j=2}^{M} \Big\{  \infden(\vecc_1, \randvecy) \leq \infden(\vecc_j, \randvecy)\Big\} \bigg| Y^N\right]}\label{eq:ML-error-1}  \\
&\leq& \Ex{}{ \min\Big\{ 1, (M-1)\prob\lefto[ \infden(\vecc_1, \randvecy) \leq \infden(\vecc_2, \randvecy)    \Big|Y^N \right] \Big\}}\label{eq:ML-error-2} \\
&=& \Ex{}{ \min\Big\{1, (M-1)\prob\lefto[ \imath_{N}(x_0; Y^N)  \leq \imath_{N}(x_0; \hat{Y}^N)  \given Y^N\right] \Big\}}. 
 \label{eq:ML-error-3}
\end{IEEEeqnarray}
Here,~\eqref{eq:ML-error-0} follows because all codewords have the same error probability under maximum likelihood decoding;~\eqref{eq:ML-error-2} follows by choosing the tighter bound between 1 and the union bound;~\eqref{eq:ML-error-3} follows because  $\vecc_1 = [\underbrace{x_0,\ldots,x_0}_{N},\underbrace{0,\ldots,0}_{N},0,\ldots]$ and $\vecc_2=  [\underbrace{0,\ldots,0}_{N},\underbrace{x_0,\ldots,x_0}_{N},0,\ldots] $, and because, under $P_{\randvecy\given \randvecx=\vecc_1}$, the sequence $Y_{N+1}^{2N}$ has the same distribution as  $\hat{Y}^N \sim P_{Y|X=0}^{\otimes N}$. Furthermore, $Y_{N+1}^{2N}$ is independent of $Y^N$ since the channel is stationary and memoryless.
\end{IEEEproof}

On the converse side, we have the following result, which follows by applying the meta-converse theorem~\cite[Th.~31]{polyanskiy10-05} with $Q_{\randvecy} = P_{\randvecy\given\randvecx=\mathbf{0}}$.
\begin{thm}
\label{thm:general-converse-trivial}
Consider a channel $(\setX, P_{Y|X}, \setY)$ that has a zero-cost input symbol.
Every $(E,M,\error)$-code with codewords satisfying the cost constraint~\eqref{eq:cost-constraint} satisfies
 \begin{IEEEeqnarray}{rCl}
M\leq \sup\limits_{\vecx:\sum\limits_{i=1}^{\infty}b[x_i]\leq E} \frac{1}{\beta_{1-\error}(P_{\randvecy\given\randvecx =\vecx},P_{\randvecy\given\randvecx=\mathbf{0}} )}\, .
\label{eq:meta-converse-general-trivial}
\end{IEEEeqnarray}
\end{thm}

The bound~\eqref{eq:meta-converse-general-trivial} is in general not computable because it involves an optimization over infinite-dimensional codewords.
As we shall see in the next section, in the MIMO Rayleigh block-fading case  it is possible to reduce this infinite-dimensional optimization problem to a three-dimensional one, which can be solved numerically.

We would like to remark that the general bounds developed in this section apply to both the no-CSI case and the perfect-CSIR case. 
For the perfect-CSIR case, we view the pair  $(\randmatV, \randmatH)$ as the channel output, and  identify the channel law with $P_{\randmatV, \randmatH | \randmatU} = P_{\randmatH}P_{\randmatV \given \randmatH,\randmatU}$. For the no-CSI case, we view $\randmatV$ as the output  and identify the channel law with $P_{\randmatV| \randmatU}$, which is obtained by averaging $P_{\randmatV \given \randmatH,\randmatU}$ over the fading matrix $\randmatH$. In both cases, the channel is stationary and memoryless. 

\subsection{Nonasymptotic Bounds: the No-CSI Case}
\label{sec:nonasy-bounds}
 Particularizing Theorems~\ref{thm:kappa-beta-cpuc} and~\ref{thm:RCU-cpuc} to the channel~\eqref{eq:MIMO-channel-io} for the no-CSI case,  we obtain the achievability~bounds given below in Corollaries~\ref{thm:non-asy-inf-bw-Verdu} and~\ref{thm:ML-lb-inf-bw}.
\begin{cor}
\label{thm:non-asy-inf-bw-Verdu}
For every $E>0$ and every $0<\error<1$, there exists an $(E,M,\error)$-code for the MIMO Rayleigh block-fading channel~\eqref{eq:MIMO-channel-io} for the case of no CSI satisfying
\begin{IEEEeqnarray}{rcl}
M -1 \geq \sup\limits_{0<\tau<\error, \, N \in \natunum }\frac{\tau }{\prob[G_{N} \geq (1+E/N) \xi]}
\label{eq:verdu-lb-cor}
\end{IEEEeqnarray}
where $G_{N} \sim \mathrm{Gamma}(\rxant N,1)$ and $\xi$ satisfies
\begin{IEEEeqnarray}{rCl}
\prob[G_{N} \leq \xi] = \error -\tau.
\end{IEEEeqnarray}
\end{cor}
\begin{IEEEproof}
Every code for the memoryless SIMO  Rayleigh-fading channel ($\txant =\cohtime=1$) can be used on a MIMO Rayleigh block-fading channel with $\txant>1$ and $\cohtime >1$.
Indeed, it is sufficient to switch off all transmit antennas but one, and to limit transmissions to the first channel use in each coherence interval.
Therefore, it is sufficient to prove that~\eqref{eq:verdu-lb-cor} is achievable for the memoryless SIMO Rayleigh-fading channel~\eqref{eq:channel-io}.
In the SIMO case, we have (see~\eqref{eq:channel-io})
\begin{IEEEeqnarray}{rCl}
\log\frac{dP^{\otimes N}_{\randvecy|X = x_0}}{dP^{\otimes N}_{\randvecy|X=0}}\lefto(\vecy^N\right) &=& \frac{x_0\log e}{1+x_0}\sum\limits_{r=1}^{\rxant}\sum\limits_{i=1}^{N}y_{r,i} - \rxant N\log(1+x_0). \IEEEeqnarraynumspace
\label{eq:info-density-def}
\end{IEEEeqnarray}
Let $x_0 = E/N$ for some $N\in\natunum$.
Then, under $P^{\otimes N}_{\randvecy|X = x_0}$, the random variable $\log\frac{d P^{\otimes N}_{\randvecy|X = x_0}}{dP_{\randvecy|X=0}^{\otimes N}}(\randvecy^N)$ has the same distribution as
\begin{IEEEeqnarray}{rCl}
\frac{E}{N}G_N \log e - \rxant N\log(1+E/N)
\label{eq:info-den-under-p-th-verdu}
\end{IEEEeqnarray}
where $G_N \sim \mathrm{Gamma}(\rxant N,1)$, and, under $P^{\otimes N}_{\randvecy|X = 0}$, it has the same distribution as
\begin{IEEEeqnarray}{rCl}
&&\frac{E}{N}\frac{ G_N \log e }{1+E/N} - \rxant N\log(1+E/N).
\label{eq:info-den-under-q-th-verdu}
\end{IEEEeqnarray}
The proof of~\eqref{eq:verdu-lb-cor} is concluded by using~\eqref{eq:info-den-under-p-th-verdu} and~\eqref{eq:info-den-under-q-th-verdu} in~\eqref{eq:verdu-lb-general} together with the Neyman-Pearson lemma~\cite{neyman33a}, and by optimizing over $N\in\natunum$.
\end{IEEEproof}

\begin{cor}
\label{thm:ML-lb-inf-bw}
For every $M>0$ and every $0<\error<1$, there exists an $(E,M,\error)$-code for the MIMO Rayleigh block-fading channel~\eqref{eq:MIMO-channel-io} for the case of no CSI satisfying
\begin{equation}
\error \leq  \min\limits_{N\in\natunum} \Ex{}{ \min\Big\{1,  (M\!-\!1)\prob\mathopen{}\Big[\bar{G}_{N} \geq (1+E/N)G_{N} \Big| G_{N}\Big]\Big\}}
\label{eq:thm-rcu-cor}
\end{equation}
where $G_{N}$ and $\bar{G}_{N}$ are i.i.d. $\mathrm{Gamma}(\rxant N,1)$ random variables.
\end{cor}
\begin{IEEEproof}
We proceed first as in the proof of Corollary~\ref{thm:non-asy-inf-bw-Verdu}. Then, we use~\eqref{eq:info-den-under-p-th-verdu} and~\eqref{eq:info-den-under-q-th-verdu} in~\eqref{eq:thm-rcu}.
\end{IEEEproof}

Numerical evidence (provided in Section~\ref{sec:numerical-results}) suggests that~\eqref{eq:thm-rcu-cor} is tighter than~\eqref{eq:verdu-lb-cor}.
However,~\eqref{eq:verdu-lb-cor} is more suitable for asymptotic analyses.

We now provide a converse bound, which is based on Theorem~\ref{thm:general-converse-trivial}.

\begin{thm}
\label{thm:non-asy-converse-inf}
Let $\{S_i\}$ be i.i.d. $\expdist(1)$-distributed random variables.
Every $(E,\!M,\!\error)$-code for the MIMO Rayleigh block-fading channel~\eqref{eq:MIMO-channel-io} for the case of no CSI satisfies
\begin{IEEEeqnarray}{rCl}
\frac{1}{M} &\geq & \sup\limits_{\eta \in \realset} \frac{\inf\limits_{\vecx}\prob\bigg[\sum\limits_{i=1}^{\infty} \Big(x_i S_i \log e -\log(1+x_i) \Big)\leq \eta \bigg] -\error }{\exp(\eta)} \, . \IEEEeqnarraynumspace 
\label{eq:thm-meta-converse}
\end{IEEEeqnarray}
The infimum in~\eqref{eq:thm-meta-converse}  is over all $\vecx\in\posrealset^\infty$ taking one of the following two forms:
\begin{IEEEeqnarray}{rCl}
\vecx &=& [q_3,\underbrace{q_2,\ldots,q_2}_{N},q_1,0,0,\ldots]\label{eq:three-atoms-form}
\end{IEEEeqnarray}
or
\begin{IEEEeqnarray}{rCl}
\vecx &=& [\underbrace{\tilde{q}_2,\ldots,\tilde{q}_2}_{\widetilde{N}_2}, \underbrace{\tilde{q}_1,\ldots,\tilde{q}_1}_{\widetilde{N}_1},0,0,\ldots].
\label{eq:le-two-atoms-form}
\end{IEEEeqnarray}
%
Here, $N\in \natunum$ and $0<q_1< q_2<q_3$ satisfy~$q_1 + Nq_2 +q_3 = \rxant E$. Furthermore,
$\widetilde{N}_1, \widetilde{N}_2\in \natunum$ and $ 0\leq \tilde{q}_1\leq  \tilde{q}_2$ satisfy $\widetilde{N}_1\tilde{q}_1 + \widetilde{N}_2\tilde{q}_2 = \rxant E$.
\end{thm}

\begin{rem}
The optimization over infinite-dimensional codewords in the converse bound~\eqref{eq:meta-converse-general-trivial} is reduced in~\eqref{eq:thm-meta-converse} to a three-dimensional optimization problem. This makes~\eqref{eq:thm-meta-converse} numerically computable.
In words, the conditions in~\eqref{eq:three-atoms-form} and \eqref{eq:le-two-atoms-form}  imply that i) the entries of $\vecx$ can take at most three distinct nonzero values,
and that ii) if the entries of $\vecx$ take exactly three distinct nonzero values, then both the largest and the smallest nonzero entries must appear only once. %
\end{rem}

\begin{IEEEproof}
Without loss of generality, we can assume  that each codeword matrix $\matC_j$ satisfies the energy constraint~\eqref{eq:power-constraint} with equality.
Indeed, for an  arbitrary code $\setC$,
we can construct a new  code $\setC'$ by appending to each codeword matrix $\matC_j$ in $\setC$ an extra $\cohtime\times\txant$ block of energy $E-\fnorm{\matC_j}^2$ (recall that the number of transmitted symbols is unlimited).
The resulting code~$\setC'$ has the same number of codewords as~$\setC$ and each codeword of~$\setC'$ satisfies~\eqref{eq:power-constraint} with equality.
Moreover, the error probability of~$\setC'$ can not exceed that of~$\setC$.

We continue the proof of~\eqref{eq:thm-meta-converse} by using Theorem~\ref{thm:general-converse-trivial}, which implies
\begin{IEEEeqnarray}{rCl}
\frac{1}{M} \geq \inf \limits_{\matU^\infty \in \setA: \,\,\fnorm{\matU^\infty}^2 = E }\beta_{1-\error}(P_{\randmatV^\infty \given \randmatU^\infty = \matU^\infty } , P_{\randmatV^\infty \given \randmatU^\infty=\mathbf{0}}).
\label{eq:meta-converse}
\end{IEEEeqnarray}
For a given $\matU^\infty=\{\matU_i\}$, let $\widetilde{\matU}^\infty = \{\widetilde{\matU}_i\}$  where $\widetilde{\matU}_i \in \posrealset^{\cohtime\times\txant}$ is a diagonal matrix whose diagonal elements are the singular values of $\matU_i$.
We shall next show that
\begin{equation}
\beta_{1-\error}(P_{\randmatV^\infty \given \randmatU^\infty = \matU^\infty } , P_{\randmatV^\infty \given \randmatU^\infty=\mathbf{0}}) = \beta_{1-\error}(P_{\randmatV^\infty \given \randmatU^\infty = \widetilde{\matU}^\infty } , P_{\randmatV^\infty \given \randmatU^\infty=\mathbf{0}}).
\label{eq:symmetry-argument-mimo}
\end{equation}
This implies that to evaluate the RHS of~\eqref{eq:meta-converse}, it suffices to focus on diagonal matrices $\{\matU_i\}$.
Note also that when the input matrices $\{\matU_i\}$ are diagonal, the $\txant\times\rxant$ MIMO block-fading channel~\eqref{eq:MIMO-channel-io} decomposes into $\min\{\txant,\cohtime\}$ noninteracting memoryless SIMO fading channels with $\rxant$ receive antennas.
Therefore, exploiting the equivalence between~\eqref{eq:SIMO-channel-io} and~\eqref{eq:channel-io}, we conclude that the RHS of~\eqref{eq:meta-converse} coincides with
\begin{equation}
\inf\limits_{\vecx\in\posrealset^\infty: \|\vecx\|_1 =  E}\beta_{1-\error}(P_{\randmatY \given \randvecx = \vecx}, P_{\randmatY \given \randvecx=\mathbf{0}})
\label{eq:inf-beta-over-simo}
\end{equation}
where $P_{\randmatY\given \randvecx}$ is the conditional distribution of the output of the channel~\eqref{eq:channel-io} given the input.

To prove~\eqref{eq:symmetry-argument-mimo}, we note that given $\randmatU_i = \matU_i$, the column vectors of the output matrix $\randmatV_i$ are i.i.d. $\jpg\lefto(\mathbf{0}, \matI_{\cohtime} + \matU_i \herm{\matU_i}\right)$-distributed. Therefore, the probability distribution $P_{\randmatV_i \given \randmatU_i=\matU_i}$ depends on $\matU_i$ only through $\matU_i \herm{\matU_i}$.
In particular, it is invariant to right-multiplication of $\matU_i$ by an arbitrary $\txant\times \txant $ unitary matrix $\matG$.
Furthermore, since the noise matrix $\randmatZ_i$ is isotropically distributed~\cite[Def.~6.21]{lapidoth03-10a}, for every $\cohtime\times\cohtime$ unitary matrix $\widetilde{\matG}$ and every $\matU\in \complexset^{\cohtime\times\txant}$, the conditional distribution of $\randmatV_i$ given $\randmatU_i=\matU$ coincides with that of $\herm{\widetilde{\matG}}\randmatV_i$ given $\randmatU_i=\widetilde{\matG}\matU$.
 Therefore, for every $i\in\natunum$, and every unitary matrices $\matG$ and $\widetilde{\matG}$, we have
\begin{IEEEeqnarray}{rCl}
\beta_{1-\error}(P_{\randmatV_i\given \randmatU_i =\matU}, P_{\randmatV_i\given \randmatU_i =\mathbf{0}}) &=& \beta_{1-\error}(P_{\herm{\widetilde{\matG}}\randmatV_i\given \randmatU_i =\widetilde{\matG}\matU{\matG}}, P_{\herm{\widetilde{\matG}}\randmatV_i\given \randmatU_i =\mathbf{0}})\\
&=& \beta_{1-\error}(P_{\randmatV_i\given \randmatU_i =\widetilde{\matG}\matU{\matG}}, P_{\randmatV_i\given \randmatU_i =\mathbf{0}}).
\label{eq:rotation-invariant-beta}
\end{IEEEeqnarray}
Here, the second step follows because $\beta_{1-\error}(\cdot,\cdot)$ stays unchanged under the change of variables $\matV_i \mapsto \herm{\widetilde{\matG}}\matV_i$.
Since $\matG$, $\widetilde{\matG}$, and $i$ are arbitrary, and since the channel $P_{\randmatV^\infty\given\randmatU^\infty}$ is block-memoryless, \eqref{eq:rotation-invariant-beta} implies~\eqref{eq:symmetry-argument-mimo}.

Next, we lower-bound $\beta_{1-\error}(P_{\randmatY \given \randvecx = \vecx}, P_{\randmatY \given \randvecx=\mathbf{0}})$ in~\eqref{eq:inf-beta-over-simo} using~\cite[Eq.~(102)]{polyanskiy10-05}.
Specifically, we fix an arbitrary $\eta\in\realset$ and obtain
\begin{IEEEeqnarray}{rCl}
\IEEEeqnarraymulticol{3}{l}{
\beta_{1-\error}(P_{\randmatY \given \randvecx = \vecx } , P_{\randmatY \given \randvecx=\mathbf{0}} )}\notag\\
&\geq & \exp(-\eta) \Big(P_{\randmatY\given\randvecx = \vecx}[\infden(\vecx,\randmatY) \leq \eta ] -\error\Big)
\label{eq:lb-beta-conv}
\end{IEEEeqnarray}
where
$\infden(\cdot,\cdot)$ was defined in~\eqref{eq:def-info-density-infty}.
Under $P_{\randmatY\given \randvecx =\vecx}$, the random variable $\infden(\vecx,\randmatY)$ has the same distribution as
\begin{IEEEeqnarray}{rCl}
\sum\limits_{r=1}^{\rxant}\sum\limits_{i=1}^{\infty}\Big( x_i S_{r,i}\log e - \log(1+x_i)\Big)
\label{eq:info-den-pyx}
\end{IEEEeqnarray}
where $\{S_{r,i}\}$ are i.i.d. $\expdist(1)$-distributed.
Substituting~\eqref{eq:info-den-pyx} into \eqref{eq:lb-beta-conv}, and then~\eqref{eq:lb-beta-conv} into~\eqref{eq:meta-converse}, we obtain
\begin{IEEEeqnarray}{rCl}
\frac{1}{M} &\geq & \frac{\inf\limits_{\vecx\in\posrealset^\infty:\|\vecx\|_1= E}\prob\bigg[\sum\limits_{r=1}^{\rxant}\sum\limits_{i=1}^{\infty} \Big(x_i S_{r,i} \log e -\log(1+x_i) \Big)\leq \eta \bigg] -\error}{\exp(\eta)}  \label{eq:ub-M-conv-temp-1}\IEEEeqnarraynumspace\\
&\geq& \frac{\inf\limits_{\vecx\in\posrealset^\infty:\|\vecx\|_1 = \rxant E}\prob\bigg[\sum\limits_{i=1}^{\infty} \Big(x_i S_i \log e -\log(1+x_i) \Big)\leq \eta \bigg] -\error}{\exp(\eta)}
\label{eq:ub-M-conv-temp}
\end{IEEEeqnarray}
where $\{S_{i}\}$ are again i.i.d. $\expdist(1)$-distributed. Here,~\eqref{eq:ub-M-conv-temp} follows because the feasible region of the optimization problem in~\eqref{eq:ub-M-conv-temp-1} is contained in the feasible region of the optimization problem in~\eqref{eq:ub-M-conv-temp}.

Lemma~\ref{lem:three-atoms} below, which is proven in~Appendix~\ref{app:proof-three-atoms}, sheds light on the structure of the vectors~$\vecx^*$ that minimize the RHS of~\eqref{eq:ub-M-conv-temp}.
\begin{lemma}
\label{lem:three-atoms}
Let $\vecx^*$ be a minimizer of
\begin{IEEEeqnarray}{rCl}
\inf_{\vecx\in \posrealset^{\infty}: \|\vecx\|_1 = \rxant E}\prob\bigg[\sum\limits_{i=1}^{\infty} \Big( x_i S_i \log e -\log(1+x_i) \Big)\leq \eta \bigg].\IEEEeqnarraynumspace
\label{eq:min-outage-prob}
\end{IEEEeqnarray}
Assume without loss of generality that the entries of $\vecx^*$ are in nonincreasing order.
Then, $\vecx^*$ must be of  the form given in~\eqref{eq:three-atoms-form} or in~\eqref{eq:le-two-atoms-form}.
\end{lemma}
%
%
%
%
%

The proof of Theorem~\ref{thm:non-asy-converse-inf} is concluded by  using Lemma~\ref{lem:three-atoms} in~\eqref{eq:ub-M-conv-temp} and by maximizing the RHS of~\eqref{eq:ub-M-conv-temp} over $\eta$.
\end{IEEEproof}

\begin{rem}
The proof of Lemma~\ref{lem:three-atoms} relies on an elegant argument of Abbe, Huang, and Telatar~\cite{abbe13-05}, used in the proof of Telatar's minimum outage probability conjecture  for  MISO Rayleigh-fading channels.
 Indeed, both~\cite{abbe13-05} and Lemma~\ref{lem:three-atoms} deal with the optimization of quantiles of a weighted convolution of exponential distributions.
\end{rem}

\subsection{Asymptotic Analysis}
\label{sec:asy-analy-energy-bit}
Evaluating the bounds in Corollary~\ref{thm:non-asy-inf-bw-Verdu} and Theorem~\ref{thm:non-asy-converse-inf} in the limit $E\to\infty$, we obtain the asymptotic  closed-form expansion for $M^*(E,\error)$ provided in the following theorem.

\begin{thm}
\label{thm:main-result}
The maximum number of messages $M^*(E,\error)$ that can be transmitted with energy~$E$ and error probability $\error \in (0, 1/2)$ over the MIMO Rayleigh block-fading channel~\eqref{eq:MIMO-channel-io} for the case of no CSI admits the following expansion as $E \to\infty$
\begin{IEEEeqnarray}{rCl}
\log M^*(E,\error) &=& \rxant E \log e - V_0\cdot  \Big(\rxant E Q^{-1}(\error)\Big)^{2/3}\big(\log (\rxant E)\big)^{1/3}
+ \bigO\lefto(\frac{E^{2/3} \log\log E}{(\log E)^{2/3}}\right). \IEEEeqnarraynumspace
\label{eq:rate-inf-bandwidth}
\end{IEEEeqnarray}
Here, $V_0$ is given in~\eqref{eq:def-V0-main-thm}.
\end{thm}

\begin{IEEEproof}
See Appendix~\ref{app:proof-thm-1}.
\end{IEEEproof}

The intuition behind~\eqref{eq:rate-inf-bandwidth} is as follows.
It is well known that in the no-CSI case, to achieve the asymptotic limit $-1.59\,\mathrm{dB}$, it is necessary to use flash signalling~\cite{verdu02-06}.
If all codewords satisfy a peak-power constraint $\|\vecx\|_{\infty} \leq A$ in addition to~\eqref{eq:power-constraint}, then  $\log M (E,\error) /(\rxant E)$ converges as $E\to\infty$ to (see~\cite{sethuraman05-09} and~\cite[Eq.~(59)]{durisi10-01})
\begin{IEEEeqnarray}{rCl}
\log e - A^{-1}\log(1+A).
\label{eq:penalty-nocsi}
\end{IEEEeqnarray}
The second term in~\eqref{eq:penalty-nocsi} can be interpreted as the penalty due to bounded peakiness, which vanishes as $A \to \infty$.
When the energy $E$ is finite, as in our setup, it turns out that for large $E$
\begin{IEEEeqnarray}{rCl}
\frac{\log M (E,\error)}{\rxant E} \approx \log e - \frac{\log (1+ A)}{A}  - \sqrt{\frac{A}{ \rxant E} }Q^{-1}(\error)\log e . \IEEEeqnarraynumspace
\label{eq:log-m-sketch}
\end{IEEEeqnarray}
The second term on the RHS of~\eqref{eq:log-m-sketch} captures the fact that codewords that satisfy~\eqref{eq:power-constraint} for a finite~$E$ are necessarily peak-power limited. 
The third term captures the penalty resulting from the stochastic variations of the fading and the noise processes, which cannot be averaged out for finite~$E$.
This penalty increases with the peak power.
 Coarsely speaking, peakier codewords result in less channel averaging. To summarize, peakiness in the codewords reduces the second term on the RHS of~\eqref{eq:log-m-sketch} but increases the third term.
%
%
%
%
%
The optimal peak power $A^*$ that minimizes the sum of these two penalty terms turns out to be
\begin{IEEEeqnarray}{rCl}
A^* = \left(\frac{3}{2}Q^{-1}(\error) \log e\right)^{-\frac{2}{3}} (\rxant E)^{\frac{1}{3}} \log^{\frac{2}{3}} (\rxant E) + \littleo(E^{\frac{1}{3}}).
\label{eq:optimal-peak}
\end{IEEEeqnarray}
Substituting~\eqref{eq:optimal-peak} into~\eqref{eq:log-m-sketch} we obtain~\eqref{eq:rate-inf-bandwidth}. See Appendix~\ref{app:proof-thm-1} for a rigorous proof.

%
%
%
%
%
%
%
%
%

\subsection{The Perfect-CSIR Case}
\label{sec:perfect-csir}
In this section, we provide  achievability and converse bounds on $M^*(E,\error)$ for the case of  perfect CSIR.
To state our achievability bound, it is convenient to introduce the following complex AWGN channel
\begin{equation}
Y_i = X_i + Z_i, \quad i\in\natunum.
\label{eq:channel-io-awgn-csir}
\end{equation}
Here, $\{Z_i\}$ are i.i.d. $\jpg(0,1)$-distributed random variables.
Theorem~\ref{thm:nonasy-coh-ach} below allows us to relate the performance of optimal codes for the AWGN channel~\eqref{eq:channel-io-awgn-csir} to the performance of optimal codes for the MIMO Rayleigh block-fading channel~\eqref{eq:MIMO-channel-io}.

\begin{thm}
\label{thm:nonasy-coh-ach}
Consider an arbitrary $(\rxant E,M,\error)$-code for the AWGN channel~\eqref{eq:channel-io-awgn-csir}. There exists  a sequence of $(E,M,\error_{N})$-codes for the MIMO Rayleigh block-fading channel~\eqref{eq:MIMO-channel-io} with perfect CSIR, for which $
\lim\nolimits_{N\to\infty} \epsilon_N  \leq \epsilon$.
\end{thm}

\begin{rem}
Theorem~\ref{thm:nonasy-coh-ach}  holds also if the fading is not Rayleigh, provided that the entries $\{H_{i,j,k}\}$ of~$\randmatH_i$ are i.i.d. and satisfy $\Ex{}{|H_{i,j,k}|^2}=1$.
\end{rem}
\begin{IEEEproof}
As in the proof of Corollary~\ref{thm:non-asy-inf-bw-Verdu}, it is sufficient to consider the case $\txant =\cohtime =1$.
Take an arbitrary $(\rxant E,M,\error)$-code for the AWGN channel.
We assume without loss of generality that only the first $M$ entries of each codeword are nonzeros.
This is because, for the AWGN channel, the error probability under maximal likelihood decoding depends only on the Euclidean distance between codewords, and because we can embed the $M$ codewords in an $M$-dimensional space without changing their Euclidean distances.
%
%
%
%

%
Next, we transform the SIMO memoryless fading channel (with perfect CSIR) into an AWGN channel as follows. Fix an arbitrary $N\in\natunum$;
for every codeword $\vecu = [u_1,\ldots,u_M,0,\ldots]$ for the AWGN channel, we generate the following codeword~$\tilde{\vecu}$ for the memoryless SIMO fading channel~\eqref{eq:SIMO-channel-io}
\begin{IEEEeqnarray}{rCl}
\tilde{\vecu} \define \frac{1}{\sqrt{\rxant N}}[\underbrace{u_1,\ldots,u_{1}}_{N},\underbrace{u_2,\ldots,u_2}_{N},\ldots].
\end{IEEEeqnarray}
By construction, $\|\tilde{\vecu}\|^2_2 = \|\vecu\|^2_2/\rxant $.
For a given channel output~$\{V_{r,i}\}$ (see~\eqref{eq:SIMO-channel-io}), the receiver performs coherent combining across the $\rxant$ receive antennas and the length-$N$ repetition block:
\begin{IEEEeqnarray}{rCl}
\widetilde{V}_j &\define & \frac{1}{\sqrt{\rxant N}}\sum\limits_{r=1}^{\rxant}\sum\limits_{i=1}^{N} \bar{H}_{r,(j-1)N + i} V_{r,(j-1)N + i}
\\ &=& \frac{u_j }{\rxant N} \sum\limits_{r=1}^{\rxant}\sum\limits_{i=1}^{N} |H_{r,(j-1)N + i} |^2 + \frac{1} {\sqrt{\rxant N}}\sum\limits_{r=1}^{\rxant}\sum\limits_{i=1}^{N}\bar{H}_{r,(j-1)N + i} Z_{r,(j-1)N + i},  \IEEEeqnarraynumspace\notag\\
&&\hfill \, j=1,\ldots,M.
\label{eq:simulate-AWGN-fading}
\end{IEEEeqnarray}
%
If we let $N\to\infty$, the first term in~\eqref{eq:simulate-AWGN-fading} converges in distribution to $u_j$ by the law of large numbers, and the second term converges in distribution to $Z_j\sim \jpg(0,1)$ by the central limit theorem. Therefore, $\widetilde{V}_j$ converges in distribution to $u_j + Z_j$.
Thus, $P_{\widetilde{V}^M | U^M= u^M}$ converges in distribution to an AWGN channel law $P_{V^M \given U^M = u^M}^{\mathrm{AWGN}} = \jpg(u^M, \matI_M)$ as $N\to\infty$.

We next evaluate the error probability $\epsilon_N$ of the code that we constructed above. 
Let  $\setD_j $ denote the decoding region for message $j$, $1\leq j \leq M$, and let $\mathrm{Int}(\setD_j)$ denote the interior of $\setD_j$. 
It follows that for every $1\leq j\leq M$ 
\begin{IEEEeqnarray}{rCl}
\lim\limits_{N\to \infty}  1-\epsilon_N &=& \lim\limits_{N\to \infty} P_{\widetilde{V}^M | U^M= \vecu_j}[\setD_j] \\
 &\geq& \lim\limits_{N\to \infty} P_{\widetilde{V}^M | U^M= \vecu_j}[\mathrm{Int}(\setD_j) ]\label{eq:coh-awgn-transform-1} \\
&=& P_{V^M \given U^M = \vecu_j}^{\mathrm{AWGN}} [\mathrm{Int}(\setD_j) ] \label{eq:coh-awgn-transform-2}\\
&=& P_{V^M \given U^M = \vecu_j}^{\mathrm{AWGN}} [\setD_j ] \label{eq:coh-awgn-transform-3}\\
&\geq& 1-\error.\label{eq:coh-awgn-transform-4}
\end{IEEEeqnarray}
Here,~\eqref{eq:coh-awgn-transform-2} follows because $P_{\widetilde{V}^M | U^M= \vecu_j}$ converges in distribution to $P_{V^M \given U^M = \vecu_j} ^{\mathrm{AWGN}}$ and because $\mathrm{Int}(\setD_j)$ is open;~\eqref{eq:coh-awgn-transform-3} follows because the boundary of  the maximum likelihood decoding region~$\setD_j$ has zero probability measure under $P_{V^M \given U^M = \vecu_j}^{\mathrm{AWGN}}$.
\end{IEEEproof}

Note that  the proof of Theorem~\ref{thm:nonasy-coh-ach} above requires perfect CSIR. The approach just described does not necessarily work if only partial CSI is available at the receiver.
For example, consider the following  partial-CSI model~\cite{medard2000-05a}
\begin{equation}
V_i= (\bar{H}_i +\hat{H}_i) U_i + Z_i,\quad i \in \natunum
\end{equation}
where $\bar{H}_i \sim \jpg(0, \rho)$, $\rho \in (0,1)$, $\hat{H}_i \sim \jpg (0, 1-\rho)$, and $\{\bar{H}_i\}$ and $\{\hat{H}_i\}$ are independent. 
We assume that the receiver has perfect knowledge of $\{\bar{H}_i\}$, but  knows only the statistics of $\{\hat{H}_i\}$. The random variables  $\{\bar{H}_i\}$ and $\{\hat{H}_i\}$ can be viewed as  the estimation of the channel coefficients and the estimation errors, respectively~\cite{medard2000-05a}. 
Following   steps similar to the ones in the proof of~\cite[Th.~7]{verdu02-06}, one can show that flash-signalling is  necessary to achieve the $-1.59$ dB limit. Hence, spreading the energy as it is done in the proof of Theorem~\ref{thm:nonasy-coh-ach}  is not first-order optimal.

For the case where perfect CSI is available at both the transmitter and the receiver, and where the fading distribution has infinite support (e.g., Rayleigh distribution),  it is well known that the minimum energy per bit $\eb^*(k,\epsilon) $ converges to $0$ in the limit   $k\to\infty$ and  $\epsilon \to 0$~\cite[p.~1325]{verdu02-06}.
 Using the approach used in the proof of Theorem~\ref{thm:nonasy-coh-ach}, one  can  show  that $\eb^*(k,\epsilon) = 0 $ for \emph{every} $k$ and $\epsilon>0$. 
Indeed, since both the transmitter and the receiver have perfect CSI, they can agree to use the channel only if the fading gain $|H|^2$ is above a threshold $\Gamma$.  By doing so, we have transformed the original fading channel into a channel with a  fading distribution $\tilde{P}_{H}$ that satisfies $\randmatE_{\tilde{P}_{H}}[|H|^2] \geq \Gamma$. 
Proceeding as in the proof of Theorem~\ref{thm:nonasy-coh-ach}, we conclude that every $(E,2^k,\error)$ code for the AWGN channel can be converted into an $(E/\randmatE_{\tilde{P}_{H}} [|H|^2], 2^k, \epsilon)$ code for the fading channel with  distribution $\tilde{P}_{H}$. 
Since $\Gamma$ can be taken arbitrarily large, we conclude that the minimum energy per bit $\eb^*(k,\epsilon)$  is $0$.

Theorem~\ref{thm:nonasy-coh-ach} implies that the asymptotic  expansion~\eqref{eq:intro-second-order-awgn} with $E$ replaced by $\rxant E$ is achievable in the perfect-CSIR case.
Theorem~\ref{thm:asy-coh} below establishes that, for $0<\error<1/2$, the converse is also true.

\begin{thm}
\label{thm:asy-coh}
The maximum number of messages $M^*(E,\error)$ that can be transmitted with energy~$E$ and error probability $0<\error<1/2$ over the MIMO Rayleigh block-fading channel~\eqref{eq:MIMO-channel-io} for the case of perfect CSIR satisfies
\begin{IEEEeqnarray}{rCl}
\log M^*(E,\error) &=& \rxant E \log e - \sqrt{2\rxant E}Q^{-1}(\error)\log e +\frac{1}{2}\log (\rxant E) + \bigO(\sqrt{\log E})\IEEEeqnarraynumspace
\label{eq:second-order-coh}
\end{IEEEeqnarray}
as  $E\to\infty$.
\end{thm}
\begin{IEEEproof}
See Appendix~\ref{app:proof-asy-coh}.
\end{IEEEproof}

Unlike Theorem~\ref{thm:nonasy-coh-ach}, the converse part of Theorem~\ref{thm:asy-coh} relies on the Gaussianity of the fading coefficients and does not necessarily  hold for other fading distributions.
Indeed, consider a single-input single-output (SISO) on-off fading channel $P_{V,H |U}$ where the channel coefficients $\{H_i\}$ are i.i.d. and satisfy
\begin{IEEEeqnarray}{rCl}
\prob[H_i = 0] = \error', \quad \prob[|H_i|^2 = 1/(1-\error')] = 1-\error'
\end{IEEEeqnarray}
where $0<\error'<\error$. Such a fading distribution satisfies $\Ex{}{|H_i|^2}=1$.
Set now $N=1$, $x_0 = \sqrt{E}$, and
\begin{IEEEeqnarray}{rCl}
M &=& \left(Q\lefto(\sqrt{\frac{2E}{1-\error'}}+ Q^{-1}\lefto(1-\frac{\error-\error'}{1-\error'} + \sqrt{\frac{2(1-\error')}{E}}\right) \right)\right)^{-1}.
\end{IEEEeqnarray}
Let
\begin{IEEEeqnarray}{rCl}
\imath(u;v,h) \define \frac{dP_{V,H|U=u}}{dP_{V,H|U=0}}(v,h).
\label{eq:def-info-density-coh-counter-example}
\end{IEEEeqnarray}
By Theorem~\ref{thm:RCU-cpuc}, there exists an $(E,M,\epsilon'')$-code for which the maximal probability of error $\epsilon'' $ is upper-bounded as follows:
\begin{IEEEeqnarray}{rCl}
\epsilon'' &\leq& \Ex{}{ \min\mathopen{}\Big\{1, (M-1)\prob\lefto[  \imath(x_0; V, H)  \leq \imath(x_0; \hat{V},H)\given V, H\right]\! \Big\}}\label{eq:error-counter-example1}\\
&\leq & (1-\error') \Ex{}{ \min\mathopen{}\Big\{1, (M-1)\prob\lefto[  \imath(x_0; V, H)  \leq \imath(x_0; \hat{V},H)\Big| V, |H|^2=(1-\error')^{-1}\right]\! \Big\}} + \error'\IEEEeqnarraynumspace   \label{eq:error-counter-example2} \\
&\leq &(1-\error')\left(\frac{\error-\error'}{1-\error'}-\sqrt{\frac{2(1-\error')}{E}} + \sqrt{\frac{(1-\error')}{E}} (1+\littleo(1)) \right) + \error',\qquad\qquad E\to\infty.\label{eq:error-counter-example}
\end{IEEEeqnarray}
Here,  in~\eqref{eq:error-counter-example1} and~\eqref{eq:error-counter-example2}, $P_{HV\hat{V}}(h,v,\hat{v}) = P_H(h) P_{V|H,U} (v| h ,x_0) P_{V|H,U}(\hat{v} | h, 0)$, and~\eqref{eq:error-counter-example} follows from~\cite[Eqs.~(33)--(40)]{polyanskiy11-08b}. For sufficiently large $E$, the RHS of~\eqref{eq:error-counter-example} is less than $\error$. This implies that $M^*(E,\error) \geq M$ for sufficiently large $E$. Furthermore, by~\cite[Eqs.~(47)--(49)]{polyanskiy11-08b}, we have
\begin{IEEEeqnarray}{rCl}
\log M^*(E,\error) \geq \log M \geq   \frac{E\log e}{1-\error'} + \bigO(\sqrt{E}), \quad E\to\infty.
\label{eq:on-off-fading}
\end{IEEEeqnarray}
Clearly, the RHS of~\eqref{eq:on-off-fading} is greater than the RHS of~\eqref{eq:second-order-coh} (computed for $\rxant =1$) for large $E$.

In Theorem~\ref{thm:nonasy-conv-coh-epb} below, we present a nonasymptotic converse bound, which we shall evaluate numerically in Section~\ref{sec:numerical-results}.
\begin{thm}
\label{thm:nonasy-conv-coh-epb}
Fix $\eta>6$, $E>0$, and $0<\error<1/2$.
Let $x_1(\eta) >\eta$ be the unique solution of
\begin{IEEEeqnarray}{rCl}
\frac{1}{4\sqrt{\pi}} e^{- (x_1-\eta)^2/(4x_1)} \left(\frac{\eta}{\sqrt{x_1} } +\sqrt{x_1}\right) = Q\lefto(\frac{\eta - x_1}{\sqrt{2x_1}}\right). \IEEEeqnarraynumspace
\label{eq:def-x1-tangent}
\end{IEEEeqnarray}
Furthermore, let
\begin{IEEEeqnarray}{rCl}
g_{\eta}( x )\define \left\{
                     \begin{array}{ll}
                       Q\lefto((x-\eta)/\sqrt{2x}\right), & \hbox{$x> x_1(\eta)$,} \\
                       1-\dfrac{x}{x_1(\eta)}Q\mathopen{}\bigg( \dfrac{\eta - x_1(\eta)}{\sqrt{2 x_1(\eta)}}\bigg) , & \hbox{$x \leq x_1(\eta)$.}
                     \end{array}
                   \right.
\label{eq:convex-lb}
\end{IEEEeqnarray}
Every $(E,M,\error)$-code for the MIMO Rayleigh block-fading channel~\eqref{eq:MIMO-channel-io} for the case of perfect CSIR satisfies
\begin{IEEEeqnarray}{rCl}
\log M \leq \eta\log e - \log\lefto|g_{\eta}(\rxant E) - \error\right|^{+}.
\label{eq:converse-csir-nonasymp}
\end{IEEEeqnarray}
\end{thm}

\begin{IEEEproof}
See Appendix~\ref{app:proof-converse-csir-nonasy}.
\end{IEEEproof}

\subsection{Numerical Results}
\label{sec:numerical-results}
\begin{figure}[t]
\centering
\includegraphics[scale=0.9]{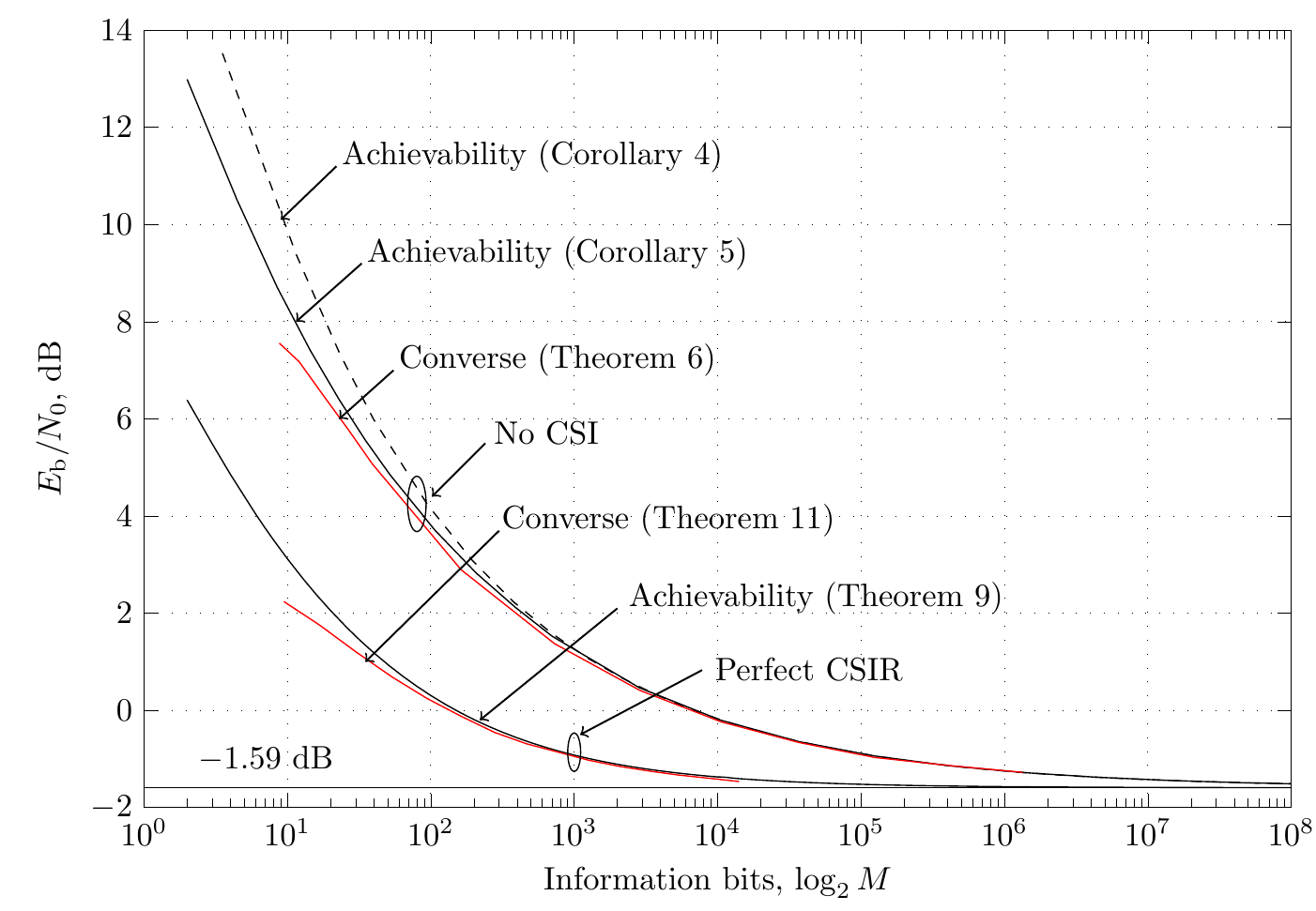}
\caption{Minimum energy per bit versus number of information bits; here $\error = 10^{-3}$ and $\rxant =1$.\label{fig:energy-info-tradeoff} }
\end{figure}

\begin{table}
\begin{center}
\caption{Minimum energy $E$ and optimal number of channel uses $N^*$ vs. number of information bits $k$ for the case $\error=10^{-3}$.\label{table:optimal-mass}}
\begin{tabular}{|c|c|c|c|c|c|c|}
  \hline
    & \multicolumn{2}{c|}{Cor.~\ref{thm:non-asy-inf-bw-Verdu} } & \multicolumn{2}{c|}{Cor.~\ref{thm:ML-lb-inf-bw}} & \multicolumn{2}{c|}{Asymptotics}\\
    \hline
   $k$ & $E/N_0$ & $N^*$ & $E/N_0$ &$N^*$ & $E/N_0$ & $N^*$\\
  \hline\hline
  $10^1$ & $98$ & $25$& $67$ & $18$ & $120$ & $50$\\
  $10^2$ & $2.6\times 10^2 $& $39$ & $2.4\times 10^2$ & $38$ & $2.9\times 10^2$& $63$ \\
  $10^3$ & $1.3 \times 10^3$ &$96$ & $1.3 \times 10^3$ & $96$ & $1.4\times 10^3$& $124$\\
  $10^4$ & $9.6 \times 10^3$ & $304$ &  $9.6\times 10^3$ & $304$ &$9.7 \times 10^3$ & $336$\\
  $10^5$ & $8.2\times 10^4$ & $1089$ & $8.2\times 10^4$ & $1090$ & $8.2 \times 10^4$ & $1137$\\
  \hline
\end{tabular}
\end{center}
\end{table}

Fig.~\ref{fig:energy-info-tradeoff} shows\footnote{The numerical routines used to obtain these results are available at https://github.com/yp-mit/spectre}
 the achievability bounds (Corollary~\ref{thm:non-asy-inf-bw-Verdu} and Corollary~\ref{thm:ML-lb-inf-bw}) and the converse bound (Theorem~\ref{thm:non-asy-converse-inf}) for the channel~\eqref{eq:MIMO-channel-io} for the no-CSI case and when $\error = 10^{-3}$ and $\rxant =1$.
Specifically, the energy per bit $E_\mathrm{b} = E/\log_2M^*(E,\error)$ is plotted against the number of information bits $\log_2 M^*(E,\error)$.
For the perfect-CSIR case, we plot the converse bound (Theorem~\ref{thm:nonasy-conv-coh-epb}) together with the achievability bound provided in~\cite[Eq.~(15)]{polyanskiy11-08b} for the AWGN case.
As proved in Theorem~\ref{thm:nonasy-coh-ach}, this bound is also achievable in perfect-CSIR case.
As expected, as the number of information bits increases, the minimum energy per bit converges to $-1.59$ dB regardless of whether CSIR is available or not.
However, for a fixed number of information bits, it is more costly to communicate in the no-CSI case than in the perfect-CSIR case.
For example, it takes $2$ dB more of energy to transmit $1000$ information bits in the no-CSI case compared to the perfect-CSIR case.
Additionally, to achieve an energy per bit of $- 1.5$ dB, we need to transmit $7\times 10^7$ information bits in the no-CSI case, but only $6\times 10^4$ bits when perfect CSIR is available.

The codebook used in both Corollary~\ref{thm:non-asy-inf-bw-Verdu} and Corollary~\ref{thm:ML-lb-inf-bw} uses only one symbol $x_0^*$ of the input alphabet in addition to $0$.
In Table~\ref{table:optimal-mass} we list the number of channel uses $N^* = E/ x_0^*$ over which the  optimal input symbol $x_0^*$ is repeated, as a function of the number of information bits $k$.
For comparison, we also list the number of repetitions $N^*\approx \left(\frac{3}{2}Q^{-1}(\error)\frac{E\log e}{\log  E}\right)^{2/3}$ predicted by the asymptotic analysis (see~\eqref{eq:def-N-star}).

\section{Conclusions}

In this paper, we established nonasymptotic bounds on the minimum energy per bit $\eb^*(k,\error)$ required to transmit $k$ information bits with  error probability $\error$ over a MIMO Rayleigh block-fading channel.
As the number of information bits $k$ goes to infinity, the ratio between $\eb^*(k,\error)$ and the noise level converges to $-1.59$ dB, regardless of whether CSIR is available or not.
However, in the nonasymptotic regime of finite $k$ and nonzero error probability $\error$, the minimum energy per bit required in the no-CSI case is larger than that in the perfect-CSIR case (see Fig.~\ref{fig:energy-info-tradeoff}).
Specifically, as $k\to\infty$ the gap to $-1.59$ dB is proportional to $((\log k)/k)^{1/3}$ in the no-CSI case, and to $1/\sqrt{k}$ in the perfect-CSIR case.

The optimal signalling strategies for the two cases are different: in the no-CSI case, the transmitted codewords must have sufficient peakiness in order to overcome the lack of channel knowledge; in the perfect-CSIR case, the energy of each codeword must be spread uniformly over sufficiently many fading blocks in order to mitigate the stochastic variations on the received-signal energy caused by the fading process.

Throughout the paper, we have focused on the scenario where the blocklength of the code is unlimited, i.e., the spectral efficiency is zero.
From a practical perspective, generalizing our analysis to the case of low but nonzero spectral efficiency is of interest. In the asymptotic regime $k\to\infty$, this can be done by approximating the spectral efficiency by an affine function of the energy per bit, and by characterizing the slope of the spectral efficiency versus energy per bit function at $-1.59$ dB (wideband slope)~\cite{verdu02-06}.
A generalization of Verd\'{u}'s wideband-slope analysis to the finite-$k$ case seems to require more sophisticated tools than the one used in the present paper (see~\cite[Sec. V.C]{yang16-inprep} for some preliminary results in this direction).

\appendices

\section{Proof of Lemma~\ref{lem:three-atoms}}
\label{app:proof-three-atoms}
The proof relies on~\cite{abbe13-05}.
In particular, we shall make repeated use of~\cite[Cor.~1 and Lem.~2]{abbe13-05}, which are restated below for convenience.
For a continuous random variable $A$, let $f_{A}$ denote its probability density function (pdf), and let $f'_{A}$ and $f''_{A}$ denote the first and the second derivatives of~$f_A$, respectively.
 Furthermore, let $S_1$ and $S_2$ be independent $\expdist(1)$-distributed random variables, which are also independent of $A$.
 Then, for every $x,q_1,q_2 \in \realset$~\cite[Lem.~2]{abbe13-05}:
\begin{IEEEeqnarray}{rCl}
\pdfa_{A + q_1 S_1}(x) - \pdfa_{A+q_2 S_2}(x) = (q_2 -q_1) \pdfa'_{A+q_1S_1 + q_2S_2}(x). \IEEEeqnarraynumspace
\label{eq:abbe-huang-lemma2}
\end{IEEEeqnarray}
This identity can be readily verified by computing the Fourier transform of both sides.
Setting $q_2 = 0$ in~\eqref{eq:abbe-huang-lemma2}, we obtain~\cite[Cor.~1]{abbe13-05}
\begin{IEEEeqnarray}{rCl}
\pdfa_{A + q_1 S_1}(x) - \pdfa_{A}(x) =  - q_1  \pdfa'_{A+q_1S_1}(x).
\label{eq:abbe-huang-cor1}
\end{IEEEeqnarray}
%

%

The proof of Lemma~\ref{lem:three-atoms} consists of four steps.
\begin{enumerate}
\item  We first restrict ourselves to the finite-dimensional setup, i.e., we assume that $\vecx\in \posrealset^m$ for some $m\in\natunum$. We shall derive a necessary condition a minimizer $\vecx^*\in\posrealset^m$ must satisfy, by deriving the Karush-Kuhn-Tucker (KKT) optimality conditions (see, e.g.,~\cite[Sec.~5.5.3]{boyd04}).

\item Building upon these conditions, we show that the entries of~$\vecx^*$ can take at most three distinct nonzero values.
\item We prove that if the entries of~$\vecx^*$ take exactly three distinct nonzero values, then the maximum and the minimum nonzero value must appear only once, i.e.,~$\vecx^*$ is of the form~\eqref{eq:three-atoms-form}.
If the entries of~$\vecx^*$ take less than three distinct nonzero values, then~$\vecx^*$ satisfies~\eqref{eq:le-two-atoms-form} trivially.
\item Finally, we take $m$ to infinity to complete the proof.
\end{enumerate}
Departing from our convention, in this appendix we shall use $\log$ to denote the natural logarithm.

\subsection{The KKT conditions}

Let 
\begin{IEEEeqnarray}{rCl}
\varphi(\vecx ,\vecs) \define \sum\limits_{i=1}^{m} \Big(x_i s_i -\log(1+x_i)\Big).
\label{eq:def-varphi-s}
\end{IEEEeqnarray}
Using~\eqref{eq:def-varphi-s}, we can express~\eqref{eq:min-outage-prob} for the case $\vecx \in \posrealset^m$ as 
\begin{equation}
\inf\limits_{\vecx \in \posrealset^m: \|\vecx\|_1 =E } \prob[\varphi(\vecx,\randvecs) \leq \eta].
 \label{eq:opt-m-over-x-rm}
 \end{equation}
By the KKT optimality conditions, if $\vecx^*$ is a minimizer of~\eqref{eq:min-outage-prob}, then there must exist a~$\lambda \in \realset$ such that
for all $k=1,\ldots,m$,
\begin{IEEEeqnarray}{rCl}
\frac{\partial \prob[\varphi(\vecx,\randvecs) \leq \eta]}{\partial x_k} \bigg|_{\vecx = \vecx^*}
\left\{
        \begin{array}{ll}
          =\lambda, & \hbox{if $x_k^* >0$} \\
          \geq  \lambda , & \hbox{otherwise.}
        \end{array}
      \right.\label{eq:KKT}
\end{IEEEeqnarray}
%
%
Let $\hat{S}_k$ be an $\expdist(1)$-distributed random variable that is independent of~$\randvecs$.
Let $\langle \vecx, \randvecs\rangle \define \sum\nolimits_{i=1}^{m} x_i S_i$, and let $\tilde{\eta} \define \eta + \sum\nolimits_{j=1}^{m}\log(1+x_j)$.
The partial derivative in~\eqref{eq:KKT} can be computed through a Fourier analysis as in the proof of~\cite[Lem.~1]{abbe13-05}. This yields
\begin{IEEEeqnarray}{rCl}
\frac{\partial \prob[\varphi(\vecx,\randvecs) \leq \eta]}{\partial x_k}
&=& \frac{f_{\langle\vecx,\randvecs\rangle } (\tilde{\eta})}{1+x_k} -\pdfa_{\langle\vecx,\randvecs\rangle + x_k\hat{S}_k}(\tilde{\eta}).  \label{eq:deriv-prob-3atoms-lemma-3}
\end{IEEEeqnarray}
%
%
 %
%
From~\eqref{eq:deriv-prob-3atoms-lemma-3}, it follows that
\begin{IEEEeqnarray}{rCl}
\IEEEeqnarraymulticol{3}{l}{\frac{\partial \prob[\varphi(\vecx,\randvecs) \leq \eta]}{\partial x_j} - \frac{\partial \prob[\varphi(\vecx,\randvecs) \leq \eta]}{\partial x_k} }\notag\\
\quad &=& f_{\langle\vecx,\randvecs\rangle + x_k\!\hat{S}_k}\!(\tilde{\eta}) -f_{\langle\vecx,\randvecs\rangle + x_j\hat{S}_j}\!(\tilde{\eta}) +  \frac{(x_k \!- \! x_j)f_{\langle\vecx,\randvecs\rangle } (\tilde{\eta})}{(1+x_k)(1+x_j)}\\
&=& (x_j\! -\! x_k)\bigg\{ f'_{\langle\vecx,\randvecs\rangle + x_k\hat{S}_k + x_j\! \hat{S}_j}\!(\tilde{\eta}) -\frac{f_{\langle\vecx,\randvecs\rangle } (\tilde{\eta})}{(1+x_k)(1+x_j)}\bigg\}\IEEEeqnarraynumspace
\label{eq:KKT-equiv}
\end{IEEEeqnarray}
where in the last step we used~\eqref{eq:abbe-huang-lemma2}.

\subsection{The entries of a minimizer can take at most three distinct nonzero values}
\label{sec:proof-three-nonzero}
As in~\cite{abbe13-05}, our proof is by contradiction. We shall assume without loss of generality that $m\geq 4$.
Let $\vecx^*$ be a minimizer of~\eqref{eq:opt-m-over-x-rm}, and assume that the entries of $\vecx^*$ take more than three distinct nonzero values, the smallest four of them being $0<x_1^*<x_2^*<x_3^*<x_4^*$.
Then, by~\eqref{eq:KKT} and~\eqref{eq:KKT-equiv},
\begin{IEEEeqnarray}{rCl}
(1+x_j^*)f'_{\langle\vecx^*,\randvecs\rangle + x_1^*\hat{S}_1 + x_j^* \hat{S}_j}(\tilde{\eta}) = \frac{f_{\langle\vecx^*,\randvecs\rangle } (\tilde{\eta})}{(1+x_1^*)},\,\,\,\, j=2,3,4. \IEEEeqnarraynumspace
\label{eq:equiv-KKT2}
\end{IEEEeqnarray}
By~\eqref{eq:abbe-huang-cor1}, the left-hand side (LHS) of~\eqref{eq:equiv-KKT2} can be expressed as follows:
\begin{equation}
 (1+x_j^*)f'_{\langle\vecx^*,\randvecs\rangle + x_1^*\hat{S}_1 + x_j^* \hat{S}_j}(\tilde{\eta}) =
 f'_{\langle\vecx^*\!,\randvecs\rangle + x_1^*\hat{S}_1 + x_j^* \hat{S}_j}\!(\tilde{\eta}) +  f_{\langle\vecx^*\!,\randvecs\rangle + x_1^*\hat{S}_1 }\!(\tilde{\eta})
 - f_{\langle\vecx^*\!,\randvecs\rangle + x_1^*\hat{S}_1 + x_j^* \hat{S}_j}\!(\tilde{\eta}).  
\label{eq:equivalent-diff-deriv}
\end{equation}
%
Since the RHS of~\eqref{eq:equiv-KKT2} does not depend on $j$, by substituting~\eqref{eq:equivalent-diff-deriv} into~\eqref{eq:equiv-KKT2} and by taking the difference between the case $j=2$ and the case $j=3$, we obtain
\begin{IEEEeqnarray}{rCl}
0 &=&  f'_{\langle\vecx^*,\randvecs\rangle + x_1^*\hat{S}_1 + x_2^* \hat{S}_2}(\tilde{\eta}) -f'_{\langle\vecx^*,\randvecs\rangle + x_1^*\hat{S}_1 + x_3^* \hat{S}_3}(\tilde{\eta})\notag \\
 && -\,   \Big(f_{\langle\vecx^*,\randvecs\rangle + x_1^*\hat{S}_1+ x_2^* \hat{S}_2}(\tilde{\eta}) - f_{\langle\vecx^*,\randvecs\rangle + x_1^*\hat{S}_1 + x_3^* \hat{S}_3}(\tilde{\eta}) \Big) \IEEEeqnarraynumspace \label{eq:difference-2-3-step1} \\
&=& (x_3^* - x_2^*) \Big( f''_{\langle\vecx^*,\randvecs\rangle + x_1^*\hat{S}_1 + x_2^* \hat{S}_2+ x_3^*\hat{S}_3}(\tilde{\eta} )
- f'_{\langle\vecx^*,\randvecs\rangle + x_1^*\hat{S}_1 + x_2^* \hat{S}_2+ x_3^*\hat{S}_3}(\tilde{\eta} )\Big)  \label{eq:difference-2-3}.
\end{IEEEeqnarray}
Here,~\eqref{eq:difference-2-3} follows by~\eqref{eq:abbe-huang-lemma2}.
Set
\begin{IEEEeqnarray}{rCl}
A \define \langle\vecx^*,\randvecs\rangle + x_1^*\hat{S}_1 + x_2^*\hat{S}_2.
\label{eq:def-a-weighted-sum-x-s}
\end{IEEEeqnarray}
Since $x_2^* \neq x_3^*$ by assumption,~\eqref{eq:difference-2-3} can be rewritten as
\begin{IEEEeqnarray}{rCl}
f''_{A + x_3^*\hat{S}_3}(\tilde{\eta} ) - f'_{A + x_3^*\hat{S}_3}(\tilde{\eta} ) = 0.
\label{eq:ratio_2_3_Equal1}
\end{IEEEeqnarray}
Following the same steps as in~\eqref{eq:difference-2-3-step1}--\eqref{eq:ratio_2_3_Equal1}, we also have that
\begin{IEEEeqnarray}{rCl}
f''_{A  +  x_4^*\hat{S}_4}(\tilde{\eta} )  - f'_{A +  x_4^*\hat{S}_4}(\tilde{\eta} )  = 0.
\label{eq:ratio_2_4_Equal1}
\end{IEEEeqnarray}

Next, we show that~\eqref{eq:ratio_2_3_Equal1} and~\eqref{eq:ratio_2_4_Equal1} cannot hold simultaneously.
This in turn implies that the entries of $\vecx^*$ must take at most three distinct nonzero values.
Let
\begin{IEEEeqnarray}{rCl}
g(t) \define f''_{A + t\hat{S}}(\tilde{\eta} ) - f'_{A  + t \hat{S}}(\tilde{\eta} )
\end{IEEEeqnarray}
where $\hat{S} \sim \expdist(1)$.
%
%
%
%
Since~\eqref{eq:ratio_2_3_Equal1} and~\eqref{eq:ratio_2_4_Equal1} imply that $g(x_3^*) = g(x_4^*) =0$,
to establish a contradiction between~\eqref{eq:ratio_2_3_Equal1} and~\eqref{eq:ratio_2_4_Equal1}, it suffices to show that the function $g(t)$ has at most one zero on $(0,\infty)$.
%
%
Observe that $g(t)$ can be rewritten as
\begin{IEEEeqnarray}{rCl}
g(t) &=& f''_{A+t\hat{S}}(\tilde{\eta}) - f'_{A+t\hat{S}} (\tilde{\eta}) \\
 &=& \Big(f''_{A} \star f_{t\hat{S}}\Big)(\tilde{\eta}) - \Big(f'_{A} \star f_{t\hat{S}}\Big)(\tilde{\eta})\\
&=&\frac{1}{t} \int\nolimits_{0}^{\tilde{\eta}} \Big(f''_{A}(\tilde{\eta} - z)-f'_{A }(\tilde{\eta} - z)\Big)e^{-z/t} dz .
\end{IEEEeqnarray}
Since the kernel $e^{-z/t}$ is \emph{strictly totally positive}~\cite[p.~11]{karlin68} on $[0,\tilde\eta]\times [0,\infty)$, it follows from~\cite[Th.~3.1(b)]{karlin68} that the number of zeros of $g(t)$ on $(0,\infty)$ cannot exceed the number of sign changes of $z\mapsto f''_{A}(z)-f'_{A}( z)$ on $(0,\tilde{\eta})$, provided that the latter number is finite.
%
Thus, to prove that $g(t)$ has at most one zero over $(0,\infty)$, it suffices to show that $f''_{A}(z)-f'_{A}( z)$ changes sign at most once on~$(0,\tilde{\eta})$.
%
%
In fact, we shall prove that it changes sign at most once over an interval that contains $(0,\tilde{\eta})$.
By~\cite[Lem.~3]{abbe13-05}, $f_{A}'(z)$ is continuous on $\realset$, and there exists a $\hat{z}>0$ such that $f_A'(z)>0$ for all $z\in(0,\hat{z})$.
Let $\bar{z} =\arg\max\limits_{z\in[0,1]} f_A'(z)$. 
Since $f_{A}'(0)=0$ (which follows because $f_A'(z)=0$ for all $z<0$ and because $f_A'(z)$ is continuous) and since $f_A'(z)>0$ for all $z\in(0,\hat{z})$, we have that $0<\bar{z}\leq 1$.
This implies that
\begin{IEEEeqnarray}{rCl}
f_A'(\bar{z}) - f_A(\bar{z}) = f_A'(\bar{z}) -\int\nolimits_{0}^{\bar{z}} f_A'(z) dz \geq f_A'(\bar{z}) - \bar{z}f_A'(\bar{z})\geq 0.
\end{IEEEeqnarray}
By Lemma~\ref{lem:strict-log-concavity} in Appendix~\ref{app:convo-exp-dist}, $f_A$ is strictly log-concave on $(0,\infty)$, which implies that $z \mapsto  f'_{A }(z)/ f_{A}(z)$ is strictly decreasing on $(0,\infty)$.
This in turn implies that there exists a unique  $z_0>0$ such that $f'_{A} (z_0) - f_{A}(z_0) = 0$. It also implies that $f'_{A} (z) - f_{A}(z) >0$ if $z\in(0,z_0)$ and $f'_{A} (z) - f_{A}(z)<0$ if $z>z_0$.
We shall now prove that
\begin{enumerate}
\item  $f''_{A}(z)-f'_{A}( z)$ changes sign at most once on~$(0,z_0)$;
 \item  $(0,\tilde{\eta}) \subset(0,z_0)$, i.e.,
\begin{IEEEeqnarray}{rCl}
\tilde{\eta} < z_0.
\label{eq:titlde-eta-leq-z0}
\end{IEEEeqnarray}
\end{enumerate}


\subsubsection{The function $f''_{A}(z)-f'_{A}(z)$ changes sign at most once on  $(0,z_0)$}
\label{app:subsection-zero-sign-change}
It suffices to prove that $f'_{A}(z)-f_{A}(z)$ is unimodal on~$(0,z_0)$.
This is done by induction.
Recall that~$f_A$ is the convolution of exponential pdfs (see~\eqref{eq:def-a-weighted-sum-x-s}), i.e., $A$  can be written as
$\sum\nolimits_{i=1}^{m'} a_i S_i$ for some $a_i>0$, $i=1,\ldots,m'$, and $2\leq m'\leq m+2$.
Let~$B_k$, $k=1,\ldots, m'$, denote the partial sum $\sum\nolimits_{i=1}^{k}a_i S_i $ and  let $z_k$, $k=2,\ldots$, denote the solution of $f'_{B_k}(z_k) - f_{B_k}(z_k) =0$.
Recall that, by the strict log-concavity of $f_{B_k}$, we have that $z_k$ is unique and that  $f'_{B_k}(z) - f_{B_k}(z)>0$ if $z\in(0,z_k)$ and $f'_{B_k}(z) - f_{B_k}(z)<0$ if $z>z_k$.
It can be verified that $f'_{B_2} -f_{B_2}$ is unimodal on $(0,z_2)$.
Assume now that~$f'_{B_k} - f_{B_k}$ is unimodal on $(0,z_k)$ for some $k>2$.
We next show that $f'_{B_{k+1}} \!\!\!- f_{B_{k+1}}$ is unimodal on $(0,z_{k+1})$.
 Note that
 \begin{IEEEeqnarray}{rCl}
 f'_{B_{k+1}} - f_{B_{k+1}} &=&  (f'_{B_k} - f_{B_k})\star f_{a_{k+1} S_{k+1}}.
 \end{IEEEeqnarray}
 Since $f'_{B_k} - f_{B_k}$ and $f_{a_{k+1}S_{k+1}}$ are smooth and strictly positive on~$(0,z_k)$, it follows that $(f'_{B_{k+1}} - f_{B_{k+1}})(z_k) > 0$. This implies that $z_{k+1} > z_{k}$.
 Since $f'_{B_k} - f_{B_k}$ is positive and unimodal on $(0, z_k)$, and since $f_{a_{k+1}S_{k+1}}$ is log-concave, it follows that $f'_{B_{k+1}} - f_{B_{k+1}}$ is positive and  unimodal on~$(0,z_k)$~\cite{ibragimov56}.
Furthermore, the strict log-concavity of $f_{B_k}$ and the definitions of $z_k$ and $z_{k+1}$ imply that, for every $z\in[z_k,z_{k+1})$,
 \begin{IEEEeqnarray}{rCl}
\big( f''_{B_{k+1}}\! -\! f'_{B_{k+1}}\big)(z)
&=& \frac{1}{a_{k+1}}\Big(\!\underbrace{f'_{B_{k}}\!(z) -\! f_{B_{k}}\!(z)}_{\leq 0} - \big(\underbrace{ f'_{B_{k+1}}\!(z) -\! f_{B_{k+1}}\! (z)}_{> 0}\big)\Big)<0\IEEEeqnarraynumspace \label{eq:diff-fb-1}
 \end{IEEEeqnarray}
The first step follows by applying~\eqref{eq:abbe-huang-cor1} twice.
The inequality~\eqref{eq:diff-fb-1} implies that $f'_{B_{k+1}} - f_{B_{k+1}}$ is unimodal on $(0,z_{k+1})$.
Hence, by induction, $f'_{A} - f_{A}$ is unimodal on $(0,z_0)$.

\subsubsection{Proof of~\eqref{eq:titlde-eta-leq-z0}}
It follows from~\eqref{eq:equiv-KKT2} that $f'_{A} (\tilde{\eta} )>0$, which implies by~\cite[Lem.~3]{abbe13-05} that $f'_{A} (t)>0$ for all $t\in(0,\tilde{\eta})$.
Therefore, we have
\begin{IEEEeqnarray}{rCl}
 f'_{A + x_3^*\hat{S}_3}(\tilde{\eta} ) = \big(f'_{A} \star f_{x_3^* \hat{S}_3}\big) (\tilde{\eta})> 0.
\end{IEEEeqnarray}
By the strict log-concavity of~$f_{A+ x_3^*\hat{S}}(\cdot)$ and by~\eqref{eq:ratio_2_3_Equal1},
\begin{IEEEeqnarray}{rCl}
\frac{f'_{A+ x^*_3 \hat{S}} (\tilde{\eta})}{f_{A+x^*_3\hat{S}} (\tilde{\eta})} > \frac{f''_{A+x^*_3\hat{S}}(\tilde{\eta}) }{f'_{A+x^*_3\hat{S}}(\tilde{\eta}) } =1.
\label{eq:ratio-d-pdf-eta}
\end{IEEEeqnarray}
Moreover, by~\eqref{eq:abbe-huang-cor1} and~\eqref{eq:ratio_2_3_Equal1},
\begin{IEEEeqnarray}{rCl}
f_{A}(\tilde{\eta}) - f_{A+x^*_3 \hat{S}_3}(\tilde{\eta})  = f'_{A}(\tilde{\eta})  - f'_{A+x^*_3 \hat{S}_3}(\tilde{\eta}) .
\label{eq:diff-f'-f}
\end{IEEEeqnarray}
Using~\eqref{eq:ratio-d-pdf-eta} in~\eqref{eq:diff-f'-f}, we conclude that
\begin{IEEEeqnarray}{rCl}
f'_{A}(\tilde{\eta}) - f_{A}(\tilde{\eta}) >0
\end{IEEEeqnarray}
This implies~\eqref{eq:titlde-eta-leq-z0}. 

\subsection{The minimum and maximum nonzero values must each appear only once}
We focus on the case when the entries of $\vecx^*$ take exactly three distinct nonzero values.
Assume without loss of generality that $\vecx^*$ has the following form
\begin{IEEEeqnarray}{rCl}
\vecx^* =[\underbrace{x_1^*,\ldots,x_1^*}_{N_1},\underbrace{x_2^*,\ldots,x_2^*}_{N_2},\underbrace{x_3^*,\ldots,x_3^*}_{N_3},0,\ldots,0]
\end{IEEEeqnarray}
where $x_1^*N_1 + x_2^*N_2 + x_3^*N_3 = E$, $0 < x_1^*<x_2^*<x_3^*$, and $N_1,N_2,N_3 >0$.
We shall prove that if $N_1 > 1$, then
\begin{IEEEeqnarray}{rCl}
\frac{\partial^2 \prob[\varphi(\vecx_\delta^* ,\randvecs) \leq \eta] }{\partial \delta^2} \bigg|_{\delta =0} < 0
\label{eq:2nd-deriv-KT-negative}
\end{IEEEeqnarray}
where $\vecx^*_{\delta} \define \vecx^* + \delta \vece_1 - \delta\vece_2$.
Since this contradicts the assumption that $\vecx^*$ is  a minimizer, we conclude that $N_1 = 1$.
Using a similar argument, one can show that $N_3 = 1$.

We first compute the LHS of~\eqref{eq:2nd-deriv-KT-negative}.
Assume $N_1>1$, so that $[\vecx^*]_1=[\vecx^*]_2=x_1^*$.
Set $\tilde{\eta}_\delta \define \eta + \sum\limits_{i=1}^{m}\log(1+[\vecx^*_\delta]_i)$.
Proceeding similarly as in the proof of~\eqref{eq:deriv-prob-3atoms-lemma-3}, we obtain
\begin{IEEEeqnarray}{rCl}
\IEEEeqnarraymulticol{3}{l}{
\frac{\partial \prob[\varphi(\vecx^*_\delta ,\randvecs) \leq \eta] }{\partial \delta} }\notag \\
 &=& - \left( f_{\langle \vecx^*_\delta,\randvecs\rangle +(x_1^*+\delta)\hat{S}_1 } (\tilde{\eta})  - f_{ \langle \vecx_\delta,\randvecs\rangle     +(x_1^* - \delta)\hat{S}_2 }  (\tilde{\eta}_\delta)   \right) \notag\\
&& +\, \frac{f_{\langle \vecx^*_\delta,\randvecs\rangle} (\tilde{\eta}_{\delta}) } {1+x_1^*+\delta}
- \frac{f_{\langle \vecx^*_\delta,\randvecs\rangle} (\tilde{\eta}_\delta) } {1+x_1^*-\delta}
\label{eq:1-order-derive-delta}
\\
&=& 2\delta f'_{\langle \vecx^*_\delta,\randvecs\rangle +(x_1^*+\delta)\hat{S}_1
+ (x_1^* -\delta) \hat{S}_2} (\tilde{\eta}_\delta)
  - \frac{ 2\delta  f_{\langle \vecx^*_\delta,\randvecs\rangle} (\tilde{\eta}_\delta) }{(1+x_1^*)^2- \delta^2} \, .\IEEEeqnarraynumspace
\label{eq:1-order-derive-delta-2}
\end{IEEEeqnarray}
Here,~\eqref{eq:1-order-derive-delta-2} follows from~\eqref{eq:abbe-huang-lemma2}.
Taking the derivative of the RHS of~\eqref{eq:1-order-derive-delta-2} with respect to $\delta$ and then setting $\delta=0$, we obtain (recall that $\tilde\eta = \eta + \sum\limits_{i=1}^{m}\log(1+[\vecx^*]_i)$)
\begin{IEEEeqnarray}{rCl}
\frac{\partial^2 \prob[\varphi(\vecx^*_\delta ,\randvecs) \leq \eta] }{\partial \delta^2} \bigg|_{\delta =0}
& =&
2 \left(  f'_{\langle \vecx^*,\randvecs\rangle +x_1^*\hat{S}_1  + x_1^*  \hat{S}_2} (\tilde{\eta})  - \frac{  f_{\langle \vecx^*,\randvecs\rangle} (\tilde{\eta}) }{(1+x_1^*)^2 } \right). \label{eq:second-order-deriv-KT}
\end{IEEEeqnarray}
From the KKT condition~\eqref{eq:equiv-KKT2}, we know that
\begin{IEEEeqnarray}{rCl}
f'_{\langle \vecx^*,\randvecs\rangle +x_1^*\hat{S}_1  + x_2^*  \hat{S}} (\tilde{\eta})  - \frac{  f_{\langle \vecx^*,\randvecs\rangle} (\tilde{\eta}) }{(1+x_1^*)(1+x_2^*)} =0
\label{eq:equiv-KKT2-copy}
\end{IEEEeqnarray}
where $\hat{S}\sim \expdist(1)$ is independent of all other random variables.
Let $T\define \hat{S}_1 + \hat{S}_2$.
Subtracting the LHS of~\eqref{eq:equiv-KKT2-copy} from~\eqref{eq:second-order-deriv-KT}, we obtain
\begin{IEEEeqnarray}{rCl}
\IEEEeqnarraymulticol{3}{l}{
\frac{1}{2}\frac{\partial^2 \prob[\varphi(\vecx_\delta^* ,\randvecs) \leq \eta] }{\partial \delta^2} \bigg|_{\delta =0} } \notag\\
&=& (x_2^*- x_1^*)\!\left(\! f''_{\langle \vecx^*\!,\randvecs\rangle +x_1^* T + x_2^*  \hat{S} }\!(\tilde{\eta})  -\frac{  f_{\langle \vecx^*,\randvecs\rangle}\!(\tilde{\eta}) }{(1+x_1^*)^2(1+x_2^*) } \right)\label{eq:second-order-deriv-KT-equiv0}\\
&=& (x_2^* -x_1^*) \!\bigg( \! f''_{\langle \vecx^*\!\!,\randvecs\rangle +x_1^* T + x_2^*  \hat{S} }(\tilde{\eta})   - \frac{ f'_{\langle \vecx^*\!\!,\randvecs\rangle +x_1^*\hat{S}_1  + x_2^*  \hat{S}} (\tilde{\eta})   }{1+x_1^*} \! \bigg) \,\,\,\quad \label{eq:second-order-deriv-KT-equiv1}\\
&=& \frac{x_2^* - x_1^*}{1+x_1^*} \left( f''_{\langle \vecx^*\!,\randvecs\rangle +x_1^* T + x_2^*  \hat{S} }(\tilde{\eta}) -   f'_{\langle \vecx^*\!,\randvecs\rangle +x_1^*T   + x_2^*  \hat{S}} (\tilde{\eta})     \right).
\label{eq:second-order-deriv-KT-equiv}
\end{IEEEeqnarray}
Here, in~\eqref{eq:second-order-deriv-KT-equiv0} we used~\eqref{eq:abbe-huang-lemma2};~\eqref{eq:second-order-deriv-KT-equiv1} follows from~\eqref{eq:equiv-KKT2-copy}; and in~\eqref{eq:second-order-deriv-KT-equiv} we used~\eqref{eq:abbe-huang-cor1}.

We shall next make~\eqref{eq:second-order-deriv-KT-equiv} depend on $x_3^*$.
Note first that by~\eqref{eq:abbe-huang-cor1},
\begin{IEEEeqnarray}{rCl}
\IEEEeqnarraymulticol{3}{l}{
f''_{\langle \vecx^*,\randvecs\rangle +x_1^* T + x_2^*  \hat{S} }(\tilde{\eta}) -   f'_{\langle \vecx^*,\randvecs\rangle +x_1^*T   + x_2^*  \hat{S}} (\tilde{\eta})} \notag\\
&+&  f'_{\langle \vecx^*,\randvecs\rangle +x_1^* T + x_2^*  \hat{S} +x_3^*\hat{S}_3 }(\tilde{\eta}) - f''_{\langle \vecx^*,\randvecs\rangle +x_1^* T + x_2^*  \hat{S} +x_3^*\hat{S}_3 }(\tilde{\eta})\notag\\
&=& x_3^* \Big( f'''_{\langle \vecx^*\!\!,\randvecs\rangle +x_1^* T + x_2^*  \hat{S} +x_3^*\hat{S}_3 }\!(\tilde{\eta}) - f''_{\langle \vecx^*\!\!,\randvecs\rangle +x_1^* T + x_2^*  \hat{S} +x_3^*\hat{S}_3 }\!(\tilde{\eta})\! \Big).
\label{eq:diff-1d-2d-negative2}
\end{IEEEeqnarray}
Also by~\eqref{eq:abbe-huang-cor1} and~\eqref{eq:ratio_2_3_Equal1},
\begin{IEEEeqnarray}{rCl}
\IEEEeqnarraymulticol{3}{l}{
f'''_{\langle \vecx^*\!,\randvecs\rangle +x_1^* T + x_2^*  \hat{S} +x_3^*\hat{S}_3 }(\tilde{\eta}) - f''_{\langle \vecx^*\!,\randvecs\rangle +x_1^* T + x_2^*  \hat{S} +x_3^*\hat{S}_3 }(\tilde{\eta})}\notag  \\
&=&\! \frac{1}{x_1^*}\Big( f'_{\langle \vecx^*\!\!,\randvecs\rangle +x_1^* T + x_2^*  \hat{S} +x_3^*\hat{S}_3 }\!(\tilde{\eta}) - f''_{\langle \vecx^*\!\!,\randvecs\rangle +x_1^* T + x_2^*  \hat{S} +x_3^*\hat{S}_3 }\!(\tilde{\eta})\! \Big).
\label{eq:diff-1d-2d-negative}
\end{IEEEeqnarray}
Combining~\eqref{eq:diff-1d-2d-negative2} and~\eqref{eq:diff-1d-2d-negative}, we conclude that
\begin{IEEEeqnarray}{rCl}
\IEEEeqnarraymulticol{3}{l}{
f''_{\langle \vecx^*,\randvecs\rangle +x_1^* T + x_2^*  \hat{S} }(\tilde{\eta}) -   f'_{\langle \vecx^*,\randvecs\rangle +x_1^*T   + x_2^*  \hat{S}} (\tilde{\eta})} \notag\\
\quad &=& \frac{x_3^*-x_1^*}{x_1^*}\left(  f'_{\langle \vecx^*\!\!,\randvecs\rangle +x_1^* T + x_2^*  \hat{S} +x_3^*\hat{S}_3 }\!(\tilde{\eta}) - f''_{\langle \vecx^*\!\!,\randvecs\rangle +x_1^* T + x_2^*  \hat{S} +x_3^*\hat{S}_3 }\!(\tilde{\eta})\!\right).
\label{eq:combin-f3d-f2d-f1d-f2d}
\end{IEEEeqnarray}
Substituting~\eqref{eq:combin-f3d-f2d-f1d-f2d} in~\eqref{eq:second-order-deriv-KT-equiv}, we obtain
\begin{IEEEeqnarray}{rCl}
\IEEEeqnarraymulticol{3}{l}{
\frac{1}{2}\frac{\partial^2 \prob[\varphi(\vecx_\delta^* ,\randvecs) \leq \eta] }{\partial \delta^2} \Big|_{\delta =0} }\notag\\
&=& \frac{(x_2^* -x_1^*)(x_3^* -x_1^*)}{x_1^*(1+x_1^*)} \Big( f'_{\langle \vecx^*,\randvecs\rangle +x_1^* T + x_2^*  \hat{S} +x_3^*\hat{S}_3 }(\tilde{\eta})\notag\\
 &&\qquad\qquad \qquad\qquad\quad  -\, f''_{\langle \vecx^*,\randvecs\rangle +x_1^* T + x_2^*  \hat{S} +x_3^*\hat{S}_3 }(\tilde{\eta})  \Big). \IEEEeqnarraynumspace
\label{eq:2nd-deriv-negative}
\end{IEEEeqnarray}
%
%
%
%
Since $(x_2^* - x_1^*)(x_3^*-x_1^*) >0$, to establish~\eqref{eq:2nd-deriv-KT-negative}, it remains to prove that
\begin{IEEEeqnarray}{rCl}
 f'_{\langle \vecx^*\!\!,\randvecs\rangle +x_1^* T + x_2^*  \hat{S} +x_3^*\hat{S}_3 }\!(\tilde{\eta}) - f''_{\langle \vecx^*\!\!,\randvecs\rangle +x_1^* T + x_2^*  \hat{S} +x_3^*\hat{S}_3 }\!(\tilde{\eta})  <0.
  \IEEEeqnarraynumspace \label{eq:sign-diff-3d-2d}
\end{IEEEeqnarray}

Let $\widetilde{A} \define \langle \vecx^*,\randvecs\rangle + x_2^*  \hat{S} +x_3^*\hat{S}_3.$
The LHS of~\eqref{eq:sign-diff-3d-2d} can be rewritten as
\begin{IEEEeqnarray}{rCl}
\IEEEeqnarraymulticol{3}{l}{
f'_{\widetilde{A}+x_1^*T}(\tilde{\eta}) - f''_{\widetilde{A} + x_1^* T} (\tilde{\eta})}\notag\\
&=& \left( f'_{\widetilde{A} + x_1^* \hat{S}_1} - f''_{\widetilde{A} + x_1^*\hat{S}_1}\right) \star f_{x_1^*\hat{S}_2}(\tilde{\eta})\\
&=&\frac{1}{x_1^*} \!\int\nolimits_{0}^{\tilde{\eta}}\!\! \Big(f'_{\widetilde{A} + x_1^*\hat{S}_1}(\tilde{\eta} - z)-f''_{\widetilde{A} + x_1^*\hat{S}_1}(\tilde{\eta} - z)\Big)e^{ -z/{x_1^*}} dz. \IEEEeqnarraynumspace
\label{eq:diff-2d-1d-positive}
\end{IEEEeqnarray}
Since $A+x_3^*\hat{S_3} \sim \widetilde{A} + x_1^*\hat{S}_1$, by~\eqref{eq:ratio_2_3_Equal1},
\begin{IEEEeqnarray}{rCl}
f'_{\widetilde{A} + x_1^*\hat{S}_1}( \tilde{\eta})-f''_{\widetilde{A} + x_1^*\hat{S}_1}( \tilde{\eta}) = 0 .
\end{IEEEeqnarray}
Note that, for every $t\in(0,\tilde\eta)$, we have
\begin{IEEEeqnarray}{rCl}
f'_{\widetilde{A} + x_1^*\hat{S}_1}(t)-f''_{\widetilde{A} + x_1^*\hat{S}_1}(t) &=&\int\nolimits_{0}^{t} (f'_{\widetilde{A} }(z)-f''_{\widetilde{A}}(z))e^{-(t-z)/x_1^*} dz\\
&=& e^{-t/x_1^*} \underbrace{\int\nolimits_{0}^{t} (f'_{\widetilde{A} }(z)-f''_{\widetilde{A}}(z))e^{z/x_1^*} dz }_{\define h(t)} .
\end{IEEEeqnarray}
In Appendix~\ref{app:subsection-zero-sign-change}, we have shown that the function $f'_{\widetilde{A} }-f''_{\widetilde{A}}$ changes sign at most once over the interval~$(0,\tilde{\eta})$.
Therefore, $h'(t)=(f'_{\widetilde{A}}(t)-f''_{\widetilde{A}}(t))e^{t/x_1^*}$ changes sign at most once over the interval~$(0,\tilde{\eta})$. But since $h(\tilde{\eta}) = e^{\tilde{\eta}/x_1^*} (f'_{\widetilde{A} + x_1^*\hat{S}_1}( \tilde{\eta})-f''_{\widetilde{A} + x_1^*\hat{S}_1}( \tilde{\eta})) =0 =h(0)$, the function $h$ does not change sign on $(0,\tilde{\eta})$.
Indeed, there are three possible cases:
\begin{enumerate}
 \item $h'(t) =0 $ for all $t\in(0,\tilde{\eta})$; in this case $h(t)=0$ for all $t \in (0, \tilde{\eta})$. 
 \item  there exists  a $t_0\in(0,\tilde{\eta})$ such that $h'(t)\leq 0$ on $(0, t_0)$, $h'(t_0)=0$, and $h'(t)\geq 0$ on $(t_0,\tilde{\eta})$; in this case $h(t)\leq 0$ for all $t \in (0,\tilde{\eta})$.
 \item there exists  a $t_0\in(0,\tilde{\eta})$ such that $h'(t)\geq 0$ on $(0, t_0)$, $h'(t_0)=0$, and $h'(t)\leq 0$ on $(t_0,\tilde{\eta})$; in this case $h(t)\geq 0$ for all $t \in (0,\tilde{\eta})$.
\end{enumerate}
In all three scenarios, $h(t)$ does not change sign on $(0,\tilde{\eta})$.
This implies that $f'_{\widetilde{A} + x_1^*\hat{S}_1}-f''_{\widetilde{A} + x_1^*\hat{S}_1}$ does not change sign on $(0,\tilde{\eta})$ either.
Furthermore,
\begin{IEEEeqnarray}{rCl}
\IEEEeqnarraymulticol{3}{l}{
 \int\nolimits_{0}^{\tilde{\eta}} f'_{\widetilde{A} + x_1^*\hat{S}_1}( z)-f''_{\widetilde{A} + x_1^*\hat{S}_1}( z) dz }\notag\\
  \quad &=& f_{\widetilde{A} + x_1^*\hat{S}_1}(\tilde{\eta})-f'_{\widetilde{A} + x_1^*\hat{S}_1}( \tilde{\eta}) +   \underbrace{f'_{\widetilde{A} + x_1^*\hat{S}_1}(0)-f_{\widetilde{A} + x_1^*\hat{S}_1}(0)}_{=0} < 0.
   \label{eq:diff-2d-1d-strict-pos}
\end{IEEEeqnarray}
Here, the first step follows because $f'_{\widetilde{A}+x_1^*\hat{S}_1}(0)= f''_{\widetilde{A}+x_1^*\hat{S}_1}(0)=0$~\cite[Lem.~3]{abbe13-05}, and the second step follows from~\eqref{eq:ratio-d-pdf-eta}.
%
We establish~\eqref{eq:sign-diff-3d-2d} by using the following chain of inequalities
\begin{IEEEeqnarray}{rCl}
 \IEEEeqnarraymulticol{3}{l}{
f'_{\widetilde{A}+x_1^*T}(\tilde{\eta}) - f''_{\widetilde{A} + x_1^* T} (\tilde{\eta})}\notag\\
\quad &\leq & \frac{e^{-\tilde{\eta}/x_1^*}}{x_1^*}  \int\nolimits_{0}^{\tilde{\eta}}\!\! \Big(f'_{\widetilde{A} + x_1^*\hat{S}_1}(\tilde{\eta} - z)-f''_{\widetilde{A} + x_1^*\hat{S}_1}(\tilde{\eta} - z)\Big) dz \label{eq:compare-f-atilde-fd1}\\
&<& 0. \label{eq:compare-f-atilde-fd2}\IEEEeqnarraynumspace
 \end{IEEEeqnarray}
Here,~\eqref{eq:compare-f-atilde-fd1} follows from~\eqref{eq:diff-2d-1d-positive} and because $e^{-z/x_1^*} \geq e^{-\tilde\eta/x_1^*}>0$ for all $z\in(0,\tilde\eta)$;~\eqref{eq:compare-f-atilde-fd2} follows from~\eqref{eq:diff-2d-1d-strict-pos}.

\subsection{Extension to  $\posrealset^{\infty}$}
Consider the following chain of equalities:
\begin{IEEEeqnarray}{rCl}
\IEEEeqnarraymulticol{3}{l}{
\inf\limits_{\vecx\in\posrealset^\infty:\|\vecx\|_1 = \rxant E}
\prob\lefto[ \sum\limits_{i=1}^{\infty } \Big( x_i S_i  -\log(1+x_i) \Big)\leq \eta \right]
}\notag\\
& = & \lim\limits_{m\to\infty} \inf_{\vecx \in \posrealset^m:\|\vecx\|_1 = \rxant E} \prob\lefto[ \sum\limits_{i=1}^{m}\!\! \Big( x_i S_i  -\log(1+x_i) \Big)\leq \eta \right]\label{eq:inf-restrict-three-mass-0}  \\
&=&\lim\limits_{m\to\infty}  \inf \prob\lefto[ \sum\limits_{i=1}^{m} \Big( x_i S_i  -\log(1+x_i) \Big)\leq \eta \right]\label{eq:inf-restrict-three-mass-1} \\
&=& \inf \prob\lefto[ \sum\limits_{i=1}^{\infty } \Big( x_i S_i -\log(1+x_i) \Big)\leq \eta \right]. \label{eq:inf-restrict-three-mass}
\end{IEEEeqnarray}
Here, both~\eqref{eq:inf-restrict-three-mass-0} and~\eqref{eq:inf-restrict-three-mass} follow from the monotone convergence theorem~\cite[Th.~2.14]{folland99};
the infimum in~\eqref{eq:inf-restrict-three-mass-1} and~\eqref{eq:inf-restrict-three-mass} is over all $\vecx$ (in~$\posrealset^{m}$ and~$\posrealset^{\infty}$, respectively) of the form~\eqref{eq:three-atoms-form} or~\eqref{eq:le-two-atoms-form}.
This concludes the proof of Lemma~\ref{lem:three-atoms}.

\section{Convolution of Exponential Distributions}
\label{app:convo-exp-dist}
In this appendix, we summarize some results about the convolution of exponential distributions that are needed in Appendices~\ref{app:proof-three-atoms},~\ref{app:proof-thm-1}, and~\ref{app:proof-asy-coh}.

The first lemma deals with the log-concavity of the convolution of exponential distributions.
Recall that a function $f$ is called log-concave if $\log f$ is concave, and it is called \emph{strictly} log-concave if $\log f$ is strictly concave.
Since the exponential distribution is log-concave, and log-concavity is preserved under convolution~\cite{ibragimov56}, it follows that the convolution of exponential distributions is also log-concave.
Lemma~\ref{lem:strict-log-concavity} below shows that this distribution is in fact \emph{strictly} log-concave.

\begin{lemma}
\label{lem:strict-log-concavity}
Fix an integer $m\geq 2 $. Let $ S_1,\ldots,S_m$ be i.i.d. $\expdist(1)$-distributed random variables, and let $a_1,\ldots,a_m $ be positive real numbers.
Furthermore, let $B\define \sum\nolimits_{i=1}^{m} a_i S_i$. Then, the pdf $f_{B}$ of $B$ is strictly log-concave on $(0,\infty)$.
\end{lemma}
\begin{IEEEproof}
The proof is based on induction.
Through algebraic manipulations, it can be verified that $f_{a_1S_1 + a_1S_2}$ is strictly log-concave on $(0,\infty)$ for every $a_1,a_2>0$.
Suppose now that the pdf of $B_k \define \sum\nolimits_{i=1}^{k} a_iS_i$ is strictly log-concave for some $k\geq 2$.
We have
\begin{IEEEeqnarray}{rCl}
f_{B_{k+1}} (t) = \int \underbrace{f_{B_k} (t-s) f_{a_{k+1}S_{k+1}} (s)}_{\define g(s,t)} ds, \quad t >0. \IEEEeqnarraynumspace
\label{eq:convolution-b-k}
\end{IEEEeqnarray}
It follows that the integrand $g(s,t)$ in~\eqref{eq:convolution-b-k} is (jointly) log-concave in $(s,t)$ on $\posrealset^2$ and it is strictly log-concave on the subspace $\{(s,t)\in \posrealset^2: s\leq t \}$.
Note that by the Pr\'{e}kopa Theorem~\cite{prekopa73},\cite[Sec.~3]{ball03} for each $a,b>0$,
\begin{IEEEeqnarray}{rCl}
f_{B_{k+1}}\lefto(\frac{a+b}{2}\right)
&=& \int g\lefto(s,\frac{a+b}{2}\right) ds  \\
&\geq& \left(\int g(s,a) ds\right)^{1/2} \left(\int g(s,b) ds\right)^{1/2}\quad \quad\label{eq:prekopa-inequality}\\
&=&\sqrt{f_{B_{k+1}}(a)f_{B_{k+1}}(b)}.
\end{IEEEeqnarray}
This implies that $f_{B_{k+1}}$ is log-concave.
%
%
%
%
Following the proof of the Pr\'{e}kopa Theorem in~\cite[Sec.~3]{ball03}, and using that the function $(s,t)\mapsto f_{B_k} (t-s) f_{a_{k+1}S_{k+1}} (s)$ is strictly positive, smooth~\cite[Lem.~3]{abbe13-05}, and strictly log-concave for $0<s<t$, we can verify that the inequality in~\eqref{eq:prekopa-inequality} is strict for every $a,b>0$.
This in turn implies that $f_{B_{k+1}}$ is \emph{strictly} log-concave on $(0,\infty)$.
By induction, $f_{B}$ is strictly log-concave on $(0,\infty)$ for every $m\geq 2$.
\end{IEEEproof}

The next lemma characterizes the optimal convex combination of exponential random variables that minimizes the probability that such combination does not exceed a given threshold.
\begin{lemma}
\label{lem:telatar-conj}
Let $n\in\natunum$, let $S_1,\ldots,S_n$ be i.i.d. $\expdist(1)$-distributed random variables, and let $\setA_n \define \{\vecx\in \posrealset^n: \|\vecx\|_1 \leq 1, x_1\geq x_2\geq \ldots\geq x_n\}$. %
Then, for every $t\in\posrealset$, there exists a $k\in\{1,\ldots,n\}$ such that
\begin{IEEEeqnarray}{rCl}
\arg\min\limits_{\vecx \in \setA_n} \prob\mathopen{}\bigg[\sum\limits_{i=1}^{n}x_iS_i \le t\bigg] = \Big[\underbrace{\frac{1}{k},\ldots,\frac{1}{k}}_{k},0,\ldots,0\Big].
\label{eq:telatar-conjecture}
\end{IEEEeqnarray}
In particular, if $t\in(0 ,1]$, then
\begin{IEEEeqnarray}{rCl}
\arg\min\limits_{\vecx \in \setA_n} \prob\mathopen{}\bigg[\sum\limits_{i=1}^{n}x_iS_i \le t\bigg] = \Big[\frac{1}{n},\ldots,\frac{1}{n}\Big].
\label{eq:telatar-conjecture-exact}
\end{IEEEeqnarray}
\end{lemma}
\begin{IEEEproof}
The equality~\eqref{eq:telatar-conjecture} follows directly from~\cite[p.~2597]{abbe13-05}. To prove~\eqref{eq:telatar-conjecture-exact}, it is sufficient to show that for every $k\in\natunum$ and every $t\in(0 ,1]$, the following inequality holds:
\begin{equation}
\prob\lefto[\sum\limits_{i=1}^k S_i \leq t k\right] \geq \prob\lefto[\sum\limits_{i=1}^{k+1} S_i \leq t (k+1)\right].
\label{eq:prob-sum-k-leq-sum-k+1}
\end{equation}
Let $f_k(x) \define x^{k} e^{-x}$.
Consider the following chain of (in)equalities
\begin{IEEEeqnarray}{rCl}
\IEEEeqnarraymulticol{3}{l}{
\prob\lefto[\sum\limits_{i=1}^k S_i \leq t k\right] - \prob\lefto[\sum\limits_{i=1}^{k+1} S_i \leq t (k+1)\right]}\notag\\
\quad &=& \int\nolimits_{0}^{tk} \frac{f_{k-1}(x)}{(k-1)!} dx -  \int\nolimits_{0}^{t(k+1)} \frac{f_k(x)}{k !} dx \label{eq:proof-telatar-conjecture-opt-k-1} \\
&=& \frac{ 1}{k!}\left( \int\nolimits_{0}^{tk} k f_{k-1}(x) dx   - \int\nolimits_{0}^{t(k+1)} f_k(x) dx \right)\label{eq:proof-telatar-conjecture-opt-k-2} \\
&=& \frac{1}{k!} \left(f_k(tk) - \int\nolimits_{tk}^{t(k+1)} f_k(x) dx \right) \label{eq:proof-telatar-conjecture-opt-k-3}\\
&\geq & \frac{1}{k!} \left(f_k(tk) - \int\nolimits_{tk}^{t(k+1)} \exp\mathopen{}\Big( \log f(tk) + \frac{f'_k(tk)}{f_k(tk)}(x-tk)\log e \Big) dx  \right) \label{eq:proof-telatar-conjecture-opt-k-4}\\
&=& \frac{f_{k}(tk)}{k!}\frac{1-t e^{1-t}}{1-t}  \label{eq:proof-telatar-conjecture-opt-k-5}\\
&\geq &0. \label{eq:proof-telatar-conjecture-opt-k}
\end{IEEEeqnarray}
Here,~\eqref{eq:proof-telatar-conjecture-opt-k-1} follows because the random variable $\sum\limits_{i=1}^{k}S_i$ is chi-squared distributed with pdf $f_{k-1}(x)/(k-1)!$; in~\eqref{eq:proof-telatar-conjecture-opt-k-3} we used integration by parts;~\eqref{eq:proof-telatar-conjecture-opt-k-4} follows because $f_{k}(x)$ is log-concave, which implies that for every $x\geq 0$
\begin{IEEEeqnarray}{rCl}
\log f_{k}(x) \leq  \log f_{k}(tk)  + \frac{f'_k(tk)}{f_k(tk)} (x-tk)\log e;
\end{IEEEeqnarray}
finally,~\eqref{eq:proof-telatar-conjecture-opt-k} follows because $t\mapsto te^{1-t}$ is monotonically increasing on $(0,1]$, and because $te^{1-t}\big|_{t=1} = 1$.
This proves~\eqref{eq:prob-sum-k-leq-sum-k+1}.
\end{IEEEproof}

The following lemma provides a uniform lower bound on the cdf of the weighted sum of exponential distributions.
\begin{lemma}
\label{lem:lb-normalized-cdf}
Let $\{S_i\}$ be i.i.d. $\expdist(1)$-distributed random variables. Let $\vecx =[x_1,x_2,\ldots]\in \posrealset^\infty$ satisfy $0<\|\vecx\|_1 <\infty$.
Furthermore, let
\begin{IEEEeqnarray}{rCl}
\weightsum(\vecx) \define \frac{1}{\|\vecx\|_2} \bigg(\sum\limits_{i=1}^{\infty} x_i S_i - \|\vecx\|_1 \bigg)
\label{eq:def-hatS}
\end{IEEEeqnarray}
and denote the cdf of   $\weightsum(\vecx)$ by $F_{\weightsum(\vecx)}(t) $.
Then, for every $t\in(-\infty,0]$,
\begin{IEEEeqnarray}{rCl}
F_{\weightsum(\vecx)}(t) \geq \left|\frac{1}{2} + t\right|^{+}.
\label{eq:lemma-bound-cdf-normalize}
\end{IEEEeqnarray}
Equivalently,
\begin{IEEEeqnarray}{rCl}
F^{-1}_{\weightsum(\vecx)}(\error) \leq  \error -\frac{1}{2} ,\quad \text{for all } \,0 < \error <\frac{1}{2}.
\label{eq:lemma-bound-cdf-normalize-inv}
\end{IEEEeqnarray}
\end{lemma}
\begin{IEEEproof}
Since $\|\vecx\|_1 >0$, we can assume without loss of generality that $x_1>0$.
Let $\vecx_n$ denote the vector that contains the first $n$ entries of $\vecx$, let
\begin{equation}
L_{n}(\vecx) \define \frac{1}{\|\vecx_n\|_2} \bigg(\sum\limits_{i=1}^{n} x_i S_i - \|\vecx_n\|_1 \bigg)
\end{equation}
and let $F_{L_{n}(\vecx)}(t)$ denote the cdf of $L_{n}(\vecx)$. Through algebraic manipulations, it can be shown that $F_{L_{n}(\vecx)}(t)$ converges pointwise to $F_{\weightsum(\vecx)}(t)$ as $n\to\infty$. Hence, to prove~\eqref{eq:lemma-bound-cdf-normalize}, it suffices to show that for every $n\in\natunum$ and every $t\in(-\infty,0]$
\begin{equation}
F_{\weightsum_n(\vecx)}(t) \geq \left|\frac{1}{2} + t\right|^{+}.
\label{eq:lemma-bound-cdf-normalize-n}
\end{equation}

We first show that~\eqref{eq:lemma-bound-cdf-normalize-n} holds when $t=0$.
Indeed, we have that
\begin{IEEEeqnarray}{rCl}
F_{\weightsum_n(\vecx)}(0) &=& \prob\lefto[\sum\limits_{i=1}^n  x_i S_i \leq \|\vecx_n\|_1 \right]\\
&\geq & \inf\limits_{\vecy \in \posrealset^n: \|\vecy\|_1 = \|\vecx_n\|_1} \prob\lefto[\sum\limits_{i=1}^n y_i S_i \leq \|\vecx_n\|_1 \right]\\
&=&  \prob\lefto[ n^{-1} \|\vecx_n\|_1 \sum\limits_{i=1}^{n} S_i  \leq  \|\vecx_n\|_1 \right]  \label{eq:identical-sum-exp-2}\\
& > & \frac{1}{2}.
\label{eq:identical-sum-exp}
\end{IEEEeqnarray}
Here,~\eqref{eq:identical-sum-exp-2} follows from~\eqref{eq:telatar-conjecture-exact};~\eqref{eq:identical-sum-exp} follows because $ \sum\nolimits_{i=1}^{n} S_i$ is chi-squared distributed, and because the median of a chi-squared distribution is smaller than its mean~\cite[Ch.~17]{johnson95-1}.

We next prove~\eqref{eq:lemma-bound-cdf-normalize-n} for the case $t<0$.
By definition, $\weightsum_n(\vecx)$ has zero mean and unit variance. Moreover, by Lemma~\ref{lem:strict-log-concavity} the pdf $f_{\weightsum_n(\vecx)}$ of $\weightsum_n(\vecx)$ is log-concave.
Hence, we have that~\cite[Lem.~5.5]{lovasz07-03}\cite[Prop.~2.1]{bobkov13-07}
\begin{IEEEeqnarray}{rCl}
\sup\limits_{ t \in \realset} f_{\weightsum_n(\vecx)}(t) \leq 1.
\label{eq:ub-pdf-shat-1}
\end{IEEEeqnarray}
Then, the bound~\eqref{eq:lemma-bound-cdf-normalize-n} holds because
\begin{IEEEeqnarray}{rCl}
F_{\weightsum_n(\vecx)}(t) &=& \underbrace{\int\nolimits_{-\infty}^{0} f_{\weightsum_n(\vecx)}(y) dy}_{\geq 1/2} - \int\nolimits_{t}^{0} \underbrace{f_{\weightsum_n(\vecx)}(y)}_{\leq 1} dy.
\geq   \frac{1}{2}  + t \IEEEeqnarraynumspace
\end{IEEEeqnarray}
The last step follows from~\eqref{eq:identical-sum-exp} and~\eqref{eq:ub-pdf-shat-1}.
\end{IEEEproof}

Consider the random variable obtained by summing finitely many independent but not necessarily identically distributed exponential random variables. The next lemma establishes that the derivative of the pdf of the resulting random variable, computed at the mean value, is negative.
Since the convolution of exponential distributions is unimodal, this implies that the  mode of this random variable is smaller than its mean, i.e., its probability distribution is right skewed.

\begin{lemma}
\label{lem:negative-derivative}
Let $m\in\natunum$, let $a_1,\ldots,a_m$ be positive real numbers, and let
 $S_1,\ldots,S_m$ be i.i.d. $\expdist(1)$-distributed random variables.
Furthermore, let $\mu \define \sum\nolimits_{i=1}^{m}a_i$, $a_{\min} \define \min\nolimits_{i}\{a_i\}$, $a_{\max}\define \max\nolimits_{i}\{ a_i\}$, and
$A \define \sum\nolimits_{i=1}^{m} a_i S_i$.
Then,
\begin{IEEEeqnarray}{rCl}
f_{A}'(\mu)\leq  - \frac{a_{\min}}{ a_{\max} } \frac{f_A(\mu)}{\mu} <0
\label{eq:lem-deriv-sum-exp-negative}
\end{IEEEeqnarray}
where $f_{A}'$ denotes the derivative of the pdf of $A$. Moreover, the first inequality in~\eqref{eq:lem-deriv-sum-exp-negative} holds with equality if and only if $a_1=\cdots =a_{m}$.
\end{lemma}

\begin{IEEEproof}
Note that the $\{S_i\}$, $i=1,\ldots,m$, have the same distribution as $\{X_i^2 + X_{m+i}^2\}$, $i=1,\ldots,m$, where $\{X_i\}$, $i=1,\ldots,2m$, are i.i.d. $\mathcal{N}(0,1/2)$-distributed.
Let $a_{m+i} \define a_i$, $i=1,\ldots,m$. Then,~$A$ has the same distribution as
\begin{IEEEeqnarray}{rCl}
\widetilde{A} \define \sum\limits_{i=1}^{2m} a_i X_i^2.
\end{IEEEeqnarray}

Next, we prove that $f'_{\widetilde{A}} (\mu) < 0$ by using~\cite[Lem.~22]{yang14-07a}, which provides expressions for the pdf and the derivative of the pdf of functions of random variables.
We first give some definitions.
Let $\randvecx \define [X_1,\ldots,X_{2m}]$, and let $f_{\randvecx}$ denote the joint pdf of $X_1,\ldots,X_{2m}$. Let $\varphi(\vecx): \realset^{2m} \to \posrealset$ be defined as
\begin{IEEEeqnarray}{rCl}
\varphi(\vecx) \define \sum\limits_{i=1}^{2m} a_i x_i^2.
\end{IEEEeqnarray}
Let $\nabla \varphi$ and $\Delta\varphi$ be the gradient and Laplacian of $\varphi$, namely,
\begin{IEEEeqnarray}{rCl}
\nabla\varphi (\vecx) \define \left[\frac{\partial}{\partial x_1}\varphi(\vecx),\ldots,\frac{\partial}{\partial x_{2m}} \varphi(\vecx) \right]
\end{IEEEeqnarray}
and
\begin{IEEEeqnarray}{rCl}
\Delta\varphi (\vecx) \define \sum\limits_{i=1}^{2m}\frac{\partial^2}{\partial x_i^2}\varphi(\vecx).
\end{IEEEeqnarray}
Finally, let $\varphi^{-1}(\mu)$ denote the preimage $\{\vecx\in\realset^{2m}: \varphi(\vecx) = \mu\}$, and let $\surform$ be the surface area form on $\varphi^{-1}(\mu)$, chosen so that $\surform(\nabla \varphi) >0$.
Note that $f_{\randvecx}$ is smooth and that the set $\varphi^{-1}(\mu)$ is bounded.
Moreover, for every $\vecx \in \varphi^{-1} (\mu)$
\begin{IEEEeqnarray}{rCl}
\|\gradient \varphi(\vecx)\|_2^2 = \sum\limits_{i=1}^{2m} 4a_i^2 x_i^2 \geq 4 \mu  \min\limits_{i=1,\ldots, m}\{ a_i\} >0.
\end{IEEEeqnarray}
Then, by~\cite[Eq.~(407)]{yang14-07a},
\begin{IEEEeqnarray}{rCl}
f'_{\tilde{A}} (\mu)  = \int\nolimits_{\varphi^{-1}(\mu)} \psi \frac{\surform}{\|\nabla \varphi\|_{2}}
\label{eq:deriv-f-tilde-a}
\end{IEEEeqnarray}
where~\cite[Eq.~(422)]{yang14-07a}
\begin{IEEEeqnarray}{rCl}
\psi \define \frac{\langle \nabla f_{\randvecx} ,\nabla\varphi \rangle  + f_{\randvecx} \cdot \Delta\varphi   }{\|\nabla \varphi\|_{2}^2}   - \frac{f_{\randvecx}      \langle \nabla \|\nabla \varphi\|_2^2 ,\nabla\varphi \rangle       }{\|\nabla \varphi\|_{2}^4} . \IEEEeqnarraynumspace \label{eq:def-psi-rie-geo}
\end{IEEEeqnarray}
The first term on the RHS of~\eqref{eq:def-psi-rie-geo} is equal to zero. Indeed,
\begin{IEEEeqnarray}{rCl}
\IEEEeqnarraymulticol{3}{l}{
\langle \nabla f_{\randvecx} ,\nabla\varphi \rangle + f_{\randvecx} \cdot \Delta\varphi }\notag\\
\quad &=& \sum\limits_{i=1}^{2m} - 2x_i f_{\randvecx} \cdot (2a_ix_i) +  f_{\randvecx}\cdot \sum\limits_{i=1}^{2m} 2 a_i \\
&=& 4 f_{\randvecx} \left( \sum\limits_{i=1}^{m} a_i -  \sum\limits_{i=1}^{2m} a_ix_i^2\right)\\
&=& 0. \label{eq:comp-1st-term-psi}
\end{IEEEeqnarray}
Here, the last step follows because for every $\vecx\in\varphi^{-1}(\mu)$
\begin{IEEEeqnarray}{rCl}
\sum\limits_{i=1}^{2m} a_ix_i^2 = \mu = \sum\limits_{i=1}^{m} a_i .
\end{IEEEeqnarray}
The second term on the RHS of~\eqref{eq:def-psi-rie-geo} can be computed as follows:
\begin{IEEEeqnarray}{rCl}
\frac{f_{\randvecx}  \langle \nabla \|\nabla \varphi\|_2^2 ,\nabla\varphi \rangle}{\|\nabla \varphi\|_2^4}    &= & f_{\randvecx}  \frac{16 \sum\nolimits_{i=1}^{2m} a_i^3 x_i^2 }{\big(\sum\nolimits_{i=1}^{2m} 4a_i^2x_i^2\big)^2}  \\
&\geq  & f_{\randvecx} \frac{ a_{\min} \sum\nolimits_{i=1}^{2m} a_i^2 x_i^2 }{ a_{\max} \mu \sum\nolimits_{i=1}^{2m} a_i^2 x_i^2 }  \label{eq:comp-2nd-term-psi-1}\\
&=&  \frac{ a_{\min} }{ a_{\max}}  \frac{f_{\randvecx}}{ \mu } .
 \label{eq:comp-2nd-term-psi}
\end{IEEEeqnarray}
Note that the inequality on the RHS of~\eqref{eq:comp-2nd-term-psi-1} holds with equality if and only if $a_1=\cdots = a_{m}$.
Finally, using~\eqref{eq:comp-1st-term-psi} and~\eqref{eq:comp-2nd-term-psi} in~\eqref{eq:deriv-f-tilde-a} we conclude that
\begin{IEEEeqnarray}{rCl}
f'_{\tilde{A}} (\mu)  \leq - \int\nolimits_{\varphi^{-1}(\mu)} \!\!  \frac{ a_{\min}}{ a_{\max}} \frac{f_{\randvecx} }{\mu}   \frac{\surform}{\|\nabla \varphi\|_{2}}  = - \frac{ a_{\min}}{ a_{\max}} \frac{f_{\tilde{A}} (\mu)}{\mu}.\IEEEeqnarraynumspace
\label{eq:deriv-f-tilde-a-final}
\end{IEEEeqnarray}
Here, the last step follows from~\cite[Lem.~22]{yang14-07a}.
The second inequality in~\eqref{eq:lem-deriv-sum-exp-negative} follows because $f_{\tilde{A}}(\mu)>0$ (see~\cite[Lem.~3]{abbe13-05}).
\end{IEEEproof}

\section {Proof of Theorem~\ref{thm:main-result}}
\label{app:proof-thm-1}

\subsection{Achievability}
\label{sec:proof-ach-infband-asy}
To prove that~\eqref{eq:rate-inf-bandwidth} is achievable,  we start from  the inequality
\begin{IEEEeqnarray}{rCl}
M-1\geq \frac{\tau}{\beta_{1-\error+\tau}(P_{\randmatY \given \randvecx = \vecc_1}, P_{\randmatY\given \randvecx = \mathbf{0}}) }
\label{eq:kappa-beta-verdu-mimo}
\end{IEEEeqnarray}
which is equivalent to Theorem~\ref{thm:kappa-beta-cpuc} (see~\eqref{eq:def-codewords-verdu} for a definition of $\vecc_1$).
%
%
First, we upper-bound $\beta_{1-\error+\tau}(P_{\randmatY \given \randvecx = \vecc_1}, P_{\randmatY\given \randvecx = \mathbf{0}})$ as~\cite[Eq.~(103)]{polyanskiy10-05}
\begin{IEEEeqnarray}{rCl}
\beta_{1-\error+\tau}(P_{\randmatY \given \randvecx = \vecc_1}, P_{\randmatY\given \randvecx = \mathbf{0}}) \leq \xi^{-1}
\label{eq:ub-beta1}
\end{IEEEeqnarray}
where $\xi>0$ satisfies
\begin{IEEEeqnarray}{rCl}
P_{\randmatY \given \randvecx = \vecc_1} [ \imath(\vecc_1,\randmatY) \leq \log \xi] = \error -\tau
\label{eq:def-xi}
\end{IEEEeqnarray}
and $\imath(\cdot,\cdot)$ was defined in~\eqref{eq:def-info-density-infty}.
The LHS of~\eqref{eq:def-xi} can be lower-bounded as follows:
\begin{IEEEeqnarray}{rCl}
\IEEEeqnarraymulticol{3}{l}{P_{\randmatY \given \randvecx = \vecc_1} [\imath_{\randvecx,\randmatY}(\vecc_1,\randmatY) \leq \log \xi]} \\
& =& \prob\lefto[ -\rxant N\log\mathopen{}\Big(1+\frac{E}{N}\Big) + \frac{E}{N}\sum\limits_{i=1}^{\rxant N} S_i\log e \leq \log\xi\right]\label{eq:evaluate-xi-1}\\
&\geq &Q\lefto(-\frac{\log\xi  + \rxant N\log(1+E/N)- \rxant E\log e}{\sqrt{\rxant N} \cdot (E/N)\log e}\right) - \frac{\constrm}{\sqrt{N}}.\IEEEeqnarraynumspace
\label{eq:evaluate-xi-2}
\end{IEEEeqnarray}
Here,~$\constrm$ denotes a positive constant\footnote{Throughout the remainder of the paper, we will use $\constrm$ to denote an arbitrary constant whose exact value is irrelevant for the analysis. Its value may change at each appearance.} independent of~$E$ and~$N$,~\eqref{eq:evaluate-xi-1} follows from~\eqref{eq:info-den-under-p-th-verdu},  and~\eqref{eq:evaluate-xi-2} follows from the Berry-Esseen Theorem (see, e.g.,~\cite[Ch.~XVI.5]{feller70b}).

Next, we set $\tau = 1/\sqrt{E}$ in~\eqref{eq:def-xi} and consider the asymptotic regime $E\to\infty$.
We shall choose $N$ as a function of $E$ so that  $N\to\infty$ as $E\to\infty$ with $N/ E \to 0 $.
Substituting~\eqref{eq:evaluate-xi-2} into~\eqref{eq:def-xi}, and solving for $\log \xi$ we obtain
\begin{IEEEeqnarray}{rCl}
\log \xi &\geq & \rxant E\log e - \rxant N\log\lefto(1+\frac{E}{N}\right) - \frac{\sqrt{\rxant} E \log e}{\sqrt{N}} Q^{-1}\!\lefto(\error - \frac{1}{\sqrt{E}} + \frac{\constrm}{\sqrt{N}}\right)\\
&=&\rxant E\log e - \rxant N\log\lefto(\!1+\frac{E}{N}\right) - \frac{\sqrt{\rxant} E \log e}{\sqrt{N}} Q^{-1}\lefto(\error\right)+ \bigO\mathopen{}\bigg(\! \Big(\frac{-1}{\sqrt{E}} + \frac{\mathrm{const}}{\sqrt{N}}\Big) \frac{E}{\sqrt{N}}\!\bigg)\IEEEeqnarraynumspace\,\,
\label{eq:compute-xi-int1}\\
&=&\rxant E \log e -  \rxant N\log\mathopen{}\lefto(\!1+\frac{E}{N}\right)- \frac{\sqrt{\rxant} E\log e}{\sqrt{N}} Q^{-1}(\error) + \bigO\mathopen{}\bigg(\frac{E}{N}\bigg).\IEEEeqnarraynumspace
\label{eq:compute-xi}
\end{IEEEeqnarray}
Here,~\eqref{eq:compute-xi-int1} follows by operating a Taylor-expansion of $Q^{-1}(\cdot)$ around $\error$, and~\eqref{eq:compute-xi} follows because $N/E\to 0$.

We next maximize the dominant terms on the RHS of~\eqref{eq:compute-xi} by choosing
\begin{IEEEeqnarray}{rCl}
N = N^* \define \arg\min\limits_{N\in \natunum} \left\{ \rxant N\log\lefto(1+\frac{E}{N}\right) + \frac{\sqrt{\rxant}E\log e}{\sqrt{N}} Q^{-1}(\error)\right\}.\IEEEeqnarraynumspace
\label{eq:N-star-min-two-penalty}
\end{IEEEeqnarray}
After some algebraic computations, we obtain
\begin{IEEEeqnarray}{rCl}
N^* 
&=& \frac{1}{\rxant}\left(\frac{3}{2}Q^{-1}(\error)\frac{(\rxant E)\log e}{\log (\rxant E)}\right)^{2/3} + \bigO\lefto(\frac{E^{2/3} \log\log E}{(\log E)^{5/3} }\right). \IEEEeqnarraynumspace
\label{eq:def-N-star}
\end{IEEEeqnarray}
Substituting~\eqref{eq:def-N-star} into~\eqref{eq:compute-xi}, then~\eqref{eq:compute-xi} into~\eqref{eq:ub-beta1}, and finally~\eqref{eq:ub-beta1} into~\eqref{eq:lb-M-verdu} we conclude that
\begin{IEEEeqnarray}{rCl}
\log M^*(E,\error)    
   & \geq & \rxant E\log e - V_0\cdot  \Big(\rxant E Q^{-1}(\error)\Big)^{2/3}\big(\log (\rxant E)\big)^{1/3} + \bigO\lefto(\frac{E^{2/3} \log\log E}{(\log E)^{2/3}}\right) \IEEEeqnarraynumspace
  \label{eq:ach-proof-end-asy}
\end{IEEEeqnarray}
where $V_0$ is given in~\eqref{eq:def-V0-main-thm}.

\subsection{Converse}
%

\label{sec:proof-conv-infband-asy}

It follows from~\eqref{eq:ub-M-conv-temp} that for every $\eta\in \realset$
\begin{IEEEeqnarray}{rCl}
\log M^*(E,\error) &\leq& \eta - \log\mathopen{}\bigg|\!\inf\limits_{\vecx}\prob\mathopen{}\bigg[\sum\limits_{i=1}^{\infty}\Big( x_iS_i \log e 
- \log(1+x_i) \Big) \leq \eta\bigg] -\error  \bigg|^{+}\IEEEeqnarraynumspace
\label{eq:upper-bd-M-asy-start}
\end{IEEEeqnarray}
 where the infimum is taken over all $\vecx$ that are of the form specified in~\eqref{eq:three-atoms-form} and~\eqref{eq:le-two-atoms-form}.

Before proceeding to further bound~\eqref{eq:upper-bd-M-asy-start}, we introduce some notation. To every $\vecx\in \posrealset^\infty$ satisfying $\|\vecx\|_1 = \rxant E$, we assign  the random variable
\begin{IEEEeqnarray}{rCl}
\weightsum(\vecx) \define \frac{1}{\|\vecx\|_2} \bigg(\sum\limits_{i=1}^{\infty}x_i S_i - \rxant E \bigg).
\label{eq:def-hatS}
\end{IEEEeqnarray}
Let $F_{\weightsum(\vecx)}(t) $ be the cdf of  $\weightsum(\vecx)$.
By construction, $\weightsum(\vecx)$ has zero mean and unit variance. 
Let $\hat{\eta}_E(\vecx):  \posrealset^\infty \to \realset $ be defined as follows: 
\begin{IEEEeqnarray}{rCl}
\hat{\eta}_E(\vecx) \define \rxant E\log e - \sum\limits_{i=1}^{\infty} \log(1+x_i) + F^{-1}_{\weightsum(\vecx)} \mathopen{}\big(\error + E^{-1/2} \big) \|\vecx\|_2 \log e
\label{eq:expansion-eta-E}. 
\end{IEEEeqnarray}
We shall choose~$\eta$ so that 
\begin{IEEEeqnarray}{rCl}
\eta =  \eta_E \define \sup_{\vecx}  \hat{\eta}_E(\vecx) 
\label{eq:def-eta-E}
\end{IEEEeqnarray}
where the supremum is again over all $\vecx$ that are of the form specified in~\eqref{eq:three-atoms-form} and~\eqref{eq:le-two-atoms-form}. 
Substituting~\eqref{eq:def-eta-E} into~\eqref{eq:upper-bd-M-asy-start}, we obtain 
\begin{IEEEeqnarray}{c}
 \log M^*(E,\error) \leq  \sup_{\vecx}  \hat{\eta}_E(\vecx)  + \bigO(\log E). \IEEEeqnarraynumspace 
\label{eq:upper-bd-M-asy-mid}
\end{IEEEeqnarray}

To conclude the proof, it remains to show that for every $\vecx$ that is of the form specified in~\eqref{eq:three-atoms-form} and~\eqref{eq:le-two-atoms-form}
\begin{IEEEeqnarray}{rCl}
\hat{\eta}_E (\vecx) \leq \rxant E\log e  - V_0\cdot  \Big(\rxant E Q^{-1}(\error)\Big)^{2/3}\big(\log (\rxant E)\big)^{1/3}
+ \bigO\lefto(\frac{E^{2/3} \log\log E}{(\log E)^{2/3}}\right). \IEEEeqnarraynumspace
\end{IEEEeqnarray}
To this end,  we consider the following three cases separately.
\begin{enumerate}
\item The vector $\vecx$ takes the form~\eqref{eq:le-two-atoms-form}, and $\widetilde{N}_2 > E^{1/6}$;
\item The vector $\vecx$ takes the form~\eqref{eq:three-atoms-form};
\item The vector $\vecx$ takes the form~\eqref{eq:le-two-atoms-form}, and $\widetilde{N}_2 \le E^{1/6}$.
\end{enumerate}

\subsubsection*{Case 1}
By assumption, $\vecx$ has at most two distinct nonzero entries $0\leq \tilde{q}_1 \leq \tilde{q}_2$, and $\widetilde{N}_2 \geq E^{1/6} $.
Suppose that we can approximate $F^{-1}_{\weightsum(\vecx)} \mathopen{}\big(\error + E^{-1/2}\big)$ by $ - Q^{-1} (\error )$ in the limit $E\to\infty$ (in a sense we shall make precise later on).
The proof is then concluded by using the result in Lemma~\ref{lem:opt-two-terms} below, together with~\eqref{eq:N-star-min-two-penalty} and~\eqref{eq:def-N-star}, in~\eqref{eq:upper-bd-M-asy-mid}.
\begin{lemma}
\label{lem:opt-two-terms}
For every positive constant $a$, we have that
\begin{IEEEeqnarray}{rCl}
\IEEEeqnarraymulticol{3}{l}{
\inf\limits_{\vecx\in\posrealset^{\infty} : \|\vecx\|_1 = \rxant E} \sum\limits_{i=1}^{\infty} \log(1+x_i) +  a \|\vecx\|_2}\notag\\
 \quad\quad &=& \min\limits_{N\in\natunum} N\log\lefto(1+\frac{\rxant E}{N}\right) + a \frac{\rxant E}{\sqrt{N}}.
\label{eq:opt-prob-lemma}
\end{IEEEeqnarray}
\end{lemma}
\begin{IEEEproof}
See Appendix~\ref{app:proof-opt-lemma-sum-two}.
\end{IEEEproof}

It remains to show that we can indeed approximate $F^{-1}_{\weightsum(\vecx)} \mathopen{}\big(\error + E^{-1/2}\big)$ by $ - Q^{-1} (\error )$.
Since $\weightsum(\vecx)$ is the normalized sum of $\widetilde{N}_1 + \widetilde{N}_2$ independent random variables, and $\widetilde{N}_2 \to \infty$ as $E\to\infty$,  it is natural to use the central-limit theorem to establish this result.
More precisely, we apply the Berry-Esseen Theorem~\cite[Ch.~XVI.5]{feller70b} to $F_{\weightsum(\vecx)} (\cdot)$ and obtain that, for an arbitrary $\xi\in\realset$,
\begin{IEEEeqnarray}{rCl}
F_{\weightsum(\vecx)}(\xi) &=& \prob\lefto[\frac{1}{\|\vecx\|_2}\left(\tilde{q}_1\sum\limits_{i=1}^{\widetilde{N}_1}S_i + \tilde{q}_2\sum\limits_{i=1}^{ \widetilde{N}_2}S_{i+\widetilde{N}_1}  - \rxant E \right) \leq \xi\right]\label{eq:case3-conv-berry-esseen-00}\\
&\geq& Q\lefto(-\xi\right) - \frac{\mathrm{const} \Big( \widetilde{N}_1 (\tilde{q}_1)^3 + \widetilde{N}_2 (\tilde{q}_2)^3\Big)}{\left(\widetilde{N}_1 (\tilde{q}_1)^2 + \widetilde{N}_2 (\tilde{q}_2)^2\right )^{3/2}}.\label{eq:case3-conv-berry-esseen-01}
\end{IEEEeqnarray}
The second term on the RHS of~\eqref{eq:case3-conv-berry-esseen-01} can be evaluated as follows
\begin{IEEEeqnarray}{rCl}
\frac{\widetilde{N}_1 (\tilde{q}_1)^3 + \widetilde{N}_2 (\tilde{q}_2)^3}{\left(\widetilde{N}_1 (\tilde{q}_1)^2 + \widetilde{N}_2 (\tilde{q}_2)^2\right )^{3/2}} &\leq& \frac{\tilde{q}_2}{\left(\widetilde{N}_1 (\tilde{q}_1)^2 + \widetilde{N}_2 (\tilde{q}_2)^2\right )^{1/2}} \label{eq:case3-conv-berry-esseen-02}\\
&\leq & \widetilde{N}_2^{-1/2}  \label{eq:case3-conv-berry-esseen-03}\\
&\leq&   E^{ -1/12}. \label{eq:case3-conv-berry-esseen-04}
\end{IEEEeqnarray}
Here,~\eqref{eq:case3-conv-berry-esseen-02}  follows because $(\tilde{q}_1)^3 \leq (\tilde{q}_1)^2\tilde{q}_2$, and in~\eqref{eq:case3-conv-berry-esseen-04} we used that $\widetilde{N}_2 > E^{\frac{1}{6}}$.
Using~\eqref{eq:case3-conv-berry-esseen-04} in~\eqref{eq:case3-conv-berry-esseen-01}, selecting $\xi$ such that the LHS of~\eqref{eq:case3-conv-berry-esseen-00} equals $\error + E^{-1/2}$, and using that the function $Q(\cdot)$ is monotonically decreasing, we conclude that
\begin{IEEEeqnarray}{rCl}
 F^{-1}_{\weightsum(\vecx)}\lefto(\error +   E^{-1/2}\right) =\xi
 &\leq& -Q^{-1}\mathopen{}\Big(\error + E^{-1/2} +  \constrm\cdot  E^{-1/12}\Big)\\
&=& - Q^{-1}(\error)  + \bigO\lefto( E^{ -1/12}\right) .\label{eq:case3-conv-berry-esseen-5}
\end{IEEEeqnarray}
Here,~\eqref{eq:case3-conv-berry-esseen-5} follows by applying Taylor's theorem to $Q^{-1}(\cdot)$ around~$\error$.

\subsubsection*{Case 2}
\label{app:case1-3atoms}
By assumption, $\vecx$ contains three distinct nonzero entries $0<q_1 <q_2 <q_3$, and $q_1$ and $q_3$ each  appear only once.
For this case, we shall use a different approach from that used in Case~1. The main differences between the two cases are as follows:
\begin{itemize}
\item In order to use the central limit theorem, we need to show that the $\vecx$ that maximizes $\hat{\eta}_E(\vecx)$ contains sufficiently many nonzero entries, and that the available energy $\rxant E$ is spread evenly over these nonzero entries as $E\to\infty$.
    These properties are satisfied in Case 1 by definition. In Case~2, however, we need to verify that they hold.

\item Intuitively, since $q_1$ and $q_3$ appear only  once in $\vecx$, we expect that they do not contribute to the dominant terms in~\eqref{eq:upper-bd-M-asy-mid}.
As a result, we can approximate the second and the third term on the RHS of~\eqref{eq:upper-bd-M-asy-mid} directly without using Lemma~\ref{lem:opt-two-terms}.

\end{itemize}
We proceed now with the proof.
%
%
The idea is to upper-bound~\eqref{eq:upper-bd-M-asy-mid} using~\eqref{eq:lemma-bound-cdf-normalize-inv} (Lemma~\ref{lem:lb-normalized-cdf} in Appendix~\ref{app:convo-exp-dist}), and then compare the resulting bound with the achievability result~\eqref{eq:ach-proof-end-asy}.
Since $0<\error < 1/2$, and since we are interested in the asymptotic regime $E\to\infty$, we can assume without loss of generality that $\error + E^{-1/2}< 1/2$.
Applying~\eqref{eq:lemma-bound-cdf-normalize-inv} to~\eqref{eq:expansion-eta-E}, we obtain
\begin{IEEEeqnarray}{rCl}
\hat{\eta}_E(\vecx)&\leq& \rxant E\log e - \sum\limits_{i=1}^{\infty} \log(1+x_i) \notag\\
&& - \Big(1/2 -\error - E^{-1/2} \Big) \|\vecx\|_2\log e + \bigO (\log E) . \IEEEeqnarraynumspace
\label{eq:converse-rough-order}
\end{IEEEeqnarray}
Since we are interested in upper-bounding $\sup_{\vecx} \hat{\eta}_E(\vecx)$,  we focus  without loss of generality on the $\vecx$ for which $\hat{\eta}_E(\vecx)$ is greater than the RHS of~\eqref{eq:ach-proof-end-asy}.  By comparing~\eqref{eq:converse-rough-order} with~\eqref{eq:ach-proof-end-asy}, 
we conclude that such $\vecx$ must satisfy
\begin{IEEEeqnarray}{rCl}
\|\vecx\|_2  \leq  \frac{V_0 \cdot  \big( Q^{-1}(\error)\big)^{2/3} }{(1/2 -\error)\log e} \big(\rxant E\big)^{2/3} \big(\log (\rxant E)\big)^{1/3} +  \littleo\lefto(E^{2/3}\right)
\label{eq:ub-x-2norm}
\end{IEEEeqnarray}
and
\begin{IEEEeqnarray}{rCl}
\sum\limits_{i=1}^{\infty} \log(1+x_i)  \leq V_0\cdot \big( \rxant E Q^{-1}(\error)\big)^{2/3} \big(\log (\rxant E)\big)^{1/3} +  \littleo\lefto(E^{2/3}\right).
\label{eq:ub-log-sum}
\end{IEEEeqnarray}

We next refine the bounds \eqref{eq:ub-x-2norm} and~\eqref{eq:ub-log-sum} by exploiting that $\vecx$ is of the form specified in~\eqref{eq:three-atoms-form}.
By~\eqref{eq:ub-x-2norm} and~\eqref{eq:three-atoms-form}, we have the following estimates
\begin{IEEEeqnarray}{rCl}
q_3  = \|\vecx\|_{\infty} \leq \|\vecx \|_2 \leq \bigO\lefto(E^{2/3}\big(\log  E\big)^{1/3} \right) \IEEEeqnarraynumspace
\label{eq:ub-x-infty}
\end{IEEEeqnarray}
and
\begin{IEEEeqnarray}{rCl}
N+2 &\geq& \frac{\|\vecx\|^2_1}{\|\vecx\|^2_2}\label{eq:-lb-n-star-1}\\
&\geq& \mathrm{const} \cdot E^{2/3}\big(\log  E\big)^{-2/3} . \IEEEeqnarraynumspace
\label{eq:-lb-n-star}
\end{IEEEeqnarray}
Here,~\eqref{eq:-lb-n-star-1} follows because $\vecx$ has $N+2$ nonzero entries and because $\|\veca\|_1 \leq \sqrt{N+2} \|\veca\|_2$ for every $(N+2)$-dimensional real vector $\veca$; in~\eqref{eq:-lb-n-star} we used~\eqref{eq:ub-x-2norm} and that $\|\vecx\|_1=\rxant E$.
Since $q_1 + q_2 N +q_3 =\rxant E$, it follows from~\eqref{eq:-lb-n-star} that
\begin{IEEEeqnarray}{rCl}
q_1 &\leq&  q_2 \leq \frac{\rxant E}{N} \leq  \bigO\lefto(E^{1/3}\big(\log E\big)^{2/3} \right). \IEEEeqnarraynumspace
\label{eq:estimate-q1-star}
\end{IEEEeqnarray}
The bound~\eqref{eq:ub-log-sum} implies that
\begin{IEEEeqnarray}{rCl}
\IEEEeqnarraymulticol{3}{l}{
V_0\cdot  \big(\rxant EQ^{-1}(\error)\big)^{2/3}\big(\log(\rxant E)\big)^{1/3}   + o\lefto(E^{2/3}\right)}\notag\\
\quad\quad  &\geq& \sum\limits_{i=1}^{\infty}\log(1+x_i) \label{eq:lb-sum-log-0}\\
&\geq& N \log(1+q_2)  \\
&=& N\log\lefto(1+ \frac{\rxant E-q_1 -q_3}{N}\right)\\
&\geq& N \log\lefto(1+ \frac{\rxant E- \bigO\lefto(E^{2/3}\big(\log E\big)^{1/3} \right) }{N}\right).
\label{eq:lb-sum-log}
\end{IEEEeqnarray}
Here, in~\eqref{eq:lb-sum-log} we used~\eqref{eq:ub-x-infty} and~\eqref{eq:estimate-q1-star}. Solving~\eqref{eq:lb-sum-log} for $N$, we obtain
\begin{IEEEeqnarray}{rCl}
N \leq \mathrm{const} \cdot E^{2/3}\big(\log E\big)^{-2/3}.
\label{eq:ub-n-star}
\end{IEEEeqnarray}
Using~\eqref{eq:-lb-n-star}  and~\eqref{eq:ub-n-star} back in~\eqref{eq:lb-sum-log} we obtain
\begin{IEEEeqnarray}{rCl}
\IEEEeqnarraymulticol{3}{l}{
\sum\limits_{i=1}^{\infty}\log(1+x_i)} \notag \\
&\geq&  N\log\mathopen{}\bigg( \! 1+ \frac{\rxant E}{N} \bigg)+ N \log\mathopen{}\bigg(1- \bigO\mathopen{}\Big(E^{-1/3}\big(\log E\big)^{1/3}\Big)  \bigg)\label{eq:lb-sum-log-x-star-temp}\\
& = & N\log\lefto(1+\frac{\rxant E}{N}\right) + \bigO\lefto(E^{1/3}\big(\log E\big)^{-1/3}\right).
\label{eq:lb-sum-log-x-star}
\end{IEEEeqnarray}
Here, the last step follows by Taylor-expanding the $\log$ function in the second term on the RHS of~\eqref{eq:lb-sum-log-x-star-temp} around $1$.
%

We are now ready to provide a refined estimate for the term $F^{-1}_{\weightsum(\vecx)}\lefto(\error + E^{-1/2}\right)\|\vecx\|_2$ on the RHS of~\eqref{eq:upper-bd-M-asy-mid}.
Let
\begin{equation}
\vecx' \define \Big[ \underbrace{\frac{\rxant E}{N+2}, \ldots, \frac{\rxant E}{N+2}}_{N+2},0,\ldots\Big].
\end{equation}
By Lemma~\ref{lem:telatar-conj} (see Appendix~\ref{app:convo-exp-dist}) and by~\eqref{eq:def-hatS},  the following inequality holds for every $\gamma\in(0,\rxant E]$:
\begin{IEEEeqnarray}{rCl}
F_{\weightsum(\vecx)}\lefto(\frac{\gamma -\rxant E}{\|\vecx\|_2}\right) &=& \prob\mathopen{}\Big[ q_1  S_1 + \sum\limits_{i=2}^{N+1}q_2 S_{i} + q_3 S_{N+2} \leq \gamma\Big]\\
  &\geq&  \prob\mathopen{}\Big[ \frac{\rxant E}{N+2}\sum\limits_{i=1}^{N+2}  S_i \leq \gamma\Big] \label{eq:conv-proof-abbe-00}\\
  &=& F_{\weightsum(\vecx')}\lefto(\frac{\gamma -\rxant E}{\|\vecx'\|_2}\right). \label{eq:conv-proof-abbe-1}
\end{IEEEeqnarray}
Since $\error + E^{-1/2}<1/2$ and since, by Lemma~\ref{lem:lb-normalized-cdf} (see Appendix~\ref{app:convo-exp-dist}), $F_{L(\vecx')}(0)\geq 1/2$,  we have $F^{-1}_{\weightsum(\vecx')}\mathopen{}\big(\error + E^{-1/2}\big) <0$. Set $\gamma=\rxant E +F^{-1}_{\weightsum(\vecx')}\mathopen{}\big(\error + E^{-1/2}\big)\|\vecx'\|_2 < \rxant E$. Then, by~\eqref{eq:conv-proof-abbe-1},
\begin{IEEEeqnarray}{rCl}
F^{-1}_{\weightsum(\vecx)}\mathopen{}\big(\error + E^{-1/2}\big)\|\vecx\|_2 \leq F^{-1}_{\weightsum(\vecx')}\mathopen{}\big(\error + E^{-1/2}\big) \|\vecx'\|_2. \IEEEeqnarraynumspace
\label{eq:lb-f-inverse}
\end{IEEEeqnarray}
Applying the Berry-Esseen central-limit theorem similarly as in~\eqref{eq:case3-conv-berry-esseen-00}--\eqref{eq:case3-conv-berry-esseen-5}, we obtain
\begin{equation}
F^{-1}_{\weightsum(\vecx') }\mathopen{}\big(\error + E^{-1/2}\big) \leq - Q^{-1}(\error ) + \bigO\lefto(\frac{1}{\sqrt{N}}\right).
\label{eq:berry-esseen-case2}
\end{equation}
Furthermore,
\begin{IEEEeqnarray}{rCl}
\|\vecx'\|_2&=& \frac{\rxant E}{\sqrt{N +2}} \\
&=&\frac{\rxant E}{\sqrt{N}} \left( 1+  \bigO\lefto( \frac{1}{N}\right)\right).  \label{eq:F-inverse-inter1}
\end{IEEEeqnarray}
Substituting~\eqref{eq:berry-esseen-case2} and~\eqref{eq:F-inverse-inter1} into~\eqref{eq:lb-f-inverse} and using~\eqref{eq:-lb-n-star} and~\eqref{eq:ub-n-star}, we obtain
\begin{IEEEeqnarray}{rCl}
F^{-1}_{\weightsum(\vecx) }\mathopen{}\big(\error + E^{-1/2}\big) \|\vecx \|_2
&\leq &- \frac{\rxant E}{\sqrt{N}} Q^{-1}(\error )  +  \bigO\lefto( E^{1/3}\big(\log E\big)^{2/3}\right) .
\label{eq:F-inverse}
\end{IEEEeqnarray}
Finally, substituting~\eqref{eq:lb-sum-log-x-star} and~\eqref{eq:F-inverse} into~\eqref{eq:upper-bd-M-asy-mid}, we conclude that
\begin{IEEEeqnarray}{rCl}
\hat{\eta_{E}}(\vecx) &\leq& \rxant E \log e - N\log\lefto(1+\frac{\rxant E}{N}\right) \notag\\
&&- \,\frac{\rxant E \log e}{\sqrt{N}}Q^{-1}(\error) +  \bigO\lefto(E^{1/3}\big(\log E\big)^{-1/3}\right).
\label{eq:proof-converse-asy-case2}
\end{IEEEeqnarray}
The proof is completed by maximizing the RHS of~\eqref{eq:proof-converse-asy-case2} over $N \in \natunum$ and by using~\eqref{eq:N-star-min-two-penalty} and~\eqref{eq:def-N-star}.

\subsubsection*{Case 3} By assumption, $\vecx$ has at most two different nonzero entries $0\leq \tilde{q}_1 \leq \tilde{q}_2$, and $\widetilde{N}_2 \leq E^{1/6} $.
Since the multiplicity of $q_2$ in $\vecx$ is less than $E^{1/6}$, it can be shown that all entries of~$\vecx$ that are equal to~$\tilde{q}_2$ do not contribute to the dominant terms in~\eqref{eq:upper-bd-M-asy-mid}. The analysis follows steps similar to the ones for Case 2.

\subsection{Proof of Lemma~\ref{lem:opt-two-terms}}
\label{app:proof-opt-lemma-sum-two}
Let
\begin{IEEEeqnarray}{rcl}
\sumlog (\vecx) \define \sum\limits_{i=1}^{\infty} \log(1+x_i) +  a \|\vecx\|_2
\end{IEEEeqnarray}
with $x_i$ standing for the $i$th entry of $\vecx$, and let $\vecx^*$ be a minimizer of
\begin{IEEEeqnarray}{rCl}
\inf_{\vecx\in\posrealset^\infty: \|\vecx\|_1 = \rxant E} \sumlog (\vecx).
\end{IEEEeqnarray}
In order to prove Lemma~\ref{lem:opt-two-terms}, it suffices to show that all nonzero entries of $\vecx^*$ must take the same value.
This is proved by contradiction.

Assume that there exist indices $i, j$ for which $0 < x_i^* < x_j^*$.
Let $b \define x_i^* + x_j^*$, $c\define \sum\nolimits_{k \neq i, k\neq j}^{n} (x_k^*)^2$, and $d\define \sum\nolimits_{k \neq i, k\neq j}^{n} \log(1+x_k^*)$. Consider now the function $f: [0,b] \to \realset$ defined as
\begin{IEEEeqnarray}{rCl}
f(t) & \define &  \log(1+t) + \log(1+b -t) \notag\\
&& +\, a\sqrt{c+t^2 + (b-t)^2} + d.
\label{eq:def-ft-opt-lemma}
\end{IEEEeqnarray}
Note that $f(t)$ is symmetric around $t = b/2$, and that $f(x_i^*) = f(x_j^*) = \sumlog(\vecx^*)$.

Standard computations reveal that  the minimum of $f(t)$ over $[0,b/2]$ is achieved at one of the boundary points, i.e.,
\begin{IEEEeqnarray}{rCl}
f( t ) > \min\{f(0) , f(b/2)\}, \,\text{for all } t \in(0,b/2).
\label{eq:ft-f0-fend}
\end{IEEEeqnarray}
Let $\vecx_1$ (resp. $\vecx_2$) be the vector obtained from $\vecx^*$ by replacing the $i$th and $j$th entries with $0$ (resp. $b/2$) and $b$ (resp. $b/2$), respectively.
Clearly, $\|\vecx_1\|_1 = \|\vecx_2\|_1 = \rxant E$.
Then,~\eqref{eq:ft-f0-fend} implies that
\begin{equation}
\sumlog(\vecx^*) > \min\{\sumlog(\vecx_1),\sumlog(\vecx_2)\}.
\end{equation}
This contradicts the assumption that $\vecx^*$ is a minimizer.
Therefore, the entries of~$\vecx^*$ cannot take more than one distinct nonzero values.

\section{Proof of Theorem~\ref{thm:asy-coh}}
\label{app:proof-asy-coh}
The achievability of~\eqref{eq:second-order-coh} follows from Theorem~\ref{thm:nonasy-coh-ach} and~\cite[Th.~3]{polyanskiy11-08b}. Next, we prove a converse.
As in the proof of Theorem~\ref{thm:non-asy-converse-inf}, we assume without loss of generality that each codeword~$\matU^\infty$ for the channel~\eqref{eq:MIMO-channel-io} satisfies the equal-energy constraint
\begin{equation}
\fnorm{\matU^\infty}^2 = E.
\label{eq:power-constraint-coh}
\end{equation}
Let $P_{\randmatV^\infty\randmatH^\infty\given \randmatU^\infty} \define P_{\randmatH^\infty}P_{\randmatV^\infty\given \randmatU^\infty\randmatH^\infty}$.
By the meta-converse theorem~\cite[Th.~31]{polyanskiy10-05} applied with $Q_{\randmatV^\infty\randmatH^\infty} = P_{\randmatV^\infty\randmatH^\infty\given \randmatU^\infty =\mathbf{0}}$, we obtain
\begin{IEEEeqnarray}{rCl}
\frac{1}{M^*(E,\error)} \geq \inf \limits_{\matU^\infty \in \complexset^{\infty\times \txant}: \fnorm{\matU^\infty}^2 = E }\beta_{1-\error}(P_{\randmatV^\infty\randmatH^\infty \given \randmatU^\infty = \matU^\infty } , Q_{\randmatV^\infty\randmatH^\infty}).
\label{eq:meta-converse-csir}
\end{IEEEeqnarray}
Proceeding similarly to the proof of Theorem~\ref{thm:non-asy-converse-inf}, we observe that the RHS of~\eqref{eq:meta-converse-csir} does not change if we
focus on diagonal input matrices. This implies that for the purpose of evaluating~\eqref{eq:meta-converse-csir}, the MIMO Rayleigh block-fading channel~\eqref{eq:MIMO-channel-io} is equivalent to the  memoryless SIMO Rayleigh-fading channel~\eqref{eq:SIMO-channel-io}.
Let now  $\vecu$ and $(\randmatV,\randmatH)$ denote the input and the output of this SIMO channel, respectively.
Then, the RHS of~\eqref{eq:meta-converse-csir} is equal to
\begin{IEEEeqnarray}{rCl}
\inf \limits_{\vecu \in \complexset^{\infty}: \|\vecu\|_2^2 = E }\beta_{1-\error}(P_{\randmatV\randmatH \given \randvecu = \vecu } , Q_{\randmatV\randmatH})
\label{eq:meta-converse-csir-simo}
\end{IEEEeqnarray}
where $P_{\randmatV\randmatH \given \randvecu = \vecu }$ denotes the conditional probability distribution of the output of the channel~\eqref{eq:SIMO-channel-io} given the input, and $Q_{\randmatV\randmatH} = P_{\randmatV\randmatH\given \randmatU=\mathbf{0}}$.
Substituting~\eqref{eq:meta-converse-csir-simo} into~\eqref{eq:meta-converse-csir}, and using the lower bound~\cite[Eq.~(102)]{polyanskiy10-05}, we obtain that for every $\eta >0$,
\begin{IEEEeqnarray}{rCl}
&&\log M^*(E,\error) \leq \eta\log e 
 - \log\mathopen{}\Bigg|\inf\limits_{\vecu \in \complexset^{\infty}: \|\vecu\|_2^2 = E}
  P_{\randmatV\randmatH\given\randvecu=\vecu}\mathopen{}\Bigg[\log\frac{dP_{ \randmatV\randmatH \given \randvecu=\vecu} }{dQ_{\randmatV\randmatH }}(\randmatV,\randmatH ) \leq \eta\log e  \Bigg] -\error\Bigg|^{+}. \notag\\
  \label{eq:conv-coh-meta-converse}
\end{IEEEeqnarray}
Under $P_{\randmatV\randmatH\given\randvecu=\vecu}$, the random variable~$\log\frac{dP_{\randmatV\randmatH\given\randvecu=\vecu} }{dQ_{\randmatV\randmatH }}(\randmatV,\randmatH ) $  in~\eqref{eq:conv-coh-meta-converse} has the same distribution as
\begin{equation}
\log e \bigg(\sum\limits_{r=1}^{\rxant}\sum\limits_{i=1}^{\infty}|u_i H_{r,i}|^2 + \!\bigg(\!2\sum\limits_{r=1}^{\rxant}\sum\limits_{i=1}^{\infty}|u_i H_{r,i}|^2\!\bigg)^{1/2}Z\bigg)
\label{eq:dist-info-den-coh-siso}
\end{equation}
where $Z\sim\mathcal{N}(0,1)$.

Let now $\epsilon_E \define \error +  c_1\sqrt{E^{-1} \log E} $, where $c_1>0$ is an arbitrary constant.
Since, by assumption,  $0<\error<1/2$, and since we are interested in the asymptotic behavior of $\log M^*(E,\error) $ as $E\to\infty$, we can assume without loss of generality that $\error_E <1/2$.
Set
\begin{equation}
\eta = \rxant E -\sqrt{2\rxant E}\,Q^{-1}(\error_E ).
\label{eq:def-eta-coh}
\end{equation}
Then, we can rewrite the minimization problem on the RHS of~\eqref{eq:conv-coh-meta-converse}  using~\eqref{eq:dist-info-den-coh-siso} and~\eqref{eq:def-eta-coh} as follows
\begin{IEEEeqnarray}{rCl}
\IEEEeqnarraymulticol{3}{l}{
\inf\limits_{\substack{\vecu\in\complexset^\infty: \\ \|\vecu\|_2^2 =E}} P_{\randmatV\randmatH\given\randvecu=\vecu}\mathopen{}\Bigg[\log\frac{dP_{ \randmatV\randmatH \given \randvecu=\vecu} }{dQ_{\randmatV\randmatH }}(\randmatV,\randmatH ) \leq \eta \log e \Bigg]
}\notag\\
&=& \inf\limits_{\substack{\vecu\in\complexset^\infty: \\ \|\vecu\|_2^2 =E}}\!\! \Ex{}{Q\lefto( \frac{\sum\limits_{r=1}^{\rxant} \sum\limits_{i=1}^{\infty}|u_iH_{r,i}|^2 - \rxant E +\sqrt{2\rxant E}Q^{-1}(\epsilon_E ) }{\sqrt{2\sum\limits_{r=1}^{\rxant}\sum\limits_{i=1}^{\infty}|u_iH_i|^2}}\right)} \label{eq:conv-coh-inf-prob1-def}\\
&\define&  q(E).\IEEEeqnarraynumspace
\label{eq:conv-coh-inf-prob1}
\end{IEEEeqnarray}
We next show that $q(E)$ admits the following large-$E$ expansion:
\begin{IEEEeqnarray}{rCl}
q(E) = \inf\limits_{\substack{\vecu\in\complexset^\infty:\\ \|\vecu\|_2^2 =1}}\prob\lefto[\sqrt{\frac{E}{2\rxant}}\bigg(\sum\limits_{r=1}^{\rxant}\sum\limits_{i=1}^{\infty}|u_iH_{r,i}|^2 -\rxant \bigg)\leq Z - Q^{-1}(\epsilon_E)\right] + \bigO\lefto(\sqrt{\frac{\log E}{E}}\right).\IEEEeqnarraynumspace
\label{eq:conv-coh-inf-prob2}
\end{IEEEeqnarray}
The key step is to replace the term $\sqrt{2\sum\limits_{r=1}^{\rxant}\sum\limits_{i=1}^{\infty}|u_iH_i|^2}$ in the denominator on the RHS of~\eqref{eq:conv-coh-inf-prob1-def} by $\sqrt{2\rxant E}$.
To this end, consider the function $t\mapsto Q((t-\eta )/\sqrt{2t})$ with $\eta$ given in~\eqref{eq:def-eta-coh}. If $t\leq \eta - 2\sqrt{ \rxant E\log E}$, we have
\begin{IEEEeqnarray}{rCl}
1&\geq& Q\lefto(\frac{t - \eta  }{\sqrt{2 t}}\right) \label{eq:bound-q-fraction-1}\\
&\geq&  Q\lefto(\frac{t-\eta}{\sqrt{2\rxant E}} \right) \label{eq:bound-q-fraction-0}\\
&\geq&  Q\lefto( -\frac{2\sqrt{\rxant E\log E} }{\sqrt{2\rxant E}}\right)  \IEEEeqnarraynumspace\\
& =& 1-\bigO\lefto(E^{-1}\right).\label{eq:bound-q-fraction-2}
\end{IEEEeqnarray}
Here,~\eqref{eq:bound-q-fraction-0} follows because $Q(\cdot)$ is monotonically decreasing.
The inequality~\eqref{eq:bound-q-fraction-2} implies that, if $t\leq \eta - 2\sqrt{ \rxant E\log E}$, then
\begin{equation}
\left|Q\lefto(\frac{t - \eta}{\sqrt{2t}}\right) -  Q\lefto(\frac{t - \eta} {\sqrt{2\rxant E}}\right) \right|\leq \bigO(E^{-1}).
\label{eq:bound-q-fraction-3}
\end{equation}
Proceeding similarly as in~\eqref{eq:bound-q-fraction-1}--\eqref{eq:bound-q-fraction-2}, we can show that~\eqref{eq:bound-q-fraction-3} holds also if $t\geq \eta+ 2\sqrt{\rxant E \log E}$.
Finally, if $|t - \eta| <2\sqrt{\rxant E \log E}$, by the mean-value theorem~\cite[p.~107]{rudin76} there exists an $a_0\in\big[(t -\eta)/\sqrt{2t} , (t - \eta)/\sqrt{2\rxant E}\big]$ such that
\begin{IEEEeqnarray}{rCl}
\IEEEeqnarraymulticol{3}{l}{
  \left|Q\lefto(\frac{t - \eta}{\sqrt{2t}}\right) -  Q\lefto(\frac{t - \eta} {\sqrt{2\rxant E}}\right) \right|}\notag\\
   \quad&=& |Q'(a_0)| \left| \frac{t-\eta}{\sqrt{2t}} - \frac{t-\eta}{\sqrt{2\rxant E}}\right|  \\
  &= & \frac{1}{\sqrt{2\pi}} e^{-{a_0^2}/{2}} \left|\frac{t - \eta }{\sqrt{2 \rxant E}} \right| \bigO\lefto( \sqrt{\frac{\log E}{E}}\right) \label{eq:bound-q-fraction-5} \\
 & = &  \frac{1}{\sqrt{2\pi}} \underbrace{e^{-\frac{(t -\eta )^2}{4 \rxant E} \cdot\left(1 + \bigO\lefto(\sqrt{\frac{\log E}{E}}\right)\right)} \left|\frac{t - \eta }{\sqrt{2 \rxant E}} \right|}_{\leq \mathrm{const}} \bigO\lefto( \sqrt{\frac{\log E}{E}}\right) \IEEEeqnarraynumspace \label{eq:bound-q-fraction-6}\\
 & =  & \bigO\lefto( \sqrt{\frac{\log E}{E}} \right). \label{eq:bound-q-fraction-4}
\end{IEEEeqnarray}
Here,~\eqref{eq:bound-q-fraction-5} and~\eqref{eq:bound-q-fraction-6} follow because
\begin{IEEEeqnarray}{rCl}
\left|a_0 - \frac{t-\eta }{\sqrt{2 \rxant E}}\right| &\leq& \left| \frac{t-\eta }{\sqrt{2t}} - \frac{t-\eta}{\sqrt{2\rxant E}}\right| \\
&=& \left|\frac{t - \eta }{\sqrt{2\rxant E}} \right| \bigO\lefto(\sqrt{ \frac{\log E}{E}}\right).\IEEEeqnarraynumspace
\end{IEEEeqnarray}
Combining~\eqref{eq:bound-q-fraction-3} and~\eqref{eq:bound-q-fraction-4}, we conclude that for every $t\in(0,\infty)$
\begin{IEEEeqnarray}{rCl}
Q\lefto(\frac{t - \eta }{\sqrt{2t}}\right) = Q\lefto(\frac{t -\eta }{\sqrt{2\rxant E}}\right) + \bigO\lefto( \sqrt{\frac{\log E}{E}}\right)
\label{eq:conv-coh-approx-q}
\end{IEEEeqnarray}
where the $\bigO(\sqrt{(\log E)/E})$ term is uniform in $t$.
This means that replacing the denominator in~\eqref{eq:conv-coh-inf-prob1} with $\sqrt{2\rxant E}$ affects the value of~\eqref{eq:conv-coh-inf-prob1} only by $\bigO\mathopen{}\big(\sqrt{(\log E)/E}\big)$.
Finally, we establish~\eqref{eq:conv-coh-inf-prob2} by using~\eqref{eq:conv-coh-approx-q} in~\eqref{eq:conv-coh-inf-prob1} and by normalizing $\vecu$ in \eqref{eq:conv-coh-inf-prob1} with respect to $E$.

Lemma~\ref{lem:optimization-exp-q} below characterizes the solution of the optimization problem in~\eqref{eq:conv-coh-inf-prob2}.
\begin{lemma}
\label{lem:optimization-exp-q}
Fix an arbitrary $a>0$, $b>0$, and $\rxant \in \natunum$.  Let $\{H_{r,i}\}$ be i.i.d. $\jpg(0,1)$-distributed and let $Z\sim\mathcal{N}(0,1)$ be independent of $\{H_{r,i}\}$.
Then, we have
\begin{IEEEeqnarray}{rCl}
\inf\limits_{\substack{\vecu\in\complexset^\infty:\\
 \|\vecu\|_2^2 =1}} \!\! \prob\lefto[a \bigg(\sum\limits_{r=1}^{\rxant}\sum\limits_{i=1}^{\infty}|u_iH_{r,i}|^2 -\rxant\bigg) \leq Z -  b\right] = Q(b).\IEEEeqnarraynumspace
\label{eq:optimization-exp-sol}
\end{IEEEeqnarray}
\end{lemma}
\begin{IEEEproof}
See Appendix~\ref{app:proof-opt-qe-lemma}.
\end{IEEEproof}

Using Lemma~\ref{lem:optimization-exp-q} in~\eqref{eq:conv-coh-inf-prob2}, we obtain
\begin{IEEEeqnarray}{rCl}
q(E)
&=&  \error_E + \bigO\lefto(\sqrt{\frac{\log E}{E}}\right). \IEEEeqnarraynumspace
\label{eq:conv-coh-inf-prob1-after-opt}
\end{IEEEeqnarray}
Finally, substituting~\eqref{eq:conv-coh-inf-prob1-after-opt} and~\eqref{eq:def-eta-coh} into~\eqref{eq:conv-coh-meta-converse}, we conclude that
\begin{IEEEeqnarray}{rCl}
\log M &\leq&
\rxant E\log e- \sqrt{2\rxant E} Q^{-1}\lefto(\error_E\right)\log e - \log\mathopen{} \bigg(\error_E -\error
+\bigO\mathopen{}\bigg(\!\sqrt{\frac{\log E}{E}}\bigg)\!\bigg) \IEEEeqnarraynumspace \\
&=& \rxant E\log e-\sqrt{2\rxant E} Q^{-1}\lefto(\error  + c_1\sqrt{\frac{\log E}{E}}\right)\log e\notag\\
&&-\, \log \lefto(c_1\sqrt{\frac{\log E}{E}} +\bigO\lefto(\!\sqrt{\frac{\log E}{E}}\right)\right) \IEEEeqnarraynumspace\\
&\leq & \rxant E\log e  - \sqrt{2\rxant E}Q^{-1}(\error )\log e +\frac{1}{2}\log E + \bigO(\sqrt{\log E}).
\end{IEEEeqnarray}
Here, the last step follows by Taylor-expanding $Q^{-1}(\cdot )$ around $\error$, and by taking $c_1$ so that
\begin{equation}
c_1\sqrt{\frac{\log E}{E}} +\bigO\lefto(\!\sqrt{\frac{\log E}{E}}\right) \geq \sqrt{\frac{\log E}{E}}.
\end{equation}
This concludes the proof.

\subsection{Proof of Lemma~\ref{lem:optimization-exp-q}}
\label{app:proof-opt-qe-lemma}
First, consider the following sequence of vectors indexed by $N$:
\begin{IEEEeqnarray}{rCl}
\vecu^{(N)} \define \frac{1}{\sqrt{N}}\big[\underbrace{1,\ldots,1}_{N},0,\ldots\big].
\label{eq:def-u-n-proof-lemma-opt}
\end{IEEEeqnarray}
Evaluating the probability on the LHS of~\eqref{eq:optimization-exp-sol} for this sequence of vectors, we establish the following upper bound
\begin{IEEEeqnarray}{rCl}
\IEEEeqnarraymulticol{3}{l}{
\inf\limits_{\substack{\vecu\in\complexset^\infty:\\
 \|\vecu\|_2^2 =1}} \!\! \prob\lefto[a \bigg(\sum\limits_{r=1}^{\rxant}\sum\limits_{i=1}^{\infty}|u_iH_{r,i}|^2 -\rxant\bigg) \leq Z -  b\right] }\notag\\
&\leq&  \lim\limits_{ N\to\infty} \!\! \prob\lefto[a \bigg(\frac{1}{N}\sum\limits_{r=1}^{\rxant}\sum\limits_{i=1}^{N}|H_{r,i}|^2 -\rxant \bigg) \leq  Z -b\right]\\
 &=& Q(b).
\label{eq:optimization-exp-ach}
\end{IEEEeqnarray}
Here, the last step follows by the law of large numbers.

Next, we prove the reverse inequality.
Suppose that for every $m\in \natunum $ and every $\vecu\in\complexset^m$ that satisfies $\|\vecu\|_2^2 =1$, the following equality holds:
\begin{IEEEeqnarray}{rCl}
\inf\limits_{a\geq 0} \prob\lefto[a \bigg(\sum\limits_{i=1}^{m}|u_iH_i|^2 -1\bigg) \leq Z -  b\right] = Q(b).
\label{eq:opt-exp-simplified}
\end{IEEEeqnarray}
Then,
\begin{IEEEeqnarray}{rCl}
\IEEEeqnarraymulticol{3}{l}{
\inf\limits_{\vecu\in\complexset^\infty:
 \|\vecu\|_2^2 =1} \prob\lefto[a \bigg(\sum\limits_{r=1}^{\rxant}\sum\limits_{i=1}^{\infty}|u_iH_{r,i}|^2 -\rxant\bigg) \leq Z -  b\right]}  \label{eq:optimization-exp-sol-2--1}\\
&= &  \inf\limits_{ \substack{ \vecu\in\complexset^\infty: 
 \|\vecu\|_2^2 =1 \\\  u_{jm_{\mathrm{r}}+1 } =\cdots = u_{(j+1)m_{\mathrm{r}}} , \forall j \in \amsbb{N}  } } \prob\lefto[ m_{\mathrm{r}} a \bigg(\sum\limits_{i=1}^{\infty}|u_iH_{i}|^2 -1\bigg) \leq Z -  b\right] \label{eq:optimization-exp-sol-2-0}\\ 
&\geq& \inf\limits_{\vecu\in\complexset^\infty:
 \|\vecu\|_2^2 =1} \prob\lefto[\rxant a \bigg(\sum\limits_{i=1}^{\infty}|u_iH_{i}|^2 -1\bigg) \leq Z -  b\right] \label{eq:optimization-exp-sol-2-1}\\
&\geq &
\inf\limits_{a\geq 0, \vecu\in\complexset^\infty:
 \|\vecu\|_2^2 =1} \prob\lefto[a \bigg(\sum\limits_{i=1}^{\infty}|u_iH_i|^2 -1\bigg) \leq Z -  b\right]
\IEEEeqnarraynumspace
\label{eq:optimization-exp-sol-0}\\
&=& \lim\limits_{m\to\infty} \inf_{\substack{ a\geq 0, \vecu\in\complexset^m:\\
 \|\vecu\|_2^2 =1}} \prob\lefto[a \bigg(\sum\limits_{i=1}^{m}|u_iH_i|^2 -1\bigg) \leq Z -  b\right]\\
&=& Q(b).\label{eq:optimization-exp-sol-2}
\end{IEEEeqnarray}
Here,~\eqref{eq:optimization-exp-sol-2-0} follows because $\{H_{r,i}\}$ are independent and identically distributed. This allows us to merge the double summation in~\eqref{eq:optimization-exp-sol-2--1} into one summation, provided that we account for the fact that each $u_i$ must now multiply $\rxant$ successive $\{H_i\}$ (see the additional constraint on~\eqref{eq:optimization-exp-sol-2-0}).
The inequality~\eqref{eq:optimization-exp-sol-2-1} follows by enlarging  the feasible region of the minimization problem on the RHS of~\eqref{eq:optimization-exp-sol-2-0}.

We next prove~\eqref{eq:opt-exp-simplified}.
Through standard algebraic manipulations, it can be verified that~\eqref{eq:opt-exp-simplified} holds when $m=1$.
Fix now an arbitrary $m\geq 2$ and an arbitrary $\vecu\in\complexset^m$ that satisfies $\|\vecu\|_2^2 =1$. Assume without loss of generality that all entries of $\vecu$ are positive (otherwise just set $m$ to be the number of positive entries in $\vecu$).
Let $B\define \sum\nolimits_{i=1}^{m}|u_i H_i|^2$, and let
\begin{IEEEeqnarray}{rCl}
g (a) \define \prob\mathopen{}\big[a (B - 1) \leq Z -  b\big] =\Ex{}{Q\mathopen{}\big(aB-a+b\big)}.
\end{IEEEeqnarray}
Since $g(0) = Q(b)$, it suffices to show that $g(a)$ is nondecreasing on $[0,\infty)$, i.e., $g'(a)\geq 0$ for all $a\in[0,\infty)$.
The derivative $g'(a)$ is given by
\begin{IEEEeqnarray}{rCl}
g'(a)& =&  \frac{d}{da} \Big( \Ex{B}{Q(aB -a + b)}\Big)\\
&=& - \int\nolimits_{-1}^{\infty}  \frac{ t}{\sqrt{2\pi}}e^{-\frac{(at)^2 + 2 a t b + b^2}{2}}  f_{B-1}(t) dt. \label{eq:compute-deriv-g-1}
\end{IEEEeqnarray}
Here,~in~\eqref{eq:compute-deriv-g-1} we used the Leibniz's integration rule~\cite{flanders73-06} and the identity  $Q'(x) = -\frac{1}{\sqrt{2\pi}} e^{-x^2/2}$.
The RHS of~\eqref{eq:compute-deriv-g-1} is equal to zero when $a=0$ because, by definition, $\Ex{}{B-1} = 0$.
When $a>0$, we have
\begin{IEEEeqnarray}{rCl}
g'(a)&\geq & - \frac{e^{-b^2/2}}{\sqrt{2\pi}}  \int\nolimits_{-1}^{\infty} t  e^{-\frac{(at)^2 }{2}}  f_{B-1}(t) dt\label{eq:compute-deriv-g-2} \\
&=& - \frac{e^{-b^2/2}}{a\sqrt{2\pi}} \int\nolimits_{-1}^{\infty} e^{-\frac{(at)^2}{2}} f_{B-1}'(t) dt \label{eq:compute-deriv-g-3}\\
&=& -\frac{ e^{-b^2/2}}{a^2} f_{B+Z/a}'(1) \label{eq:compute-deriv-g-4}.
\end{IEEEeqnarray}
Here,~\eqref{eq:compute-deriv-g-2} follows because $e^{-a b t} t \leq t$ for every $t\in\realset$; in~\eqref{eq:compute-deriv-g-3} we used integration by parts and that $f_{B-1}(-1) =0$.

It remains to show that $f_{B+Z/a}'(1) \leq 0$ for every $a > 0$.
Since $Z\sim \mathcal{N}(0,1)$, by the central limit theorem for densities (see~\cite[Th.~VII.2.7]{petrov75}), the pdf of $Z$ can be approximated  to an arbitrary precision by the pdf of a sum of i.i.d. $\expdist(1)$-distributed random variables.
Moreover,~$B$ is the convolution of finitely many exponential distributions and $\Ex{}{B+Z/a} = 1$. Hence, to prove $f_{B+Z/a}'(1) \leq 0$, it suffices to show that the derivative of the convolution of finitely many exponential pdfs computed at the mean value of the resulting distribution is nonpositive.
This follows from Lemma~\ref{lem:negative-derivative} (see Appendix~\ref{app:convo-exp-dist}).

\section{Proof of Theorem~\ref{thm:nonasy-conv-coh-epb}}
\label{app:proof-converse-csir-nonasy}
Let $\eta>0$ be an arbitrary constant and let the function $q_{\eta}(\cdot) $ be defined as follows:
\begin{IEEEeqnarray}{rCl}
q_{\eta}(x) \define Q\lefto(\frac{x-\eta}{\sqrt{2x}}\right), \quad x>0.
\label{eq:def-q-eta-appendix}
\end{IEEEeqnarray}
It follows from~\eqref{eq:conv-coh-meta-converse},~\eqref{eq:dist-info-den-coh-siso}, and~\eqref{eq:conv-coh-inf-prob1} that every $(E,M,\error)$-code for the MIMO Rayleigh block-fading channel~\eqref{eq:MIMO-channel-io} for the case of perfect CSIR satisfies
\begin{IEEEeqnarray}{rCl}
\log M \leq \eta\log e -\log\lefto|\inf\limits_{\vecu \in \complexset^\infty: \|\vecu\|_2^2 = E}\Ex{}{q_\eta\lefto(\sum\limits_{r=1}^{\rxant}\sum\limits_{i=1}^{\infty}|u_i H_{r,i}|^2 \right)} -\error\right|^{+}.
\label{eq:conv-csir-nonasy-1}
\end{IEEEeqnarray}
Suppose that the function $g_{\eta}(\cdot)$ defined in~\eqref{eq:convex-lb} is convex on $[0,\infty)$, and that
\begin{equation}
q_{\eta}(x) \geq g_{\eta}(x), \quad \text{for all } x\in[0,\infty).
\end{equation}
In other words, suppose that $g_{\eta}(\cdot)$ is a convex lower bound on $q_{\eta}(x)$.
Then,~\eqref{eq:converse-csir-nonasymp} follows because, for every   $\vecu \in \complexset^\infty$ with $\|\vecu\|_2^2 = E$,
\begin{IEEEeqnarray}{rCl}
\Ex{}{q_\eta\lefto(\sum\limits_{r=1}^{\rxant}\sum\limits_{i=1}^{\infty}|u_i H_{r,i}|^2 \right)} &\geq&
\Ex{}{g_\eta\lefto(\sum\limits_{r=1}^{\rxant}\sum\limits_{i=1}^{\infty}|u_i H_{r,i}|^2 \right)}\\
&\geq & g_{\eta}\lefto( \Ex{}{\sum\limits_{r=1}^{\rxant}\sum\limits_{i=1}^{\infty}|u_i H_{r,i}|^2 }  \right) \label{eq:avg-g-eta-jensen}\\
 &=& g_{\eta}(\rxant E).
\end{IEEEeqnarray}
Here,~\eqref{eq:avg-g-eta-jensen} follows from Jensen's inequality.

\begin{figure}[t]
\centering
\includegraphics[scale=0.9]{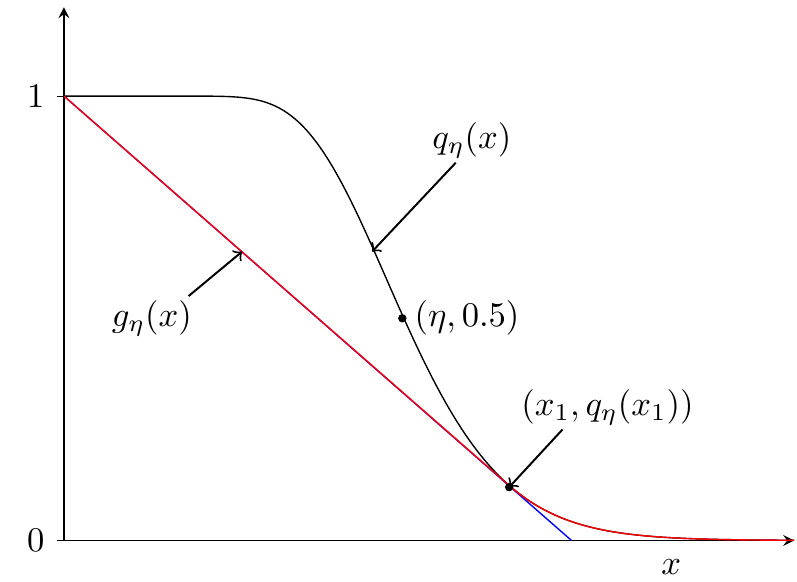}
\caption{\label{fig:convex-envelop}A geometric illustration of $q_{\eta}(\cdot)$ in~\eqref{eq:def-q-eta-appendix} (black curve), of the tangent line of $q_{\eta}(\cdot)$ (blue curve), and of $g_{\eta}(\cdot)$ in~\eqref{eq:convex-lb}  (red curve).}
\end{figure}
It remains to prove that $g_{\eta}(x)$ is indeed a convex lower bound on~$q_{\eta}(x)$.
Observe the following properties of $q_{\eta}(\cdot)$, which can be verified through standard algebraic manipulations:
\begin{itemize}
\item $q_{\eta}(\cdot)$ is monotonically decreasing;
\item $\lim\limits_{x\to 0}q_{\eta}(x) =1, \quad \lim\limits_{x\to\infty} q_{\eta}(x) =0;$
\item $q_{\eta}(\eta) =1/2$;
\item if $\eta>\pi$, then $q_{\eta}'(\eta) = -1/(2\sqrt{\eta\pi}) < -1/(2\eta)$;
\item if $\eta>6$, there exists an $0<x_0<\eta$ such that $q_{\eta}(\cdot)$ is concave on $(0,x_0)$ and convex on $(x_0,\infty)$.
\end{itemize}
Assume that $\eta>6$. Then, the above properties of $q_{\eta}(\cdot)$ imply that there exists a unique $x_1$ such that the line connecting $(0,1)$ and $(x_1,q_{\eta}(x_1))$ lies below the graph of $q_{\eta}(x)$ and is tangent to $q_{\eta}(\cdot)$ at $(x_1,q_{\eta}(x_1))$ (see~Fig.~\ref{fig:convex-envelop}).
Since the slope of the line connecting $(0,1)$ and $(x_1,q_{\eta}(x_1))$ is
\begin{equation}
 - \frac{1}{x_1} Q\lefto(\frac{\eta - x_1}{\sqrt{2x_1}}\right)
\end{equation}
 and since the derivative of $q_{\eta}(x)$ at $x=x_1$ is given by
 \begin{equation}
-\frac{1}{x_1} \frac{1}{4\sqrt{\pi}} e^{- (x_1-\eta)^2/(4x_1)} \left(\frac{\eta}{\sqrt{x_1} } +\sqrt{x_1}\right)
 \end{equation}
it follows that  $x_1$ is the solution of~\eqref{eq:def-x1-tangent}.
 Furthermore, since $q_{\eta}'(\eta) < -1/(2\eta)$, and since $ -1/(2\eta)$ is the slope of the line connecting $(0,1)$ and $(\eta,1/2)$, we have that $x_1>\eta$.
Observe now that $g_{\eta}(x)$ coincides with the line connecting $(0,1)$ and $(x_1,q_{\eta}(x_1))$ for $x\leq x_1$, and that it coincides with $q_{\eta}(x)$ if $x\geq x_1$. This proves that $g_{\eta}(x)$ is indeed a convex lower bound on $q_{\eta}(x)$.

%


\end{document}